%% file: main_part1.tex
\documentclass[aps,pre,floatfix,longbibliography,nofootinbib]{revtex4-2} 

\usepackage{booktabs}
\usepackage{amsmath}  
\usepackage{xcolor}  
\usepackage{graphicx}
\usepackage{subcaption}
\usepackage{diagbox}
\usepackage{adjustbox}
\usepackage{float}
\usepackage{lipsum}

\usepackage{hyperref}
\usepackage{tikz}
\usetikzlibrary{shapes.geometric} 
\usepackage{amssymb} 
\usepackage{adjustbox}  
\usepackage{array}
\begin{document}


\title{\textbf{General-purpose Data-driven Wall Model for Low-speed Flows \\ Part I: Baseline Model}}%



\author{Yuenong Ling}
\email{Contact author: lingyn@mit.edu}
\affiliation{%
 Department of Aeronautics and Astronautics,
 Massachusetts Institute of Technology,
 Cambridge, MA 01239, USA 
}%

\author{Imran Hayat}
\affiliation{%
 Graduate Aerospace Laboratories,
 California Institute of Technology, 
 Pasadena, CA 91125, USA
}%

\author{Konrad Goc}
\affiliation{%
The Boeing Company, Everett, 98204 WA, USA
}%

\author{Adrian Lozano-Duran}%
\affiliation{%
  Department of Aeronautics and Astronautics, Massachusetts Institute of Technology,
  Cambridge, MA 02139, USA
}%
\affiliation{
 Graduate Aerospace Laboratories,
 California Institute of Technology, 
 Pasadena, CA 91125, USA
}
%


\begin{abstract}
We present a general-purpose wall model for large-eddy simulation
(LES). The model builds on the building-block flow
principle~\citep{lozano-duranMachineLearningBuildingblockflow2023},
leveraging essential physics from simple flows to train a
generalizable model applicable across complex geometries and flow
conditions. The model addresses key limitations of traditional
equilibrium wall models (EQWM) and improves upon shortcomings of
earlier building-block--based approaches. The model comprises four
components: (i) a baseline wall model, (ii) an error model, (iii) a
classifier, and (iv) a confidence score. The baseline model predicts
the wall-shear stress, while the error model estimates epistemic
errors (due to missing or unrepresented physics) and aleatoric errors
(stemming from unpredictable physics), both used for uncertainty
quantification. When appropriate, the error model corrects the
baseline prediction, enabling additive learning without retraining or
overfitting. The classifier identifies the flow regime and the
training data used to make predictions, improving the explainability
of the model. Finally, the confidence score quantifies prediction
reliability, flagging instances of extrapolation or operation outside
the range of applicability of the baseline model.
In Part~I of this work, we present the baseline model, while the
remaining three components are introduced in Part~II. The baseline
model is designed to capture a broad range of flow phenomena,
including turbulence over curved walls and zero, adverse, and
favorable mean pressure gradients (PGs), as well as flow separation
and laminar flow. The problem is formulated as a regression task to
predict wall shear stress using a feedforward artificial neural
network. Model inputs are localized in space and dimensionless, with
their selection guided by information-theoretic criteria. Training
data include, among other cases, a newly generated direct numerical
simulation dataset of turbulent boundary layers under favorable and
adverse PG conditions.  Validation is carried out through both
\textit{a priori} and \textit{a posteriori} tests. The \textit{a
  priori} evaluation spans 140 diverse high-fidelity numerical
datasets and experiments (67 training cases included), covering turbulent boundary layers,
airfoils, Gaussian bumps, and full aircraft geometries, among
others. We demonstrate that the baseline wall model outperforms the EQWM in 90\% of test scenarios, while maintaining errors below 20\% for 98\% of the cases.
\end{abstract}


\maketitle


\section{Introduction}
\label{Sec:Intro}

Wall-modeled large-eddy simulation (WMLES) is a key tool for
simulating high–Reynolds-number turbulence across diverse scientific
and engineering applications. Compared with wall-resolved LES (WRLES)
and direct numerical simulation (DNS), WMLES is computationally
tractable: it resolves large energy-containing eddies while modeling
the near-wall dynamics, enabling accurate predictions of complex flows
at competitive cost suitable for industrial
design~\cite{slotnick2014cfd, Mani2023}. However, many wall models
rely on strong equilibrium assumptions that break down in separated
flows, under strong pressure gradients, or during laminar-to-turbulent
transition, which limits their accuracy and
generalizability~\cite{deardorffNumericalStudyThreedimensional1970,
  schumannSubgridScaleModel1975, piomelliWalllayerModelsLargeeddy2002,
  larssonLargeEddySimulation2016,
  boseWallModeledLargeEddySimulation2018}.

To address these challenges, non-equilibrium wall models have been
developed to account for departures from the idealized equilibrium
behavior of near-wall turbulence. Here, we use the term
\emph{non-equilibrium} to denote any deviation from the canonical
zero-pressure-gradient (ZPG) turbulent boundary layer.  One class of
non-equilibrium wall models consists of Reynolds-Averaged
Navier–Stokes (RANS)-based partial differential equation (PDE)
formulations, which have proven effective in capturing non-equilibrium
effects~\cite{balarasTwolayerApproximateBoundary1996,
  cabot2000approximate, wangDynamicWallModeling2002a,
  parkImprovedDynamicNonequilibrium2014}. However, their use in WMLES
is limited because they require a secondary mesh with full
connectivity, complicating implementation and reducing solver
efficiency. Moreover, despite their sophistication, RANS-based-PDE
wall models are not a panacea: they often require ad hoc tuning of
eddy-viscosity closures across flow scenarios~\cite{lozano2020non},
while still underperforming in favorable-pressure-gradient (FPG)
regions and separation~\cite{hayat2024wall}.

Ordinary differential equation (ODE) non-equilibrium wall models relax
the secondary-mesh connectivity requirement by estimating one or more
terms in the wall-shear-stress balance (convective, pressure-gradient,
and viscous)~\cite{hoffmann1995approximate, hickel2013parametrized,
  tamakiPhysicsModelingTrailingedge2020a,
  catalanoNumericalSimulationFlow2003, kamogawa2023ordinary,
  fowlerMultitimescaleWallModel2023}. Other approaches include dynamic
slip-based models that adjust a slip length to infer wall shear stress
($\boldsymbol{\tau}_w$) predictions~\cite{boseDynamicSlipBoundary2014a,
  baeDynamicSlipWall2019a,raje2025new}. Mixed viscous and
pressure-gradient scalings have also been leveraged to incorporate
pressure-gradient effects~\cite{manhart2008near,
  dupratWalllayerModelLargeeddy2011}, and analytical non-equilibrium
models have been derived from asymptotic
solutions~\cite{gonzalezLargeEddySimulationCompressible2018}. More
recently, sensor-based augmentation has been proposed to enhance
stresses where pressure gradients dominate near
separation~\cite{agrawalNonequilibriumWallModel2024}.  While these
strategies perform well in the cases for which they were developed,
many still depend on case-specific assumptions, limiting their ability
to generalize across the broad range of flows encountered in WMLES.

Another critical challenge is the need for laminar wall models, which
are essential for accurately predicting the leading-edge region of
wings and the subsequent transition to
turbulence~\cite{slotnick2014cfd}.  Because the laminar boundary layer
is extremely thin, resolving it within the standard WMLES framework is
computationally prohibitive. Moreover, without special treatment,
deploying a wall model designed for turbulent flow can induce large
errors, degrading lift and drag predictions and even triggering
spurious leading-edge
separation~\cite{larssonLargeEddySimulation2016}. One remedy is to use
sensors to switch between a wall model and a no-slip boundary
condition—effectively resolving the near-wall region—when turbulent
activity is low~\cite{bodart2012sensor}; however, this still demands
fine near-wall resolution. To address these limitations, several
laminar wall-modeling strategies have been
proposed. \citet{gonzalez2020wall} and \citet{dauricio2023wall}
derived analytical solutions based on Falkner–Skan boundary-layer
theory to predict local wall shear stress, and
\citet{dujardin2023wallmodel} showed that coupling a laminar and a
turbulent wall model via a sensor can improve airfoil simulations.
Other approaches have proposed the used of parabolized stability
equations for the laminar region~\cite{Lozano2018_PSE}.  While
promising, most prior approaches rely on boundary-layer edge
(outer-layer) quantities as inputs, which can be difficult to obtain
or even ill-defined in simulations over complex geometries
\cite{griffin2021general}.

Over the past decade, data-driven wall models have increasingly relied
on machine learning (ML) to infer mappings from the resolved field to
the wall shear stress using high-fidelity data, typically
parameterized by artificial neural networks (ANNs) of varying
complexity. Table~\ref{tab:mlwm} summarizes existing ML-based wall
models, including their input structure, physics considered, and
reported generalization capability.  Early developments focused on
channel-flow datasets and leveraged the law of the wall to train
neural networks for $\boldsymbol{\tau}_w$
prediction~\cite{yangPredictiveLargeeddysimulationWall2019,
  zhouWallModelBased2021, zhou2023wall}. These efforts were later
extended to include rotation
effects~\cite{huangWallmodeledLargeeddySimulations2019} and
high-speed-flow regimes~\cite{zangenehDatadrivenModelImproving2021}.
Beyond supervised learning, reinforcement-learning strategies have
also been explored to infer optimal wall-modeling
policies~\cite{baeScientificMultiagentReinforcement2022a,
  zhouWallModelingTurbulent2024,
  vadrotLoglawRecoveryReinforcementlearning2023}.  In addition, graph
neural networks have been employed to model wall shear
stress~\cite{dupuyDatadrivenWallModeling2023,
  dupuyModelingWallShear2023, dupuy2024using}, and non-local
convolutional neural-network (CNN) wall models have been evaluated
through both \textit{a priori} and \textit{a posteriori}
analyses~\cite{tabejamaatPrioriAssessmentNonlocal2023,
  tabejamaatPosterioriStudyWall2024, boxho2024development,
  boxho2025wall}. A different direction was introduced by
\citet{lozano-duranMachineLearningBuildingblockflow2023}, who proposed
a building-block strategy designed to capture multiple near-wall
physical mechanisms while providing prediction confidence. This
approach has since demonstrated strong applicability to complex
geometries~\cite{lozano-duranBuildingBlockFlowModelLargeEddy2023,
  mabuilding, arranzWallmodeledBasedBuildingblock2023,
  lingBuildingBlockFlowModelLargeEddy2024,
  arranzBuildingblockflowComputationalModel2024}, and it is the one
adopted in the present work.
%
\setlength{\tabcolsep}{5.5pt}
\begin{table}[h!]
    \centering
\caption{Overview of machine-learning wall models, ordered
  chronologically. Columns report (i) input locality, (ii) input
  structure, (iii) physics considered, (iv) applicability to arbitrary
  geometries, (v) training database, and (vi) untrained test cases
  that differ substantially from the training data. \textbf{Input
    Locality} refers to the classification as local (\checkmark),
  non-local (\(\times\)), or semi-local (--). \textbf{Input Structure}
   specifies the data organization: \emph{Points} (one or more spatial
  samples), \emph{Graph} (graph-neural-network topology),
  \emph{Integral} (integrals of known quantities), or \emph{Plane}
  (full planar fields).  \textbf{Physics Considered} includes
  turbulence under zero/adverse/favorable mean-pressure gradients
  (ZPG/APG/FPG), mean-flow three-dimensionality (3D),
  Rotation, wall Curvature, Laminar flow, Separation, and
  statistically unsteady turbulence (Unsteady). \textbf{Arbitrary
    Geometry} (\checkmark) indicates applicability to arbitrary
  geometries. \textbf{Training cases} list datasets used for model
  development, while \textbf{Untrained Test Cases} enumerate distinct
  flows used to assess generalization as reported in the original
  sources. Unless otherwise noted, all flows are turbulent.}
    \label{tab:mlwm}
    \footnotesize
    \begin{tabular}{p{3.2cm}p{0.8cm}p{1.2cm}p{1.4cm}p{1.8cm}p{1.2cm}p{2.8cm}p{2.7cm}}
        \toprule
        & \textbf{Year} & \textbf{Input Locality} & \textbf{Input Structure} & \textbf{Physics Considered} & \textbf{Arbitrary Geometry} & \textbf{Training Case} & \textbf{Untrained Testing Cases} \\
        \midrule
        \citet{yangPredictiveLargeeddysimulationWall2019} & 2019 & $\times$ & Points & ZPG & $\times$ & Channel &  Channel with spanwise acceleration\\
        \midrule
        \citet{huangWallmodeledLargeeddySimulations2019} & 2019 & -  & Points &  ZPG \newline  Rotation & $\times$ & Channel with spanwise rotation & N/A \\ 
        \midrule
        \citet{zangenehDatadrivenModelImproving2021} & 2021 & $\times$ & Points &  ZPG \newline  APG/FPG & $\times$ &  Supersonic TBL\newline  Expansion/\newline Compression Corner & N/A \\
        \midrule
        \citet{zhouWallModelBased2021} & 2021 & $\times$ & Points &  APG/FPG \newline  Curvature & $\times$ & Periodic hill & Channel \\
        \midrule
        \citet{baeScientificMultiagentReinforcement2022a} & 2022 & - & Points & ZPG & $\checkmark$ & Channel & ZPG TBL \\
        \midrule
        \citet{lozano-duranMachineLearningBuildingblockflow2023} & 2023 & $\checkmark$ & Points &  Laminar \newline  ZPG \newline  Separation \newline  APG/FPG \newline  3D \newline   Unsteady & $\checkmark$ &  Laminar Channel \newline  Channel \newline  Channel with PG \newline  3D Channel &  Laminar BL \newline  Turbulent BL \newline  Pipe \newline  NASA CRM-HL \newline  NASA Juncture Flow \\
        \midrule
        \citet{dupuyDatadrivenWallModeling2023} & 2023 & $\times$ & Points &  ZPG \newline  APG/FPG \newline Separation \newline  3D & $\times$ &  Channel \newline  Backward Step \newline  3D Diffuser & N/A \\ 
        \midrule
        \citet{dupuyModelingWallShear2023} & 2023 & $\times$ & Graph &  ZPG \newline  APG/FPG \newline  Curvature \newline  Separation \newline  3D & $\checkmark$ &  Channel \newline  Blade \newline  Backward Step \newline  Smooth Ramp \newline  3D Diffuser & N/A \\
        \midrule
        \citet{tabejamaatPrioriAssessmentNonlocal2023} & 2023 & $\times$ & Plane & ZPG & $\times$ & Channel & N/A \\
        \midrule
        \citet{leeArtificialNeuralNetworkbased2023} & 2023 & $\times$ & Integral & ZPG & $\times$ & Channel & Separation Bubble \\
        \midrule
        \citet{vadrotLoglawRecoveryReinforcementlearning2023} & 2023 & $-$ & Points & ZPG & $\checkmark$ & Channel & N/A \\
        \midrule
        \citet{zhou2023wall} & 2023 & $\times$ & Points &  APG/FPG \newline  Curvature & $\checkmark$ &  Periodic Hill \newline  Law of the Wall & N/A \\
        \midrule
        \citet{maejimaPhysicsinformedMachinelearningSolution2024} & 2024 & $\times$ & Points & ZPG & $\checkmark$ & ZPG TBL & N/A \\
        \midrule
        \citet{radhakrishnanDatadrivenWallModeling2024} & 2024 & $\checkmark$ & Points &  ZPG \newline  APG/FPG \newline  3D & $\checkmark$ &  Channel \newline  3D Diffuser &  Juncture \newline  NASA Hump \\
        \midrule
        \citet{zhouWallModelingTurbulent2024} & 2024 & - & Points &  APG/FPG \newline  Curvature & $\checkmark$ & Periodic Hill & Gaussian Bump \\
        \midrule
        \citet{boxho2025wall} & 2025 & $\times$ & Points &  ZPG \newline APG/FPG \newline Curvature & $\times$ & Channel \newline Periodic Hill & N/A \\
        \midrule
        \citet{zhang2025knowledge} & 2025 & $\times$ & Points &  APG/FPG \newline  Curvature & $\times$ & Periodic Hill & Gaussian Bump \\
        \midrule
        \citet{zhouWallModelSeparated2025} & 2025 & $\times$ & Points & APG/FPG \newline  Curvature & $\times$ & Periodic Hill & Wavy Wall \\
        \bottomrule
    \end{tabular}
\end{table}

Paradoxically, despite the promise of ML to enable general-purpose
wall models, most existing approaches remain fundamentally limited in
scope.  By construction, many models cannot be applied broadly given
the required inputs, and even in the cases where they are applicable,
they often underperform relative to simple equilibrium wall models. As
a result, only a few ML wall models have seen adoption in practical
simulations. The key challenges are:
\begin{enumerate}
\item \emph{Arbitrary geometries.} Many models rely on non-local
  inputs that are not directly tied to near-wall quantities,
  complicating implementation or even preventing their use in general
  geometries. For example, while CNNs can capture non-local effects,
  adapting planar or grid-aligned inputs to complex surfaces often
  limits their practicality. Other models do not comply with rotation,
  translation, or Galilean invariance, which are fundamental physical
  requirements that any model must satisfy.

\item \emph{Insufficient physics} Practical applications involve
  superposed effects (mean PG effects, flow separation, wall
  curvature, laminar flow, etc.), yet most models are trained on a
  handful of canonical cases where crucial regimes are often
  missing. Even when relevant physics are present (e.g., periodic
  hills with pressure gradients, separation, and curvature), studies
  typically target a single geometry and a narrow range of Reynolds
  numbers, encouraging memorization rather than learning transferable
  physics.

  \item \emph{Limited validation.} Many data-driven models are tested
    on one or few cases closely resembling their training set. As
    summarized in Table~\ref{tab:mlwm}, with the notable exceptions of
    \citet{lozano-duranMachineLearningBuildingblockflow2023} (and
    follow-up versions) and
    \citet{radhakrishnanDatadrivenWallModeling2024}, most works do not
    report performance on flows that differ substantially from
    training. Systematic benchmarking across diverse, out-of-training
    distributions is essential to avoid over-optimistic conclusions
    and to guide iteration.

\item \emph{Lack of explainability.} Most ML wall models behave as
  black boxes that map inputs to $\boldsymbol{\tau}_w$ without
  exposing which mechanisms drive their predictions. This opacity
  limits physical interpretation. Thus, a model can perform accurately
  on benchmarks yet ``get the right answer for the wrong reasons,''
  due to error cancelation, a risk that may go unnoticed in
  practice. Explainability tools are needed to reveal the basis of the
  predictions and make model behavior more transparent; however, they
  remain rarely used in wall modeling with few
  exceptions~\cite{lozano-duranBuildingBlockFlowModelLargeEddy2023}.

\item \emph{Need for uncertainty quantification (UQ).} Most ML wall
  models return estimates of $\boldsymbol{\tau}_w$ without a
  confidence measure or a decomposition of \emph{epistemic} versus
  \emph{aleatoric} uncertainty. In realistic WMLES settings---where
  ground truth is unavailable---UQ is vital to detect
  out-of-distribution inputs, prioritize additional data collection,
  trigger mesh/SGS/wall-model adaptation, and support risk-aware
  decisions. Accordingly, models should report predictive
  distributions or credible intervals with reliability metrics, not
  just a mere point prediction.
  
  \item \emph{Effect of external errors.} Wall models are susceptible
    to (i) \emph{external} errors propagated from the outer LES field
    and its subgrid-scale (SGS) model, and (ii) \emph{internal} errors
    due to the wall-modeling assumptions
    themselves~\cite{Lozano2019_Error, lozano2022performance}. Even a
    physically exact wall model would inherit outer-layer
    inaccuracies; in practice, it has been shown that external errors
    can completely degrade the performance of accurate wall
    models~\cite{lozano-duranBuildingBlockFlowModelLargeEddy2023}. Thus,
    improving the wall model alone is insufficient and advances in SGS
    modeling and outer-layer accuracy are also required to improve WMLES
    performance.

\end{enumerate}

\begin{figure}
    \centering
    \includegraphics[width=.95\linewidth]
    {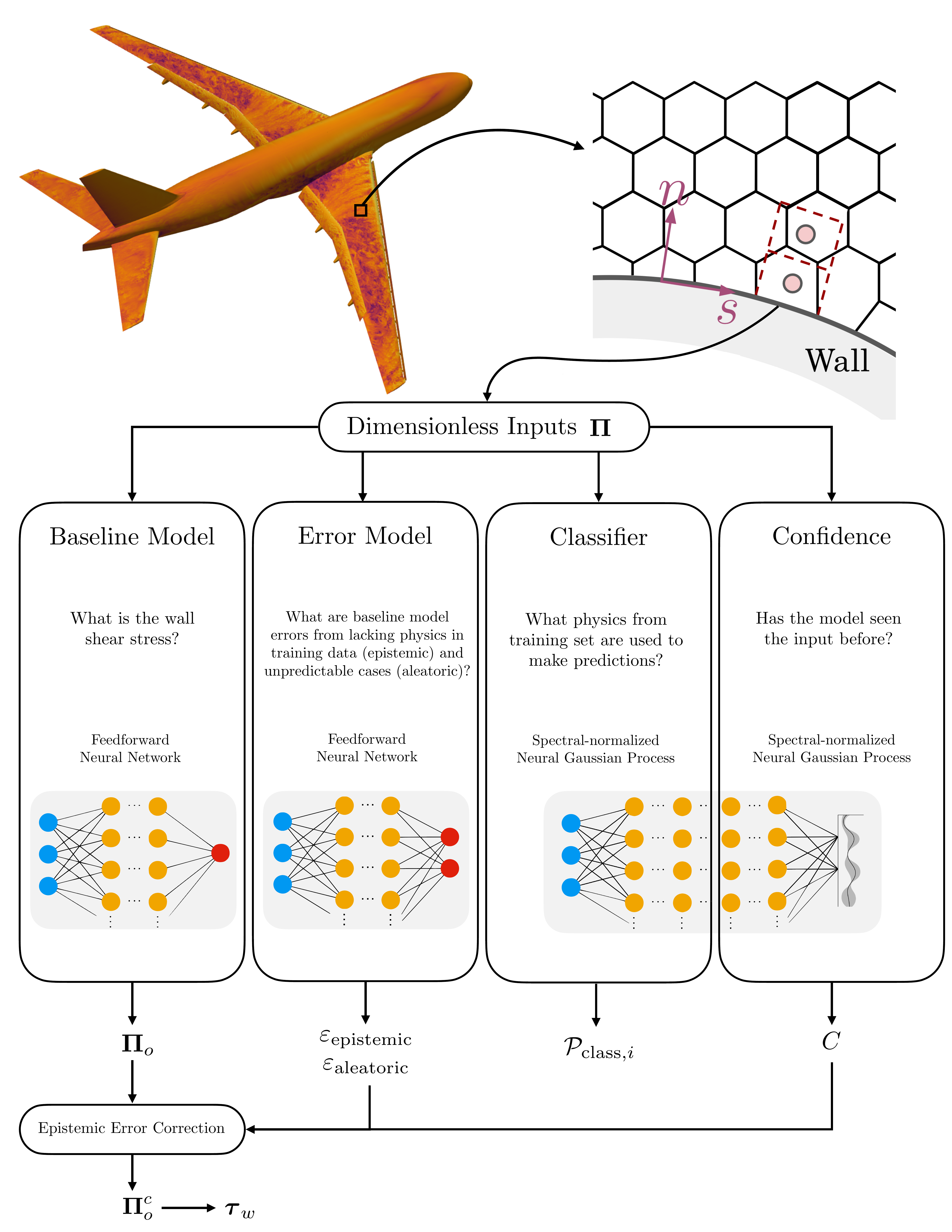}
    \caption{Overview of the building-block flow wall model version 2
      (BFM-WM-v2). The model comprises four components: (i) a baseline
      wall model, (ii) an error model, (iii) a classifier, and (iv) a
      confidence score. Each panel contains the question the module is
      aiming to answer. When appropriate, the error model corrects the
      baseline prediction. The model inputs and output are formulated
      in dimensionless form. In Part I of this work, we present the baseline model. The other three components are presented in Part II.}
    \label{fig:CompSche}
\end{figure}

We develop a wall model that addresses the six challenges outlined
above. The overall approach is summarized in
Figure~\ref{fig:CompSche}. In Part~I, we tackle Challenges~1--3;
Challenges~4 and~5 are treated in Part~II. Challenge 6 is deferred to future work to be addressed by the SGS model. An overview of all the
modeling components is shown in Figure~\ref{fig:CompSche}.  The
approach is rooted in the building-block-flow principle---extracting
essential physics from simple canonical cases---and is implemented
with an artificial neural network. The formulation facilitates
incorporating diverse cases into the training set, constructed from
mean-flow statistics. Among these, we leverage a new DNS dataset
spanning turbulent boundary layers (TBLs) subjected to a range of
mean-pressure-gradient conditions. Model performance is validated
against a comprehensive suite of experimental and high-fidelity
numerical benchmarks, covering varied flow physics including pressure
gradients, wall curvature, and laminar–turbulent transition, among
others.

The remainder of the paper is organized as
follows. Section~\ref{Sec:Method} introduces the wall-model
formulation, the training database, and outlines the non-dimensional
input-selection methodology. Section~\ref{Sec:Results} presents
comprehensive \textit{a priori} assessments and \textit{a posteriori}
WMLES results. Finally, Section~\ref{Sec:Discussion} discusses the
limitations of the current model and outlines directions for future
extensions. 

\section{Methodology}
\label{Sec:Method}

\subsection{Overview of building-block flow model \label{Sec:BFM}}

The wall model follows the building-block flow model (BFM) principle
of \citet{lozano-duranMachineLearningBuildingblockflow2023}. The
central assumption is that the unresolved physics in complex scenarios
can be locally mapped to the small-scale dynamics of simpler
flows. Accordingly, we posit a finite set of \emph{building-block
flows} (BBFs) that encapsulate the essential physics needed to
construct generalizable wall models. This design prioritizes learning
the governing mechanisms rather than memorizing cases---for example,
avoiding correct predictions over a particular wing merely because
similar wings appeared in the training set—and thus aims to capture
the underlying flow physics faithfully.

The building-block idea has been used for both subgrid-scale (SGS)
models (to represent unresolved motions away from the wall) and wall
models (to represent the missing scales at solid boundaries). The
first implementation (version 1), BFM-v1, was introduced in our prior
work~\cite{lingWallmodeledLargeeddySimulation2022,
  lozano-duranBuildingBlockFlowModelLargeEddy2023,
  arranzBuildingblockflowComputationalModel2024}, which comprised
coupled SGS model (BFM-SGS-v1) and wall model component (BFM-WM-v1)
and extensions accounting for wall roughness in incompressible and
compressible flow regimes~\cite{mabuilding,
  maMachinelearningWallmodelLargeeddy2025}. BFM-v1 was assessed across
flows ranging from low geometric complexity (turbulent channels,
pipes, boundary layers) to moderate (airfoils, Gaussian bump) and
fully complex configurations (aircraft-like geometries). While BFM-v1
improves upon traditional wall models, it still underperforms in
several regimes. Here, we present the second-generation wall model
component, referred to as BFM-WM-v2. The updated SGS model component
(BFM-SGS-v2) will be reported in a subsequent work.

We summarize the model requirements we aim to satisfy, following the
guidelines from our previous
work~\cite{lozano-duranMachineLearningBuildingblockflow2023}:
\begin{enumerate}
\item The wall model should provide accurate predictions of the
  \emph{mean} (e.g., time-averaged) wall shear stress, rather than
  attempting to capture its time-dependent dynamics.
\item The model must account for different flow regimes (e.g., laminar
  flow, wall-attached turbulence, separated flow, \dots) in a unified
  manner, i.e., the input and output structure must remain identical
  across all cases.
\item The formulation must be scalable, allowing additional flow
  physics (building-block flows) to be incorporated in future
  versions.
\item The model must provide a confidence score and an uncertainty
  estimate for its prediction at each wall location.
\item The formulation must be directly applicable to complex
  geometries with unstructured grids without requiring any
  case-specific modifications.
\item The input set should exclude wall-normal gradients. This choice
  is motivated by the fact that, in WMLES, the near-wall grid is
  deliberately coarse. As a result, wall-normal gradients depend
  strongly on the numerical scheme used, and excluding them improves
  portability across solvers.
\item All model inputs and outputs must be expressed in dimensionless
  form to ensure dimensional homogeneity, i.e., invariance under
  changes in the units of the variables.
\item The model must satisfy Galilean invariance and remain invariant
  under constant spatial and temporal translations, as well as
  rotations of the reference frame.
\end{enumerate}

\begin{figure}
    \centering
    \includegraphics[width=\linewidth]{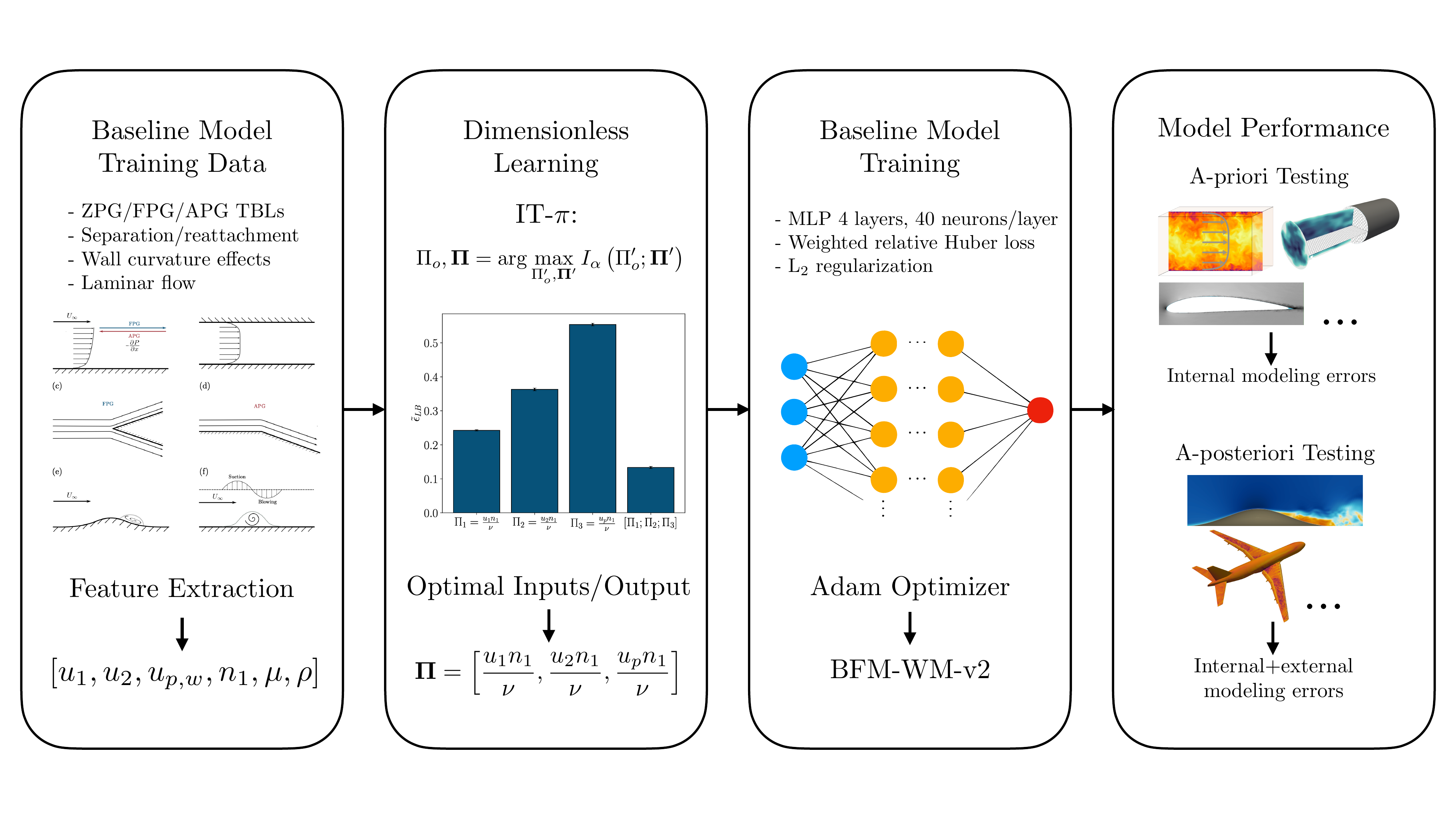}
    \caption{Workflow of the training data, dimensionless input/output
      design, training process, and testing of the baseline model.}
    \label{fig:workflow}
\end{figure}

Figure~\ref{fig:workflow} shows an overview of the preparation of
training data, the design of dimensionless input and output variables,
the training procedure, and the evaluation of the baseline model.

\subsection{Main model assumptions}\label{Subsec:Assp}

We discuss the key assumptions underlying the model. In addition to
the standard WMLES assumptions, the main assumptions specific to
BFM-WM-v2 are:
\begin{itemize}
\item \emph{Building-block-flow assumption:} This is the main
  assumption of the wall model, namely that there is a finite set of
  simple flows (building-block flows) that contain the essential flow
  physics to formulate generalizable wall models.
\item \emph{Local state sufficiency:} We assume that the local flow
  state provides the necessary information for accurately predicting
  the wall shear stress. This assumption is supported by recent
  findings showing that history effects are small in the near-wall
  region of statistically unsteady wall turbulence and PG turbulent
  boundary layers~\cite{lozano2020non,baxerres2024evidence}. Although
  incorporating global information could further improve predictions,
  restricting the inputs to local quantities ensures applicability to
  arbitrarily complex geometries.
\item \emph{Ensemble-averaged inputs for training:} The model is
  trained on ensemble-averaged values of near-wall DNS data rather
  than instantaneous samples. This choice is motivated by the
  objective of predicting the mean wall shear stress accurately,
  without reproducing instantaneous dynamics. The approach offers
  several advantages: First, it simplifies data preparation and
  reduces the complexity of the training process. Second, it enables
  the inclusion of a wider range of flow cases by leveraging mean
  quantities without requiring full flow fields.  Finally, it avoids
  the underfitting issues reported when training deterministic models
  with complex instantaneous
  inputs~\cite{dupuyDatadrivenWallModeling2023}. When deployed, a
  time-averaging filter is applied to the instantaneous WMLES inputs
  during inference to approximate the averaged
  value~\cite{yangIntegralWallModel2015,
    lozano-duranMachineLearningBuildingblockflow2023}.
\item \emph{Wall-shear stress direction:} We assume that the
  wall-shear stress vector, $\boldsymbol{\tau}_w$, is aligned with the
  velocity vector at the first grid point, $\boldsymbol{u}_1$, which
  corresponds to the velocity in the WMLES grid closest to the wall
  and serves as an input to the wall model. As a result, the primary task
  of the model is to predict only the magnitude of the wall-shear
  stress, $\tau_w$. This assumption is supported by evidence that, at
  high Reynolds numbers and for typical WMLES wall-normal grid
  spacing, $\boldsymbol{\tau}_w$ and $\boldsymbol{u}_1$ remain aligned
  on average within a reasonable margin of error~\cite{lozano2020non}.
\end{itemize}

\subsection{Baseline model formulation}

\begin{figure}
    \centering
    \begin{subfigure}[b]{0.52\textwidth}
        \includegraphics[width=0.84\linewidth]{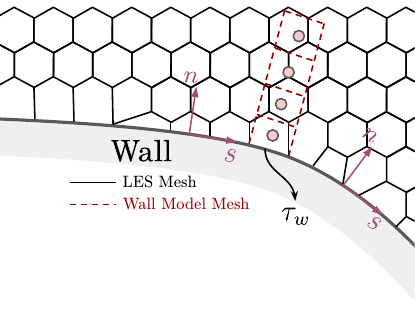}
        \caption{}
        \label{fig:sub:local}
    \end{subfigure}
    \hfill
    \begin{subfigure}[b]{0.42\textwidth}
        \includegraphics[width=.8\linewidth]{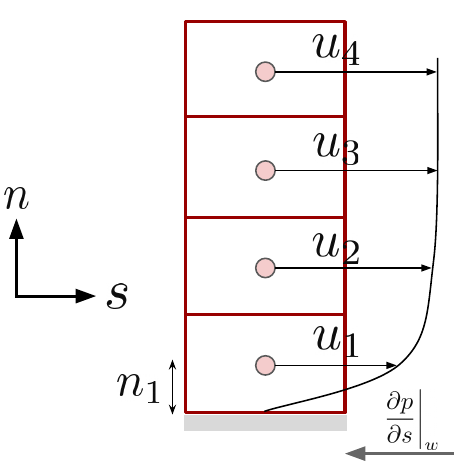}
        \caption{}
        \label{fig:sub:model_inputs}
    \end{subfigure}
    \caption{(a) Wall-model mesh (dashed red) defined in the local
      $(n,s)$ coordinates over a curved wall; the surrounding
      unstructured LES mesh is shown in black. (b) Dimensional inputs
      investigated to construct the wall model, in addition to fluid
      properties such as density and viscosity. Subscripts $1$–$4$
      denote the $i$-th off-wall grid point, and the subscript $w$
      indicates wall values.  }
    \label{fig:model_schem}
\end{figure}

The wall model predicts the wall-shear stress vector
$\boldsymbol{\tau}_w$ based on local near-wall flow features, which is
imposed as boundary conditions for the LES equations. The procedure is
illustrated in Figure~\ref{fig:sub:local}. The prediction of
$\boldsymbol{\tau}_w$ is formulated as
\begin{equation}
  \boldsymbol{\tau}_w = \text{ANN}(\boldsymbol{q}; \boldsymbol{\theta})\, \frac{\boldsymbol{u}_1}{u_1},
  \label{Eq:WMProblem_General}
\end{equation}
where $\boldsymbol{q}$ is the input feature vector composed of
near-wall flow quantities, and ANN denotes a feedforward artificial
neural network with trainable parameters $\boldsymbol{\theta}$. The
direction of $\boldsymbol{\tau}_w$ is assumed to align with the
wall-parallel velocity at the first off-wall grid point,
$\boldsymbol{u}_1$, and $u_1 = \|\boldsymbol{u}_1\|$, where
$\|\cdot\|$ denotes the Euclidean norm.

Figure~\ref{fig:sub:model_inputs} illustrates the set of local input
features included in $\boldsymbol{q}$. These comprise the
wall-parallel velocities at various distances from the wall, $u_1 =
u|_{n=n_1}$, and $u_2 = u|_{n=3n_1}$, where $n_1$ is the wall-normal
distance of the first grid point. These quantities characterize the
shape of the local velocity profile. We also examined an extended
stencil that includes $u_3 = u|_{n=5n_1}$ and $u_4 =
u|_{n=7n_1}$. However, the results indicate that the added complexity
yields only marginal improvements.

To capture the effect of streamwise pressure gradients, the model also
includes a pressure-gradient/acceleration-based velocity at the wall:
\begin{equation}
  u_{p} = \operatorname{sign}\left(\frac{1}{\rho}\left.\frac{\partial p}{\partial
    s}\right|_w + \left.\frac{\partial u_{||}}{\partial t} \right|_w\right)
  \left| \mu \left( \frac{1}{\rho}\left.\frac{\partial p}{\partial
      s}\right|_w + \left.\frac{\partial u_{||}}{\partial t} \right|_w \right) \right|^{1/3},
\end{equation}
where $\partial p / \partial s |_w$ is the wall-parallel pressure
gradient projected along the local flow direction ($s$), 
\begin{equation}
  \left.\frac{\partial p}{\partial s}\right|_w = \nabla_{\boldsymbol{x}_{\parallel}} p \cdot \boldsymbol{u}_{1 \parallel},
\end{equation}
with $\nabla_{\boldsymbol{x}_{\parallel}}$ the spatial gradient
operator and $(\cdot)_{\parallel}$ denotes projection onto the
wall-parallel plane.  The term $\partial u_{||}/\partial t|_w $ is the
wall-parallel acceleration of the wall with respect to an inertial
frame of reference. The model also takes as input the local kinematic
viscosity $\mu$ and density $\rho$. In summary, the input feature
vector is
\begin{equation}
  \boldsymbol{q} = [u_1, u_2, u_{p}, n_1, \mu, \rho],
\end{equation}
where all variables are dimensional. Their dimensionless form is
discussed in \S\ref{Sec:ITPI}.

\subsection{Training database}
\label{Sec:Data}

\begin{figure}
    \centering
    \includegraphics[width=0.9\linewidth]{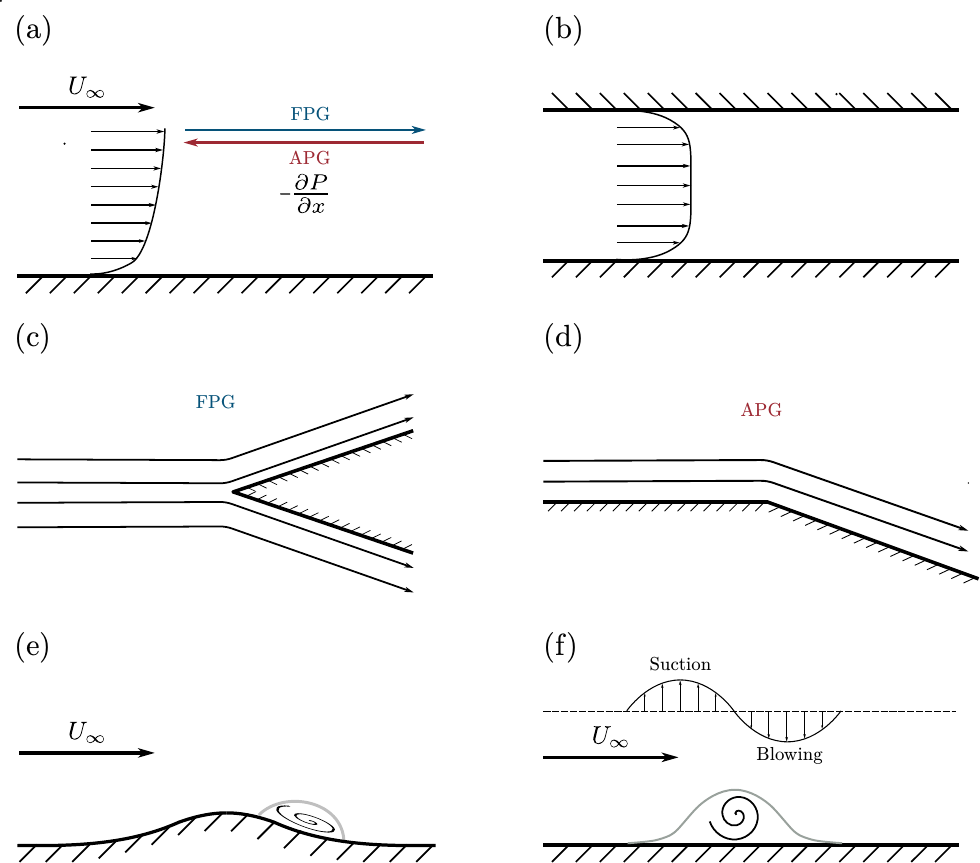}
    \caption{Schematics of representative training cases. (a) APG and
      FPG turbulent boundary layers; (b) turbulent channel flows; (c)
      FPG Falkner--Skan laminar flow; (d) APG Falkner--Skan laminar
      flow; (e) spanwise-periodic Gaussian bumps; (f) pressure-induced
      turbulent separation bubbles.  Regions enclosed by grey lines
      indicate separated flow.}
    \label{fig:trainingcases}
\end{figure}

The training database used to develop the wall model is summarized in
Table~\ref{tab:trainingcases}, with representative flow cases shown in
Figure~\ref{fig:trainingcases}. Following the building-block
principle, we select simple flow configurations that isolate key
physical mechanisms relevant for wall modeling. Our aim is to assemble
a database that is both diverse and physically rich. The dataset
includes turbulent channel flows, ZPG/FPG/APG turbulent boundary
layers, separation bubbles with subsequent reattachment,
wall-curvature effects, laminar flows, and high-Reynolds-number
conditions.
%
\begin{table}
    \centering
    \caption{Summary of training cases. Details of flow parameters can be found in the text.}
    \label{tab:trainingcases}
    \begin{tabular}{p{7cm}p{2.8cm}lp{1.2cm}c}
        \toprule
        \textbf{Case} & \textbf{Dominant \newline Flow Physics} & \textbf{Type} & \textbf{Subcase Count} & \textbf{Symbol} \\
        \midrule
        APG \& FPG turbulent boundary layers \newline (\citet{Arranz_2025})   &  ZPG/APG/FPG \newline  Separation & DNS & 16 & $\bigtriangledown$\\
        \midrule
        Turbulent channel flows (\citet{lozano-duran_effect_2014}, \citet{lee2015direct}, \citet{hoyas_wall_2022}) & Akin ZPG & DNS & 6 & $\mathbf{c}$ \\
        \midrule
        Synthetic data by the log law \cite{pope2001turbulent}  & High-Re ZPG & Analytical & 3 & $\mathbf{s}$ \\
        \midrule
        Falkner-Skan laminar boundary layers \newline (\citet{FalknerSkan1931}) &  Laminar APG/FPG & Analytical/ODE & 40 & $\mathbf{l}$\\
        \midrule
        Spanwise-periodic Gaussian bump ($Re_L=2M$) \newline (\citet{uzun_high-fidelity_2022}) &  Wall curvature  \newline  Separation & DNS/WRLES & 1 & $\bigstar$\\
        \midrule
        Pressure-induced turbulent separation bubbles \newline (\citet{kamogawa2023ordinary}) & Separation \newline w/ reattachment & WRLES & 1 & \adjustbox{valign=c}{\input{Figures/Symbols/Right}} \\
        \bottomrule
    \end{tabular}
\end{table}

\begin{figure}
    \centering
    \includegraphics[width=0.5\linewidth]{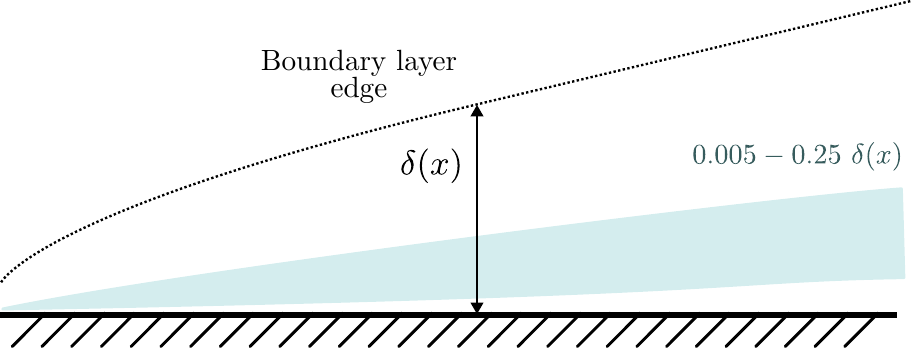}
    \caption{The sampling region for the training data covering the
      range 0.005-0.25 $\delta(x)$, where $\delta(x)$ is boundary
      layer thickness.}
    \label{fig:sampling}
\end{figure}

The training data include all quantities required to construct the
input--output pairs $(\boldsymbol{q}, \tau_w)$ to train the wall
model. Sampling is performed over a broad range of wall-normal
matching locations along streamwise directions, as illustrated in
Figure~\ref{fig:sampling}. Specifically, the wall-model matching
height $h_{\text{wm}}$ is normalized by the local boundary layer
thickness $\delta$, and varies from $h_{\text{wm}}/\delta = 0.005$ to
$0.25$. This corresponds to spatial resolutions ranging from
approximately 200 to just 4 grid points per $\delta$. The range spans
from the finest grid resolutions feasible in complex geometries to
highly under-resolved settings, where only a few grid points lie
within the boundary layer~\cite{Lozano2019_Error,
  lozano2022performance}. This ensures the training data reflect both
idealized and practically relevant WMLES conditions. A total of $N
\sim 10^6$ input--output pairs are collected for the training dataset.

In the following, we discuss the rationale behind including each
specific case in the training database and describe the corresponding
flow parameters.
\begin{enumerate}
\item \emph{ZPG, APG, and FPG turbulent boundary layers.} We employ a
  new collection of DNS of TBLs spanning zero-, adverse-, and
  favorable-pressure-gradient conditions by prescribing a ceiling with
  a virtual straight ramp that deflects the freestream either upward
  (APG) or downward (FPG)~\cite{Arranz_2025}. This configuration
  generates statistically stationary, spatially developing TBLs with
  controlled pressure-gradient histories, including attached,
  incipiently separated, and reattaching states—thus covering the
  range of behaviors encountered in practical aerodynamic
  applications. The inclusion of this dataset is motivated by the need
  to capture non-equilibrium near-wall physics induced by streamwise
  pressure gradients—effects that are poorly represented in canonical
  ZPG databases and often absent or too limited in existing
  datasets. The ramp deflection angle is varied over $\alpha \in
  \{-4^\circ, -3^\circ, -2^\circ, -1^\circ, 5^\circ, 10^\circ,
  15^\circ, 20^\circ\}$, producing favorable ($\alpha<0$), zero
  ($\alpha\approx 0$), and adverse ($\alpha>0$) pressure
  gradients. The pressure-gradient intensity is characterized by the
  Clauser parameter $\beta = (\delta^\ast/\tau_w)\,(\mathrm
  dP_e/\mathrm dx)$, where $\delta^\ast$ is the displacement thickness
  and $\mathrm dP_e/\mathrm dx$ is the edge-pressure gradient. We have
  $\beta \approx 0$ for ZPG, $\beta<0$ for FPG, and $\beta>0$ for APG,
  with $\beta \to \infty$ in the limit of separation as $\tau_w \to
  0$. The cases cover the range $\beta \in [-0.6,\infty)$. Two inflow
    Reynolds numbers, $Re_{\text{in}} = 300$ and $670$, are used,
    yielding a momentum-thickness Reynolds-number range of $300 <
    Re_\theta < 10^4$. Additional details on the case setup and
    simulation parameters are provided in Appendix~\ref{sec:app:tbl}.
  \item \emph{Turbulent channel flow.} This case represents fully
    developed turbulent flow between two parallel plates, driven by a
    constant pressure gradient under statistically steady
    conditions. Its simple geometry and absence of mean streamwise
    pressure gradients make it a canonical configuration for isolating
    near-wall turbulence physics.  The data considered several
    friction Reynolds numbers, defined as $Re_{\tau}=u_\tau h/\nu$,
    where $u_{\tau}=\sqrt{\tau_w/\rho}$ is the friction velocity, $h$
    is the channel half-height, and $\nu$ is the kinematic
    viscosity. Specifically, we include $Re_{\tau}=550$, $950$,
    $2000$, $4200$, $5200$, and
    $10000$~\cite{lozano-duran_effect_2014, lee2015direct,
      hoyas_wall_2022}.
  \item \emph{Synthetic log-law for high Reynolds numbers.} This
    dataset consists of analytically generated velocity profiles based
    on the classical logarithmic law of the
    wall~\cite{schumannSubgridScaleModel1975,pope2001turbulent},
    representative of ZPG turbulent boundary layers at high Reynolds
    numbers. These profiles emulate the asymptotic behavior of
    wall-bounded turbulence in regimes currently inaccessible to DNS.
    This guides the model to predict correct physics when deployed in
    simulations where no high-fidelity training data exist.  We
    include three representative friction Reynolds numbers in this
    category: $Re_{\tau} = 5 \times 10^4$, $10^6$, and $5 \times
    10^6$.
   \item \emph{Falkner--Skan laminar boundary layers.} This dataset
     comprises similarity solutions of laminar flat-plate boundary
     layers with streamwise acceleration or deceleration, obtained by
     numerically solving the Falkner--Skan~\cite{FalknerSkan1931} ODE
     via a shooting method. The cases are computed on smooth flat
     surfaces with no curvature so that pressure-gradient effects are
     isolated in a clean laminar setting.  The rationale for including
     these cases is to train the wall model with an explicit laminar
     regime and a physically consistent response to streamwise
     pressure gradients prior to transition.  The edge velocity is
     prescribed as $U_e = C x^m$, where $x$ is the streamwise
     coordinate, $C$ is a constant, and $m$ controls the pressure
     gradient: negative $m$ yields decelerating flow with possible
     incipient separation, while positive $m$ produces accelerating
     flow.  The aim is to avoid spurious leading-edge separation or
     incorrect shear in low-Re zones, and to encode how acceleration
     ($m>0$) steepens the near-wall profile while deceleration ($m<0$)
     thickens it toward separation.  The parameter range follows $m
     \in \{-0.0904,-0.08,-0.05,0,0.1,0.6,1.0,2.0\}$, covering
     scenarios from incipient separation through strong favorable
     pressure gradients. Four freestream-based Reynolds numbers are
     used spanning $10^3$ to $10^5$, resulting in $40$ subcases in
     total.
 \item \emph{2-D Gaussian bump.} This case employs the
   spanwise-periodic Gaussian bump of \citet{uzun_high-fidelity_2022}
   to introduce controlled surface curvature on an otherwise simple
   flat-plate configuration. The purpose of including this case is to
   isolate the influence of curvature on near-wall turbulence and wall
   shear stress without introducing confounding effects associated
   with complex geometries. Instead of prescribing curvature as an
   explicit input feature, we encode its effects implicitly by
   sampling velocities along multiple wall-normal directions. This
   strategy enables the model to infer curvature-induced modifications
   directly from local flow signatures while remaining applicable to
   flat walls, curved surfaces, and rotating systems.  The incoming
   boundary layer encounters a smooth convex rise followed by
   recovery, producing streamwise variations of curvature, pressure
   gradient, and thickness with minimal geometric complexity. The
   computational setup is three-dimensional and periodic in the span,
   and the reference Reynolds number is $Re_L = 2\times 10^6$ based on
   the bump length scale, yielding a fully turbulent approach flow and
   a broad range of local non-equilibrium responses.  The
   parameterization includes a curvature range summarized by $\delta/R
   \in [0, 0.07]$, representative of many practical applications,
   together with the streamwise evolution from mild APG on the
   upstream face to FPG over the downstream recovery. The local
   Clauser parameter $\beta = (\delta^\ast/\tau_w)(\mathrm d
   P_e/\mathrm d x)$, varies smoothly across the bump, capturing
   attached flow, potential incipient separation, and recovery without
   relying on equilibrium assumptions. Although not as exhaustively
   parametric as the APG/FPG flat-plate suite, this configuration
   supplies high-quality training data at $Re_L=2\times 10^6$ with
   realistic curvature levels and coupled PG effects. As detailed in
   Appendix~\ref{Sec:App:Curve}, it is necessary to include this
   dataset to improve the model performance for cases with curved
   geometry.
 \item \emph{Blowing-suction-induced separation bubbles.}  This
   dataset comprises simulations of a flat-plate TBL that undergoes
   controlled separation and reattachment by imposing a
   suction–blowing distribution at the freestream. The forcing
   generates a strong APG followed by a FPG, producing a separated
   shear layer that reattaches downstream and a recovery region with
   pronounced non-equilibrium dynamics~\cite{kamogawa2023ordinary}.
   The rationale for including these cases is to expose the wall model
   to the physics of shear-layer development and
   reattachment. Training on these flows teaches the model how wall
   stress behaves as $\tau_w$ weakens and approaches zero at
   separation, and how the outer shear layer interacts with and
   impinges on the wall during reattachment.  This informs the
   behavior of the model in extended reattaching zones---regimes where
   canonical log-law scaling breaks down and where equilibrium-based
   closures systematically fail~\cite{eaton1981review,
     simpson1981review}. The flow setup is characterized by an
   upstream $Re_{\theta}\approx 2000$ and a streamwise evolution of
   $\beta$ that becomes large and positive approaching the separation
   point (as $\tau_w\to 0$) and turns negative in the FPG-driven
   recovery after reattachment. The case thus samples a broad range of
   states from attached flow through separation to reattached
   recovery, providing critical coverage for wall-stress prediction
   under strong APG/FPG transients.

\end{enumerate} 

\subsection{Dimensionless learning by information-theoretic Buckingham-$\pi$
\label{Sec:ITPI}}

The model operates internally with dimensionless input and output
variables. This ensures dimensional homogeneity---i.e., invariance
under changes in the units of all quantities---which is a necessary
requirement for constructing generalizable models. Violating this
principle leads to models that are physically inconsistent and lack
robustness across different flow conditions. Here, we examine the
optimal dimensionless formulation of the input and output variables to
achieve maximum predictive performance.

The construction of dimensionless variables, as guided by the
Buckingham-$\pi$ theorem~\cite{buckingham1914physically}, is not
uniquely determined.  Traditional approaches to identifying
dimensionless variables rely heavily on physical intuition; however,
such choices are not guaranteed to be optimal, in the sense that they
may not produce a set of dimensionless variables with the highest
predictive power. Here, we adopt a more systematic strategy using the
Information-Theoretic Buckingham-$\pi$ (IT-$\pi$) approach introduced
by \citet{yuan2025dimensionless}. This method leverages
information-theoretic principles to identify the most predictive input
and output dimensionless variables. The core insight of the approach
is that the predictive capability of any model is fundamentally
bounded by the amount of information the inputs carry about the
outputs~\cite{lozano2022information}, where information is defined in
the formal sense of information theory~\cite{shannon1948}. A key
feature of IT-$\pi$ is that the identification of optimal
dimensionless variables is independent of the model structure, whether
it involves analytical expressions, neural networks, or other
functional forms. This methodology has previously been applied to
derive optimal dimensionless inputs for wall models incorporating
compressibility and surface roughness effects~\cite{mabuilding,
  yuan2025dimensionless}.

The optimal dimensionless inputs $\boldsymbol{\Pi} =\ \left[\Pi_1,
  \Pi_2, \ldots, \Pi_{l}\right]$ and dimensionless output $\Pi_o$ can
be found by solving the optimization problem:
\begin{equation}
\label{eq:IT_pi}
\Pi_o, \boldsymbol{\Pi} = \arg \min_{\Pi'_o, \boldsymbol{\Pi}'} \tilde{\epsilon}_{LB},
\end{equation}
where $\tilde{\epsilon}_{LB} = e^{-I[\Pi_o;\mathbf{\Pi}]}$ denotes the
normalized irreducible error and $I[\cdot;\cdot]$ is the mutual
information~\cite{shannon1948}. As shown by
\citet{yuan2025dimensionless}, $\tilde{\epsilon}_{LB}$ quantifies the
portion of the prediction error that cannot be reduced any further
regardless of the underlying modeling approach. Its values lie in
$[0,1]$, with smaller values indicating lower irreducible error.  We
use this approach to determine both the required number of inputs and
their corresponding dimensionless form.  The optimization problem from
Eq.~(\ref{eq:IT_pi}) is solved by employing the covariance matrix
adaptation evolution strategy (CMA-ES)\cite{hansen2003reducing} using
\textit{pycma} package~\cite{hansen2019pycma}. The reader is referred
to \citet{yuan2025dimensionless} for additional details.

\begin{figure}
    \centering
    \includegraphics[width=\linewidth]{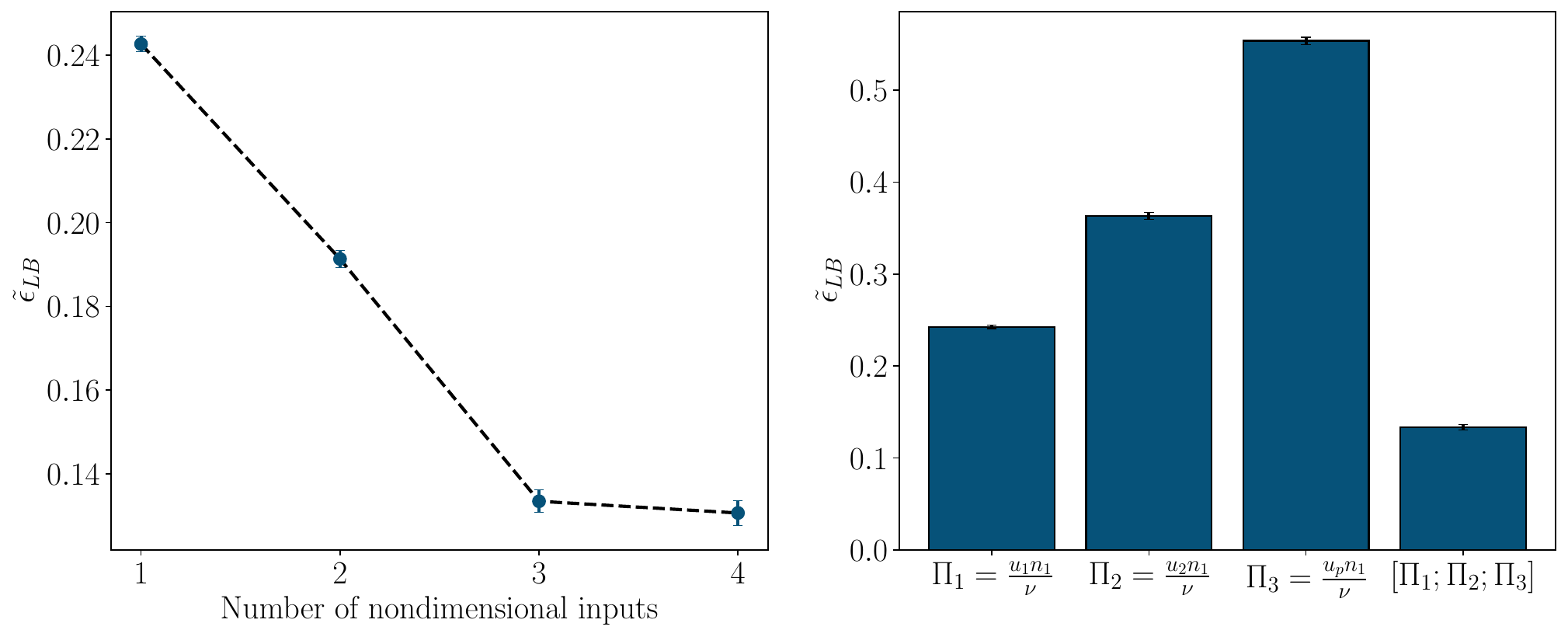}
    \caption{(Left) Optimal normalized irreducible error as a function
      of the number of dimensionless inputs considered. (Right)
      Normalized irreducible error for the optimal three-input
      solution, $\boldsymbol{\Pi} = [\Pi_1, \Pi_2, \Pi_3]$, evaluated
      using each individual dimensionless input and using all three
      inputs jointly. Error bars denote the estimated uncertainty in
      $\tilde{\epsilon}$.}
    \label{fig:irrep_error}
\end{figure}

The dataset used to identify the optimal dimensionless variables via
IT-$\pi$ corresponds to the training set introduced in
\S\ref{tab:trainingcases}, which is also employed later to train the
wall model. The optimization process was limited to a maximum of 400
iterations, with a population size of 200 and a convergence tolerance
of $10^{-5}$. We progressively increased the number of required input
features until the normalized irreducible error
$\tilde{\epsilon}_{LB}$ ceased to decrease. The results, shown in the
left panel of Figure~\ref{fig:irrep_error}, indicate that the optimal
number of inputs is three. when the number of inputs increases from 3
to 4, the reduction in the normalized irreducible error is marginal.
The three optimized dimensionless inputs and the optimal dimensionless
output are
\begin{equation}
  \boldsymbol{\Pi} = \left[ \frac{u_1 n_1}{\nu}, \frac{u_2 n_1}{\nu}, \frac{u_p n_1}{\nu} \right],
  \quad \Pi_o = \frac{u_{\tau} n_1}{\nu}.
\end{equation}
These inputs can be interpreted as local Reynolds numbers based on the
first and second off-wall grid points, as well as the
pressure-gradient velocity-scale Reynolds number, while the output
corresponds to the local friction Reynolds number. The irreducible
errors associated with each individual variable from the optimal
solution along with their combination are summarized in the right
panel of Figure~\ref{fig:irrep_error}. The predictive capabilities of
the inputs follow the ordering $\tilde{\epsilon}_{LB}(u_1 n_1/\nu) <
\tilde{\epsilon}_{LB}(u_2 n_1/\nu) < \tilde{\epsilon}_{LB}(u_p
n_1/\nu)$. This explains the strong predictive performance of EQWMs,
which typically rely on $u_1 n_1/\nu$. Incorporating $u_2 n_1/\nu$
enables detection of changes in the shape of the mean velocity profile
(due to PG effects, laminar flow, wall curvature, etc.). Finally, $u_p
n_1/\nu$ further assists in capturing pressure-gradient effects.

\subsection{Model architecture and training
\label{Sec:modeltraining}}

The regression task in terms of the optimal dimensionless variables is
defined as
\begin{equation}
\label{eq:IT_Regression}
\Pi_{\mathrm{o}} = \text{ANN} \left( \boldsymbol{\Pi} ; \boldsymbol{\theta} \right),
\end{equation}
where the input vector is given by $\boldsymbol{\Pi} = [u_1 n_1 /
  \nu,\; u_p n_1 / \nu,\; u_2 n_1 / \nu]$, and the output is
$\Pi_{\mathrm{o}} = u_{\tau} n_1 / \nu$ as determined in the
\S\ref{Sec:ITPI}.  We employ a multilayer perceptron (MLP)
architecture consisting of four hidden layers, each with forty
neurons. The activation function used in all layers is the Rectified
Linear Unit (ReLU). The selected network size has proven sufficient to
accurately fit the training data while avoiding overfitting.

To ensure balanced training across different flow regimes, weights
inversely proportional to the number of samples in each subcase from
Table~\ref{tab:trainingcases} are applied.  Specifically, let
$\mathcal{C} = \left\{ C_1, C_2, \ldots, C_n \right\}$ denote the
collection of $n = 67$ distinct flow datasets from subcases in
Table~\ref{tab:trainingcases}, where each dataset $C_i$ contains $N_i$
samples and $\sum_{i=1}^n N_i = N$. The weight $w_i$ assigned to each
sample $s \in C_i$ is defined as
\begin{equation}
\label{eq:weights}
w_i = \frac{N}{N_i}, \quad \text{if } s \in C_i.
\end{equation}
A weighted relative Huber loss is employed to reduce the influence of
outliers during training~\cite{huber1992robust}. The loss for a single
sample is given by
\begin{equation}
\label{eq:loss}
L_h\left(\Pi_{\mathrm{o}}^\text{true}, \text{ANN}(\boldsymbol{\Pi})\right) = 
\begin{cases}
\displaystyle \frac{1}{2} w \left( \frac{\Pi_{\mathrm{o}}^\text{true} - \text{ANN}(\boldsymbol{\Pi})}{|\Pi_{\mathrm{o}}^\text{true}|+\epsilon} \right)^2, 
& \text{if } \frac{|\Pi_{\mathrm{o}}^\text{true} - \text{ANN}(\boldsymbol{\Pi})|}{|\Pi_{\mathrm{o}}^\text{true}|+\epsilon} \leq \sigma, \\[12pt]
\displaystyle w \sigma \left( \frac{|\Pi_{\mathrm{o}}^\text{true} - \text{ANN}(\boldsymbol{\Pi})|}{|\Pi_{\mathrm{o}}^\text{true}|+\epsilon} - \frac{1}{2}\sigma \right),
& \text{otherwise},
\end{cases}
\end{equation}
where $\Pi_{\mathrm{o}}^\text{true}$ is the true  output, $\sigma = 1$ is
the threshold hyperparameter, $w$ is the sample weight defined in
Eq.~\eqref{eq:weights}, and $\epsilon=10^{-8}$ is a small constant
added to prevent division by zero. To prevent overfitting, an $L_2$
regularization term is added to the total loss function, defined as
\begin{equation}
\label{eq:loss_avg}
L = \frac{1}{N} \sum_{i=1}^{N} L_h\left(\Pi_{\mathrm{o},i}^\text{true}, \text{ANN}(\boldsymbol{\Pi}_i)\right)
+ \frac{\lambda}{2} \sum_{l=1}^{L} \left\| \mathbf{W}^{(l)} \right\|_F^2,
\end{equation}
where the index $i$ denotes the $i$-th training sample,
$\mathbf{W}^{(l)}$ is the weight matrix at the $l$-th layer of the
MLP, $\|\cdot\|_F$ denotes the Frobenius norm, and $\lambda = 10^{-5}$
is the regularization hyperparameter.

\begin{figure}
    \centering
    \includegraphics[width=.93\textwidth]{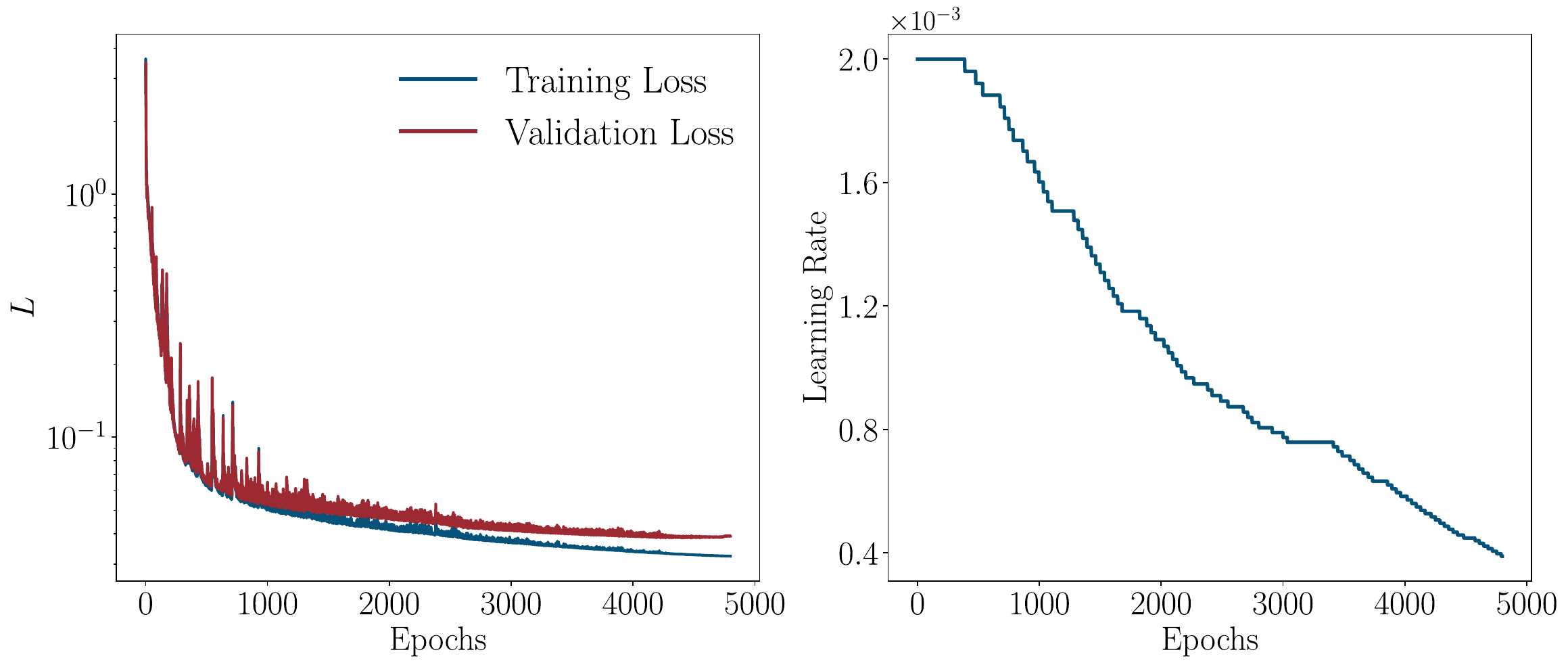}
    \caption{Left: Convergence history of the loss function $L$ as
      defined in Eq.~(\ref{eq:loss_avg}). Right: Learning rate history
      due to the learning rate scheduler.}
    \label{fig:hist}
\end{figure}

\begin{figure}
    \centering
    \includegraphics[width=.95\textwidth]{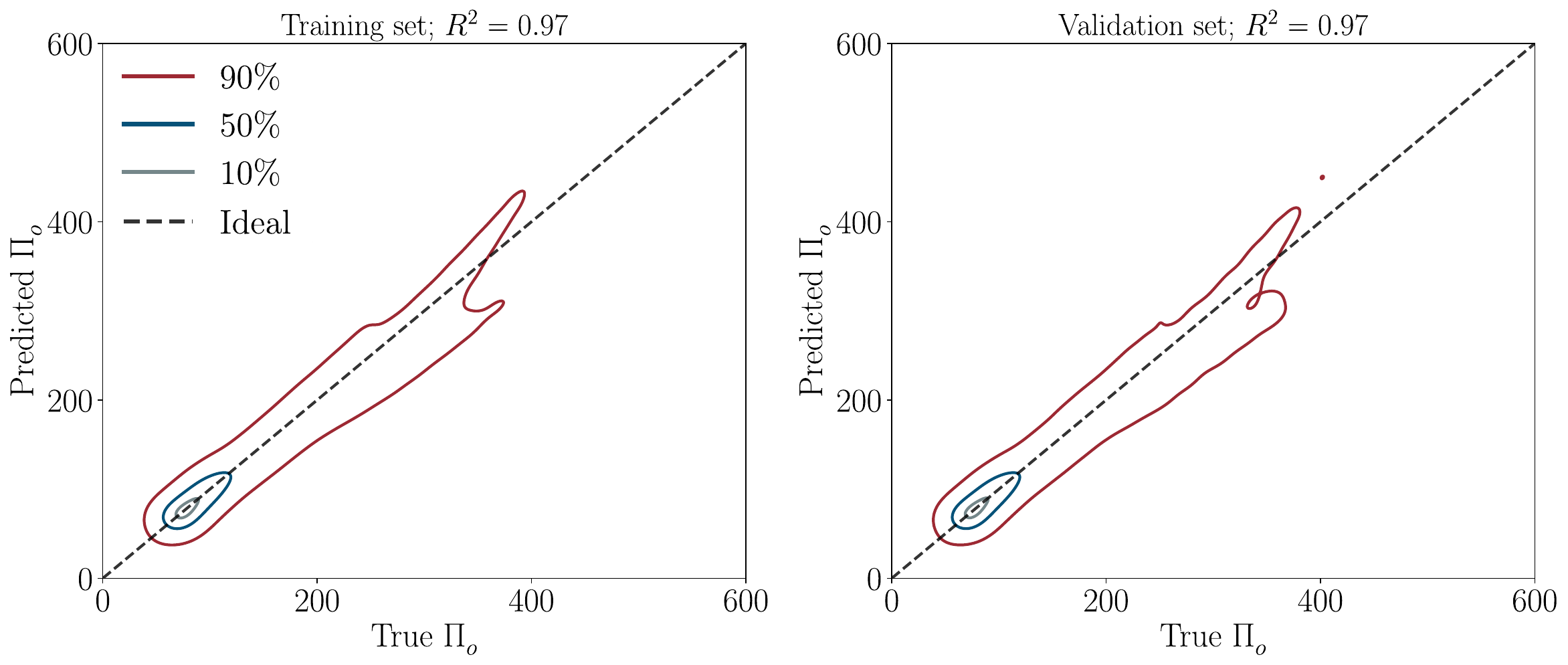}
    \caption{Joint probability density contours of predicted versus
      true values of $\Pi_o$.  Contours enclose 10\%, 50\%, and 90\%
      of the data for the training set (left) and validation set
      (right). In both cases, the high coefficient of determination
      ($R^2 = 0.97$) demonstrates strong agreement between the
      predicted and true outputs.  }
    \label{fig:reg_cont}
\end{figure}

The complete dataset was partitioned into training, validation, and
testing sets using a 7:1.5:1.5 ratio. Training was performed using the
Adam optimizer~\cite{kingma2014adam}, with decay rates $0.9$ and
$0.999$, and an initial learning rate of 0.002. A learning rate
scheduler was employed to reduce the learning rate by a factor of 0.95
when the validation loss failed to improve for more than 25
consecutive epochs.  To avoid overfitting, an early stopping criterion
was implemented, terminating training if the validation loss did not
decrease for 100 consecutive epochs. The model was trained using the
open-source PyTorch framework~\cite{paszke2019pytorch}. The
convergence and learning rate histories are shown in
Figure~\ref{fig:hist}. The performance of the model on both training
and validation sets are shown in Figure~\ref{fig:reg_cont}.

\section{Results}
\label{Sec:Results}

The collection of test cases comprises flow scenarios drawn from both
experimental studies and high-fidelity numerical simulations. These
cases have been compiled over the past 50 years through the collective
efforts of researchers worldwide. In total, 140 cases (training cases
included) are considered for evaluation. The selection is designed to
span a broad range of flow physics---including regimes for which the
model was not explicitly trained.

To evaluate the performance of our model, we adopt two complementary
testing methodologies: \textit{a priori} and \textit{a posteriori}
analyses. The \textit{a priori} test assesses model performance using
input–output pairs derived directly from high-fidelity data, bypassing
the need to run full WMLES. In contrast, the \textit{a posteriori}
test involves coupling the wall model to an SGS model and conductin
WMLES. In this work, we place greater emphasis on the \textit{a
  priori} assessment. This decision is motivated by the distinct
nature of the errors inherent in wall
modeling~\cite{lozano2022performance}, which can be broadly
categorized into two types, internal errors and external errors:
\begin{itemize}

\item \emph{Internal errors} represent the intrinsic physical
  limitations of the wall model. Even in the presence of exact (i.e.,
  error-free) input data, the wall model may produce inaccurate
  results when its physical assumptions do not hold. For example,
  internal errors can arise from the inadequacy of the model to
  accurately model the physics of the near-wall region (e.g.,
  compressibility effects, separation patterns not captured in the
  training database, three-dimensional turbulent boundary layer
  effects, etc.).  One example of internal error is the erroneous
  prediction of wall-shear stress by the classic equilibrium wall
  model when using the mean velocity profile from a separation bubble
  obtained via DNS as input. Although the input is error-free, the
  prediction may still exhibit significant inaccuracies, as the model
  is not designed to account for separated flows.

\item \emph{External errors} are introduced by exogenous factors
  unrelated to the wall model itself. These errors originate from
  imperfect input data in actual WMLES. A dominant contributor is the
  performance of the SGS model: underperformance of the SGS model
  leads to inaccurate mean profiles, which in turn propagate errors
  into the wall model predictions. Additional sources of external
  error include history effects (e.g., imperfect inflow conditions),
  inadequate grid resolution (e.g., too few points per boundary-layer
  thickness), and uncertainties in other boundary conditions, to name
  a few. These errors are classified as external because they persist
  even when the wall model provides an exact physical representation
  of the near-wall region.  The combined external and internal errors
  are collectively referred to as the total error.  Continuing the
  example above, the equilibrium wall model will yield external errors
  in the prediction of wall-shear stress even when applied to a
  flat-plate ZPG TBL, if the input mean velocity profiles are
  inaccurate. This can occur even if the model was previously
  calibrated to produce nearly exact predictions in ZPG TBLs.

\end{itemize}

The distinction between internal and external errors plays a critical
role in the design and evaluation of wall models. Internal errors
arise from intrinsic limitations of the model itself, whereas external
errors originate from factors beyond the model. In a practical WMLES
setup (i.e., \textit{a posteriori} testing), both types of errors are
present, making it challenging to isolate and assess the contribution
of each.  The situation is further complicated by potential error
cancellation between internal and external sources, which can be
misleading and even counterproductive, leading to erroneous
conclusions about the true performance of the wall model. Since the
primary objective of this work is to evaluate the wall model, we focus
on \textit{a priori} testing as the principal means of performance
assessment.

While our primary objective is to minimize internal errors, we also
aim to evaluate the performance of our wall model within a realistic
solver environment using \textit{a posteriori} testing. To this end,
we conduct the analysis using two modalities: \textit{nudged a
  posteriori} testing and \textit{true a posteriori} testing.  In the
former, we employ the nudged simulation approach proposed by
\citet{lingNumericallyConsistentDataDriven2025}, which mitigates the
influence of the SGS model to reduce external wall-modeling
errors. This method consists of performing \textit{a posteriori}
simulations in which the flow fields are nudged toward the correct
statistical state of the flow. Specifically, a forcing term is added
to the momentum equation to drive the velocity field toward the target
mean velocity profile.  In contrast, the \textit{true a posteriori}
test evaluates the total modeling error without explicitly separating
internal and external contributions as encountered in practical
WMLES. Addressing these external errors in WMLES requires advances in
the SGS model, as demonstrated by
BFM-v1~\cite{arranzBuildingblockflowComputationalModel2024}.  The
development of its successor, BFM-SGS-v2, is presented in a subsequent
work.

\subsection{\textit{A priori} testing}
 \label{Sec:Apriori}

 %
The set of cases used for \textit{a priori} testing is summarized in
Table~\ref{tab:testcases}. Each entry in the table typically
represents a family of flow conditions, yielding a total of 70
distinct cases, none of which appear in the training set. Providing a
full account of all individual variations would be cumbersome;
therefore, we do not detail them here. Interested readers are referred
to the original sources (cited in the table) for detailed descriptions
of the corresponding simulations or experiments.  Instead, we provide
a high-level overview of the scenarios considered. The test set
includes internal flows (pipe, square duct, rotating channel),
turbulent boundary layers under zero, favorable, and adverse pressure
gradients, and a variety of separated flows---such as pressure-induced
separation bubbles with and without sweep, converging–diverging
channels, backward-facing steps, Gaussian bumps, smooth ramps,
axisymmetric separated boundary layers, rounded steps, and
wall-mounted humps. The set also comprises strongly three-dimensional
and curvature-dominated flows, including concave, convex, and curved
walls; periodic hills; laterally strained and spinning cylinders;
swept wings; a 3D diffuser; and airfoils spanning laminar,
transitional, and near-stall regimes. In addition to these cases, the
training set in Table~\ref{tab:trainingcases}---which is not included
in Table~\ref{tab:testcases}---is also used to evaluate the wall
model, contributing an additional 67 cases. For each case, we examine
multiple spatial positions along the walls as well as a range of
matching locations (corresponding to different WMLES grid resolutions)
spanning $0.005$--$0.25\,\delta$. Between 10 and 100 matching
locations are evaluated at each spatial position, yielding a total of
$\mathcal{O}(1000)$ case scenarios.
%
\begin{table}[h!]
    \centering
    \caption{Summary of the cases used for \textit{a priori}
      testing. \textbf{Symbols} identify each case consistently across
      subsequent figures, and \textbf{Count} denotes the number of
      subcases with distinct flow conditions included under each main
      case.}
    \label{tab:testcases}
    \begin{tabular}{p{7.3cm}p{4cm}lcc}
        \toprule
        \textbf{Case} & \textbf{Dominant Flow Physics} & \textbf{Type} & \textbf{Symbol} & \textbf{Count} \\
        \midrule
        Pipe (\citet{pirozzoliOnepointStatisticsTurbulent2021}) & ZPG & DNS & $\mathbf{p}$ & 6 \\
        \midrule
        Square duct (\citet{pirozzoli_turbulence_2018, modesti_role_2018}) & Secondary & DNS & $\mathbf{d}$ & 4 \\
        \midrule
        Channel with (spanwise) rotation (\citet{andersson1995turbulence}) & Rotation & DNS & $\mathbf{r}$ & 7 \\
        \midrule
        TBLs subjected to mild APG (\citet{hirt1986measurement}) & PG & EXP & \adjustbox{valign=c}{\begin{tikzpicture}
  \node[regular polygon, regular polygon sides=6, draw] at (0,0) {}; 
\end{tikzpicture}} & 1 \\
        \midrule
        TBL subjected to FPG followed by APG (\citet{volino_non-equilibrium_2020}) & PG & EXP & $\triangleright$ & 8\\
        \midrule
        TBLs subjected to mild APG (\citet{bobkeHistoryEffectsEquilibrium2017}) & PG & WRLES & $\circ$ & 5 \\
        \midrule
        TBLs subjected to FPG and APG (\citet{gungor2022energy,gungor2024turbulent}) & PG & DNS & \textbf{g} & 3 \\
        \midrule
        Backward facing step (\citet{driverFeaturesReattachingTurbulent1985}) & Separation & EXP & $+$ & 2 \\
        \midrule
        Pressure-induced separation bubble (\citet{colemanNumericalStudyTurbulent2018}) & Separation & DNS & \adjustbox{valign=c}{\input{Figures/Symbols/Right}} & 3 \\
        \midrule
        Separation bubble with sweep (\citet{coleman_numerical_2019}) & Separation + 3D TBL & DNS & $|$ & 1 \\
        \midrule
        Converging-diverging channel (\citet{marquillieInstabilityStreaksWall2011}) & Separation + PG & DNS & \adjustbox{valign=c, scale=0.7}{\input{Figures/Symbols/ThinDiamond}} & 1 \\
        \midrule
        Axisymmetric separated boundary layer (\citet{driverFeaturesReattachingTurbulent1985}) & Separation + PG & EXP & \adjustbox{valign=c}{\input{Figures/Symbols/Left}}  & 1 \\
        \midrule
        Convex curvature boundary layer (\citet{smitsEffectShortRegions1979}) & Curvature + PG & EXP & \adjustbox{valign=c}{\input{Figures/Symbols/Up}}  & 1 \\
        \midrule
        Concave bend (\citet{barlow_structure_1988}) & Curvature + PG & EXP & $C$ & 1 \\
        \midrule
        Smooth ramp (\citet{simmonsExperimentalCharacterizationSmooth2022,uzun2024direct}) & Curvature + PG & EXP/DNS & \adjustbox{valign=c}{\begin{tikzpicture}
  \node[regular polygon, regular polygon sides=5, draw] at (0,0) {}; 
\end{tikzpicture}} & 1 \\
        \midrule
        Curved boundary layer (\citet{appelbaumSystematicDNSApproach2025}) & Curvature + PG & DNS & $\triangle$ & 2 \\
        \midrule
        Gaussian bump ($Re_L=1M$) (\citet{uzun2021simulation}) & Curvature + PG & DNS & $\star$ & 2 \\
        \midrule
        NACA 0012/4412 airfoils (\citet{tanarroEffectAdversePressure2020}) & Curvature + PG & DNS & $\mathbf{n}$  & 2 \\
        \midrule
        Periodic hill (\citet{balakumarDNSSimulationsSeparated2015, gloerfeltLargeEddySimulation2019}) & Curvature + PG & DNS &  \adjustbox{valign=c}{\input{Figures/Symbols/Down}}  & 1 \\
        \midrule
        Rounded Step (\citet{bassi2016development}) & Curvature + PG & DNS & $\square$  & 1 \\
        \midrule
        A-airfoil near stall (\citet{tamakiWallResolvedLargeEddySimulation2023}) & Curvature + PG & WRLES & $\mathbf{a}$ & 2 \\
        \midrule
        NASA wall-mounted hump (\citet{uzun2018large}) & Curvature + PG & WRLES & \adjustbox{valign=c}{\begin{tikzpicture}
  \node[regular polygon, regular polygon sides=6, rotate=30,draw] at (0,0) {}; 
\end{tikzpicture}} & 1 \\
        \midrule
        Laminar NACA0012 airfoil (\citet{swanson2016comparison}) & Laminar + Curvature + PG & DNS & $\mathbf{0}$ & 1 \\
        \midrule
        Transitional boundary layer (\citet{roach1990influence}) & Transition & EXP & \adjustbox{valign=c}{\input{Figures/Symbols/Cross_Filled}}  & 8 \\
        \midrule
        Transitional boundary layer (\citet{lee_large-scale_2017}) & Transition & DNS & \adjustbox{valign=c}{\input{Figures/Symbols/Plus_Filled}}  & 1 \\
        \midrule
        35deg yawed wing (\citet{vandenbergMeasurementsIncompressibleThreedimensional1975}) & 3D TBL & EXP & \adjustbox{valign=c}{\begin{tikzpicture}
  \node[regular polygon, regular polygon sides=8, draw] at (0,0) {}; 
\end{tikzpicture}} & 1 \\
        \midrule
        Laterally strained boundary layers (\citet{pompeo1993laterally}) & PG + 3D TBL & EXP & $\triangleleft$ & 3 \\
        \midrule
        3D TBL driven by a spinning cylinder (\citet{driver1987experimental}) & PG + 3D TBL & EXP & $\times$ & 5 \\
        \midrule
        3D diffuser (\citet{ohlssonDirectNumericalSimulation2010,lehmkuhl2019low}) & PG + 3D TBL & DNS & $\mathbf{3}$ & 1 \\
        \bottomrule
    \end{tabular}
\end{table}

We evaluate the performance of the wall model using two complementary
error metrics, chosen based on the proximity of the flow to incipient
separation. The rationale for using two metrics is to avoid spuriously
large relative errors when $\tau_{w,\mathrm{true}}\!\to 0$, a
situation that typically arises near separation. The \emph{relative
error} is defined as
\begin{equation}
  \label{eq:relative_error}
\varepsilon(\%) =
\dfrac{|\tau_{w,\text{true}} - \tau_{w,\text{pred}}|}
      {|\tau_{w,\text{true}}|},
\end{equation}
and the \emph{dimensionless error} is defined as
\begin{equation}
    \label{eq:dimensionless_error}
\varepsilon =
\dfrac{|\tau_{w,\text{true}} - \tau_{w,\text{pred}}|}
      {\rho U_o^2}.
\end{equation}
Here, $U_o$ is an outer velocity scale
used for normalization (e.g., freestream velocity for external flows
and bulk inlet velocity for internal flows). In what follows, when the
flow is far from incipient separation---quantified by the condition
$|u_{\tau,\text{true}}|\delta/\nu > 50$---we report the relative error
$\varepsilon(\%)$. Conversely, when the flow is close to incipient
separation, i.e., $|u_{\tau,\text{true}}|\delta/\nu \le 50$, we assess
performance using the dimensionless error $\varepsilon$.

For reference, we compare the performance of BFM-WM-v2 against a
standard algebraic equilibrium wall model (EQWM). The friction
velocity $u_{\tau}$ in the EQWM is obtained by solving
\begin{equation}
\label{eq:eqwm}
u^{+}(n_1^{+}) =
\begin{cases}
n_1^{+} + a_1 \left(n_1^{+}\right)^2, & \text{for } n_1^{+} < 23, \\[6pt]
\dfrac{1}{\kappa} \ln n_1^{+} + B, & \text{otherwise},
\end{cases}
\end{equation}
where $n_1$ is the wall-normal matching location, $(\cdot)^+$ denotes
normalization by the friction velocity $u_{\tau}$ and kinematic viscosity
$\nu$, $\kappa = 0.41$ is the von Kármán constant, and $B = 5.0$ is the
log-law intercept~\cite{deardorffNumericalStudyThreedimensional1970,
piomelliWalllayerModelsLargeeddy2002,
larssonLargeEddySimulation2016}.

\subsubsection{Overview of model performance}

\begin{figure}
    \centering
    \includegraphics[width=0.95\linewidth]{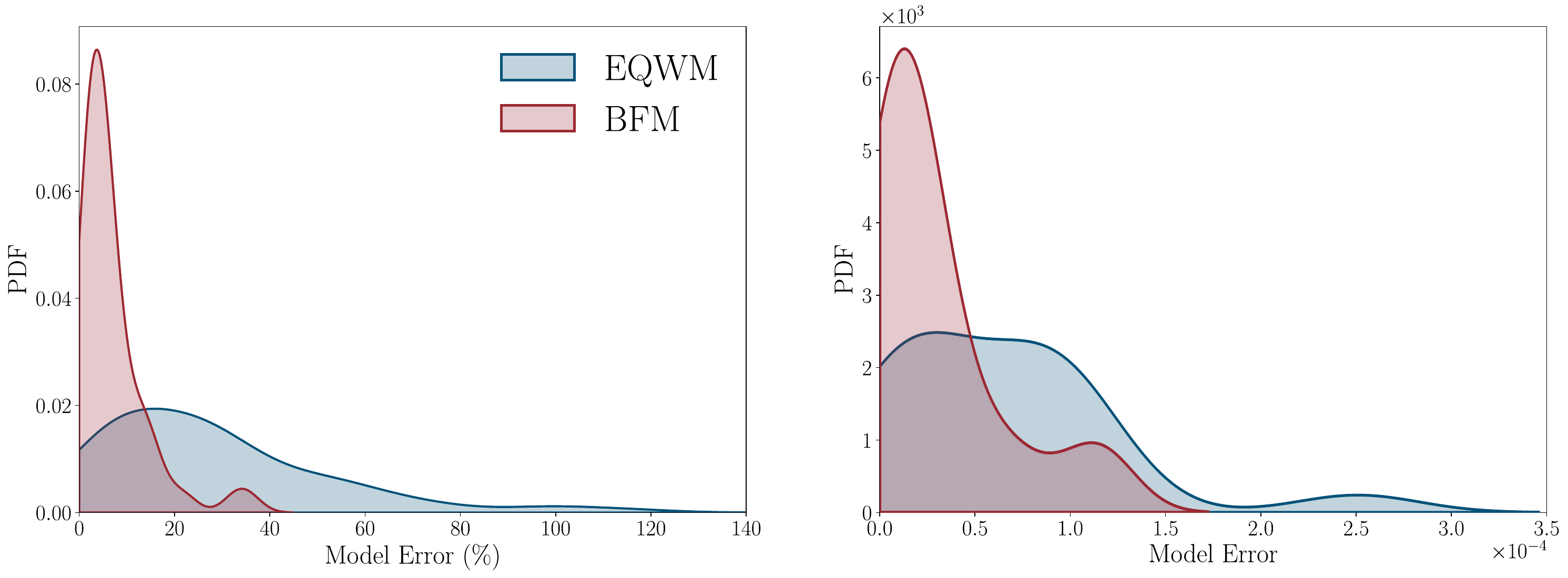}
    \caption{\textit{A priori} error distribution for BFM and EQWM for
      all cases considered.  (a) Relative error
      (Eq.~\ref{eq:relative_error}) for sample points far from
      incipient separation, i.e., $|u_{\tau,\text{true}}|\delta/\nu >
      50$.  (b) Dimensionless error (Eq.~\ref{eq:dimensionless_error})
      for sample points close to incipient separation, i.e.,
      $|u_{\tau,\text{true}}|\delta/\nu \le 50$.}
    \label{fig:err_all}
\end{figure}

We begin by providing an overview of the overall performance of
BFM-WM-v2 compared with EQWM across all training and testing
cases. For brevity in the notation, we refer to BFM-WM-v2 simply as
BFM from this point onward.   For each case, the errors are averaged
over all spatial positions and matching locations, and the resulting
mean values are reported.  Figure~\ref{fig:err_all} presents the
probability density function (PDF) of these mean errors for all cases,
comparing BFM with EQWM. The results show that BFM consistently
outperforms EQWM, in most instances providing substantially improved
accuracy.

\begin{figure}
  \centering
  \begin{subfigure}[b]{.9\linewidth}
      \includegraphics[width=\linewidth]{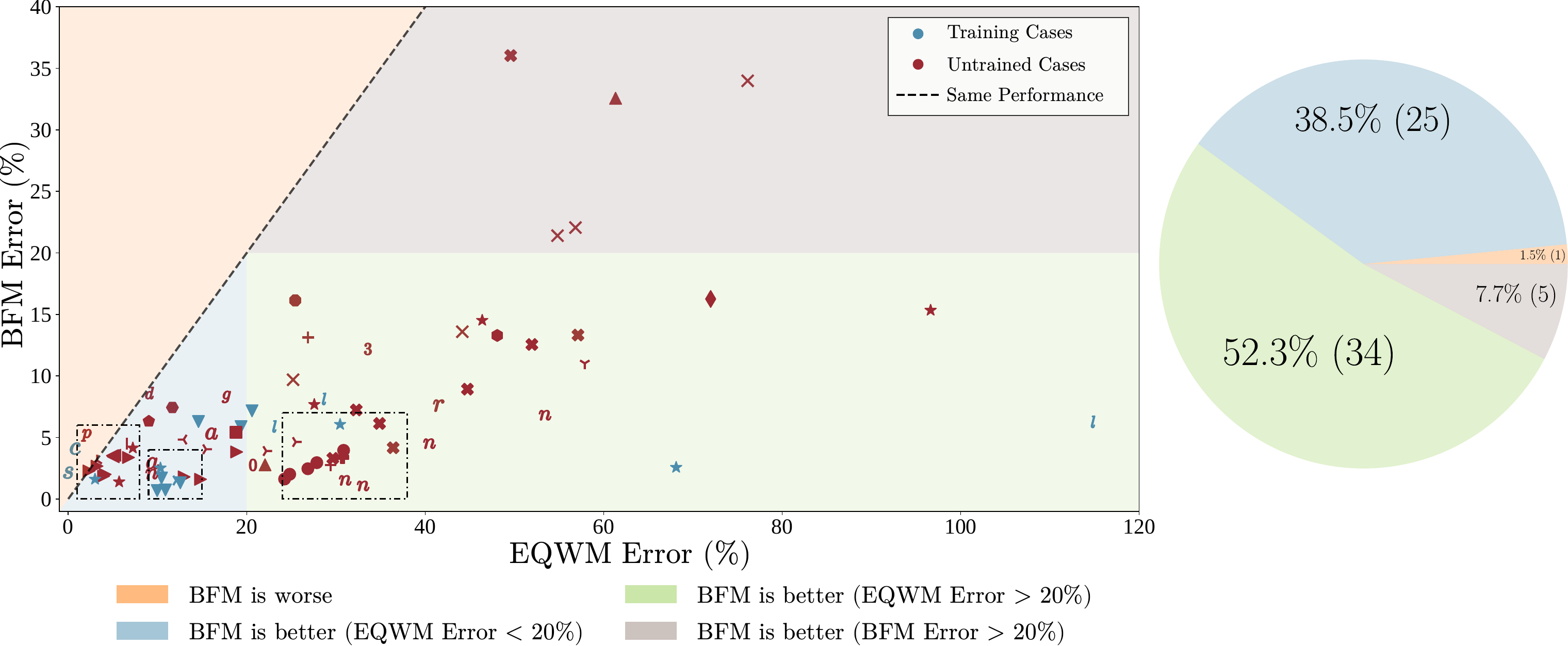}%
    \caption{Relative errors for BFM vs.  EQWM}
      \label{fig:sub:comp_rele}
  \end{subfigure}
  \hfill
  \begin{subfigure}[b]{.33\linewidth}
      {\includegraphics[width=1\linewidth]{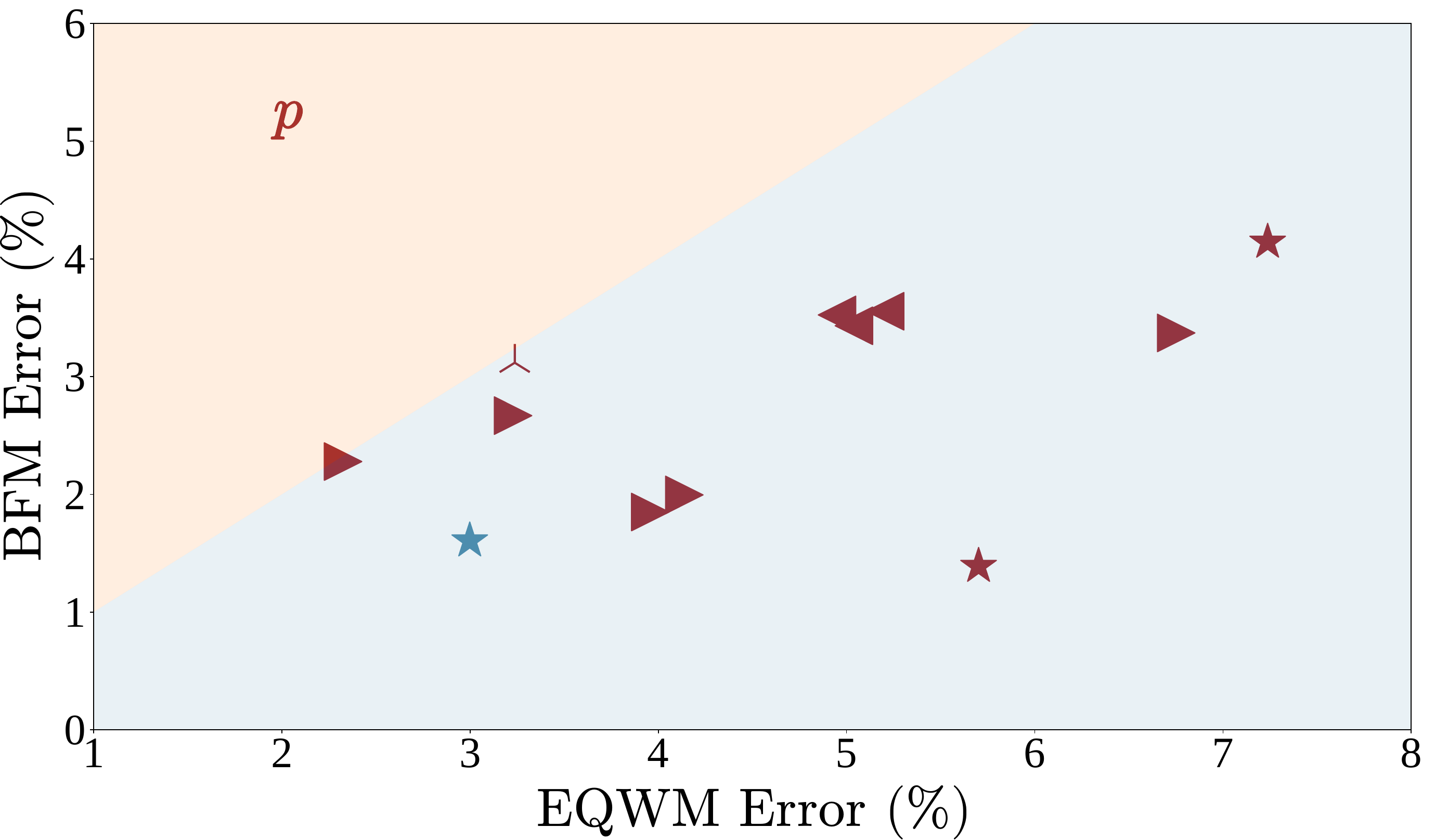}}
    \caption{Zoomed-in region left}
  \end{subfigure}%
  \hfill
  \begin{subfigure}[b]{.325\linewidth}
      {\includegraphics[width=1\linewidth]{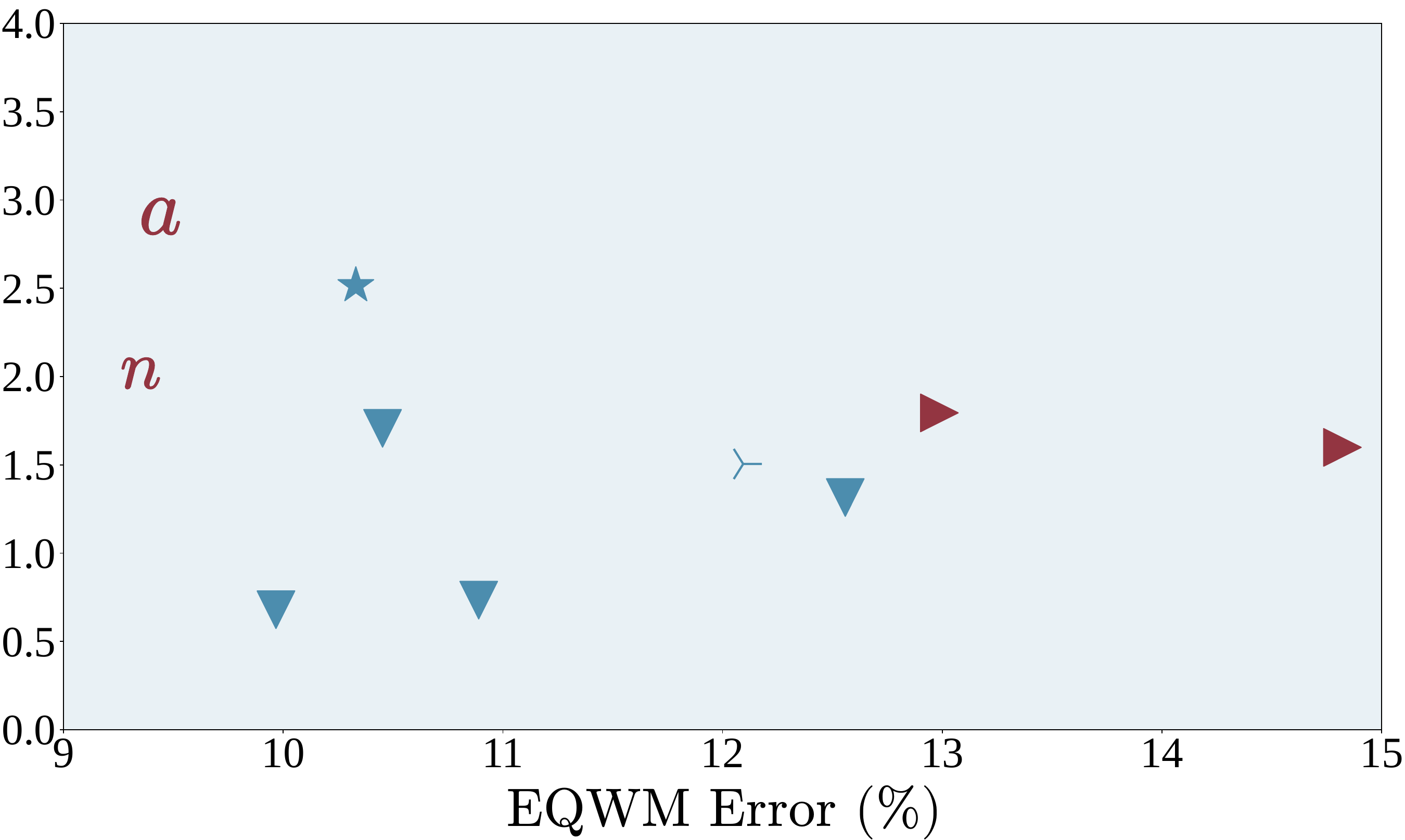}}
    \caption{Zoomed-in region middle}
  \end{subfigure}%
  \hfill
  \begin{subfigure}[b]{.325\linewidth}
      {\includegraphics[width=1\linewidth]{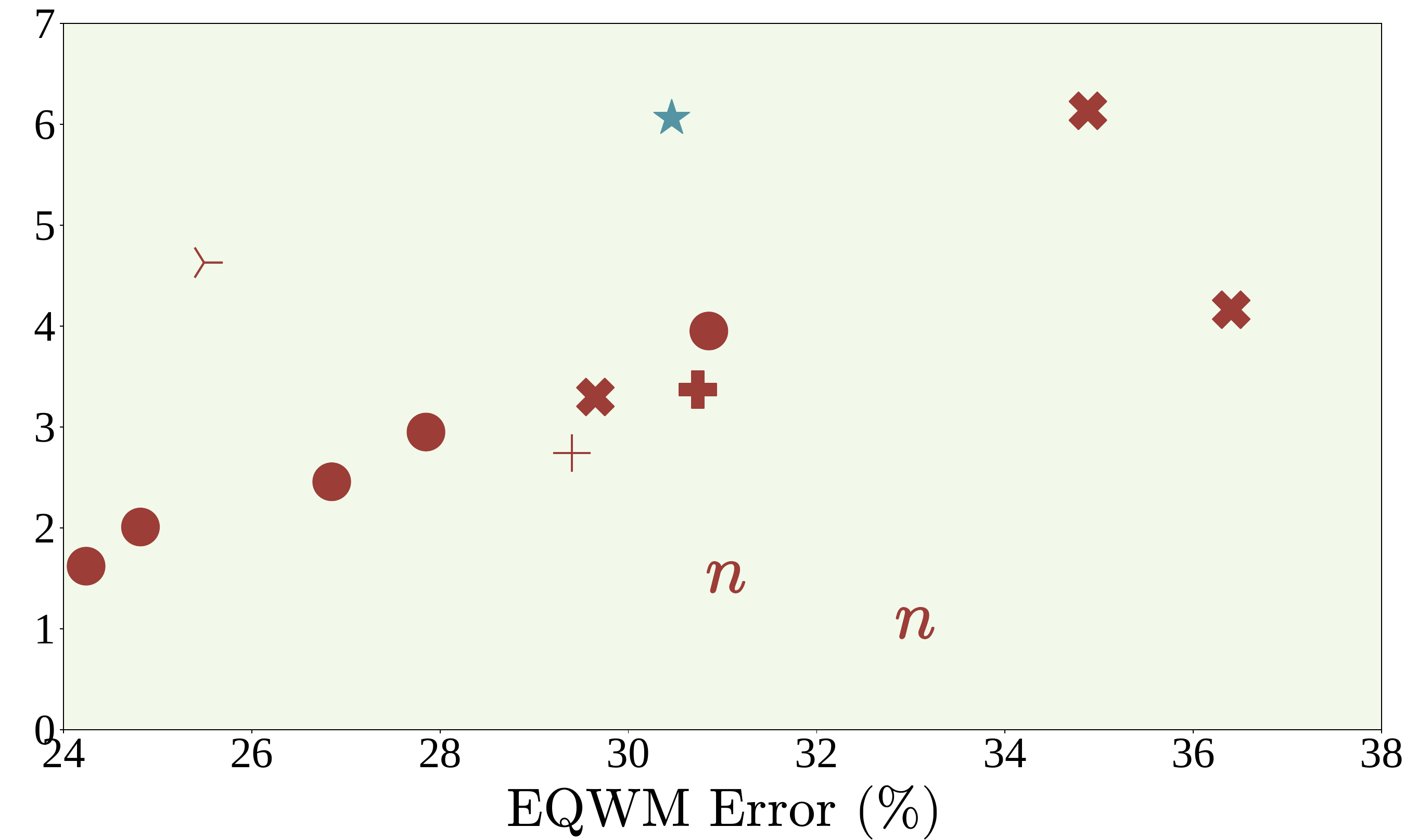}}
    \caption{Zoomed-in region right}
  \end{subfigure}%
  \caption
      { (a) \textit{A priori} relative errors
        (Eq.~\ref{eq:relative_error}) of BFM versus EQWM for sample
        points far from incipient separation. Symbols follow the
        conventions defined in Tables~\ref{tab:trainingcases} and
        \ref{tab:testcases}. Each symbol correspond to the mean error of a subcase in testing cases or training case. Cases with only change in Reynolds numbers are merged to fewer symbols to save space.
        Colored regions indicate the outcome
        categories described in the legend. The accompanying pie chart
        reports the percentage and number (in parentheses) of subcases
        falling within each category. Panels (b--d) show enlarged
        views of three regions with dense symbol overlap, highlighted
        by black dash-dotted boxes in the main plot for
        clarity.  \label{fig:AprioriResults}%
    }%
\end{figure}

\begin{figure}[h]
    \centering
    \includegraphics[width=0.7\linewidth]{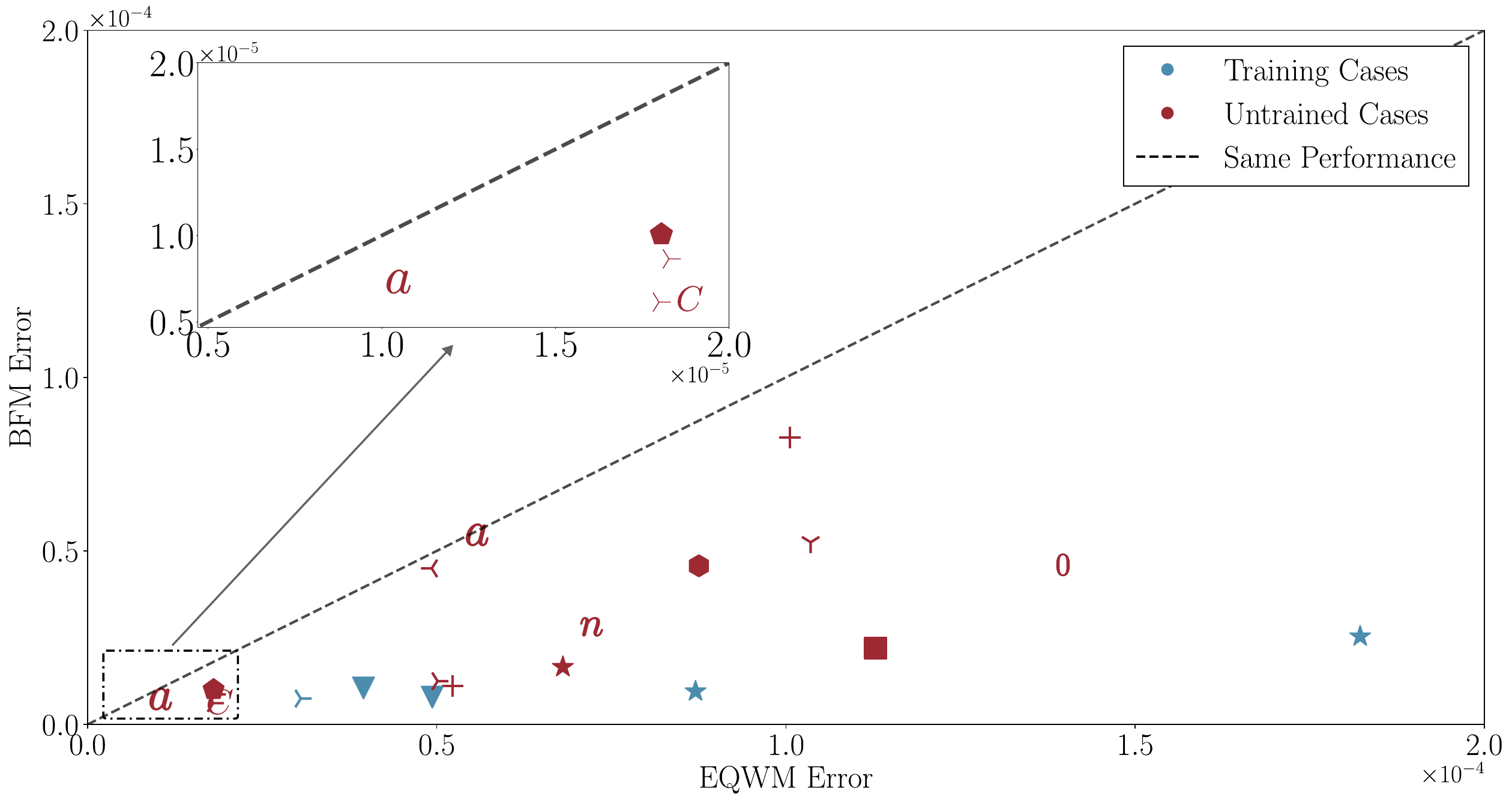}
    \caption{ \textit{A priori} dimensionless errors
      (Eq.~\ref{eq:dimensionless_error}) of BFM versus EQWM for sample
      points near incipient separation. Symbols follow the conventions
      defined in Tables~\ref{tab:trainingcases} and
      \ref{tab:testcases}. The inset provides a zoomed-in view of the
      region near the origin. One outlier—omitted from the main
      plot—corresponds to an APG Falkner--Skan laminar flow case, with
      $\varepsilon = 5.12\times 10^{-4}$ and $\varepsilon = 3.2\times
      10^{-5}$ for EQWM and BFM, respectively.}
      \label{fig:AprioriResults_abs}
\end{figure}

Further insight into the model performance is provided in
Figure~\ref{fig:AprioriResults} and
Figure~\ref{fig:AprioriResults_abs}.  The former compares the average
errors of EQWM and BFM for cases far from incipient separation (i.e.,
$|u_{\tau,\text{true}}|\,\delta/\nu > 50$), while the latter presents
results for cases close to separation (i.e.,
$|u_{\tau,\text{true}}|\,\delta/\nu \le 50$). The symbols used for
each case follow the conventions listed in Table~\ref{tab:testcases}
and Table~\ref{tab:trainingcases}. Multiple occurrences of the same
symbol indicate subcases corresponding to different flow conditions
within the same main case.  Based on these results, we can draw the
following conclusions:
\begin{enumerate}
    \item \emph{For training cases}, BFM demonstrates high accuracy,
      confirming that the model has been properly constructed.  This
      is reflected by the blue symbols in the figure, none of which
      exhibit relative errors exceeding 10\%. For cases near incipient
      separation, the dimensionless errors remain small---as low as
      $\mathcal{O}(10^{-5})$---which is an order of magnitude lower
      than the typical errors produced by EQWM at the same sample
      points.
    \item \emph{For untrained cases}, BFM outperforms EQWM across all
      scenarios considered, as indicated by the symbols lying below
      the dashed line representing equal performance. For more than
      half of the cases, the error is reduced from above 40\% to below
      20\%. From the chart summarizing the number of cases in each
      region (Figure~\ref{fig:sub:comp_rele}), we observe that more than
      50\% of the cases undergo major improvements.
    \item An exception to the observations above occurs in three cases
      that favor EQWM---turbulent channel flow, turbulent pipe flow,
      and synthetic data generated from the law of the wall. These
      results are expected, as EQWM is specifically calibrated for
      these canonical flows. Nonetheless, the performance of BFM is
      only marginally worse, with errors remaining close to 1\%.
   \item \emph{Underperforming cases} are identified in the right gray
     region, where BFM yields errors above 20\%. Even so, BFM still
     significantly outperforms EQWM. These cases include the curved
     TBL by \citet{appelbaumSystematicDNSApproach2025}, the
     transitional boundary layer by \citet{roach1990influence}, and
     the 3D TBL driven by a spinning cylinder by
     \citet{driver1987experimental}. A detailed analysis of the
     potential causes of BFM underperformance in these cases is
     provided in the next section.
   \item \emph{For cases near separation}, BFM provides an
     order-of-magnitude improvement, with errors on the order of
     $10^{-5}$ compared to $10^{-4}$ for EQWM. This highlights the
     inaccuracy of EQWM near incipient separation and demonstrates the
     advantage offered by BFM.
   \item \emph{For laminar flows}, BFM also outperforms EQWM, with the
     latter exhibiting relative errors exceeding 100\% in some
     instances. An exception is the Falkner--Skan laminar separation
     case, which is omitted from the plot because its relative errors
     are extremely large. The corresponding dimensionless errors are
     $\varepsilon = 5.12\times 10^{-4}$ for EQWM and $\varepsilon =
     3.2\times 10^{-5}$ for BFM, demonstrating an order-of-magnitude
     improvement with BFM.
\end{enumerate}
%

\subsubsection{Detailed analysis for selected cases}

We select five representative cases for detailed analysis. These
include: (i) Falkner--Skan laminar flow, to compare model behavior in
laminar regimes; (ii) a TBL subjected with mild pressure gradient, to
examine sensitivity to pressure-gradient effects; (iii) square duct
flow, to evaluate performance in flows with strong mean-flow
three-dimensionality; (iv) the Aerospatiale A-airfoil, as a realistic
high-Reynolds-number configuration; and (v) a two-dimensional Gaussian
bump at $Re_L = 1{,}000{,}000$, to assess the ability of the model to
handle relaminarization phenomena not present in the training
database.  In addition, we investigate the causes of underperformance
of BFM in cases exhibiting relative errors greater than 20\%, i.e.,
those falling within the gray region of
Figure~\ref{fig:AprioriResults}.

In the following analysis, we project the model error at each wall
location and for all matching heights onto two of the three
dimensionless inputs: $\Pi_1 = u_1 n_1 / \nu$ and $\Pi_3 = u_p n_1 /
\nu$. As discussed in \S\ref{Sec:ITPI}, $\Pi_1$ represents a local
near-wall Reynolds number, while $\Pi_3$ captures the influence of
pressure-gradient effects. This projection therefore enables a more
granular assessment and facilitates interpretation of the error
distribution as a function of local Reynolds number and pressure
gradient. Each scatter point is color-coded by its relative error.
Above each scatter plot, a relative frequency plot is provided to summarize the distribution of relative
errors, ranging from 0\% and capped at 100\%. For cases containing
samples with prediction errors above 100\%, a small inset box reports
the percentage of these outliers. For reference, EQWM results are also
shown within the $(\Pi_1,\Pi_3)$ input space, although EQWM formally
depends only on $\Pi_1$.

\begin{figure}
    \centering
    \begin{subfigure}[b]{0.6\textwidth}
        \includegraphics[width=\linewidth]{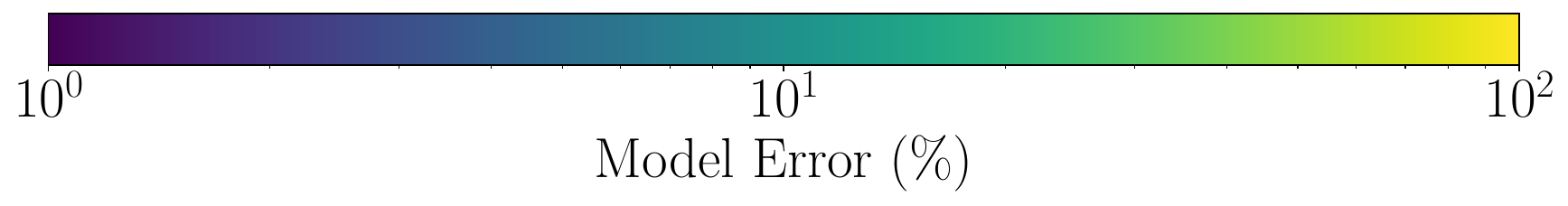}
        \label{fig:colorbar}
    \end{subfigure}
    \hfill
    \begin{subfigure}[b]{0.45\textwidth}
        \includegraphics[width=\linewidth]{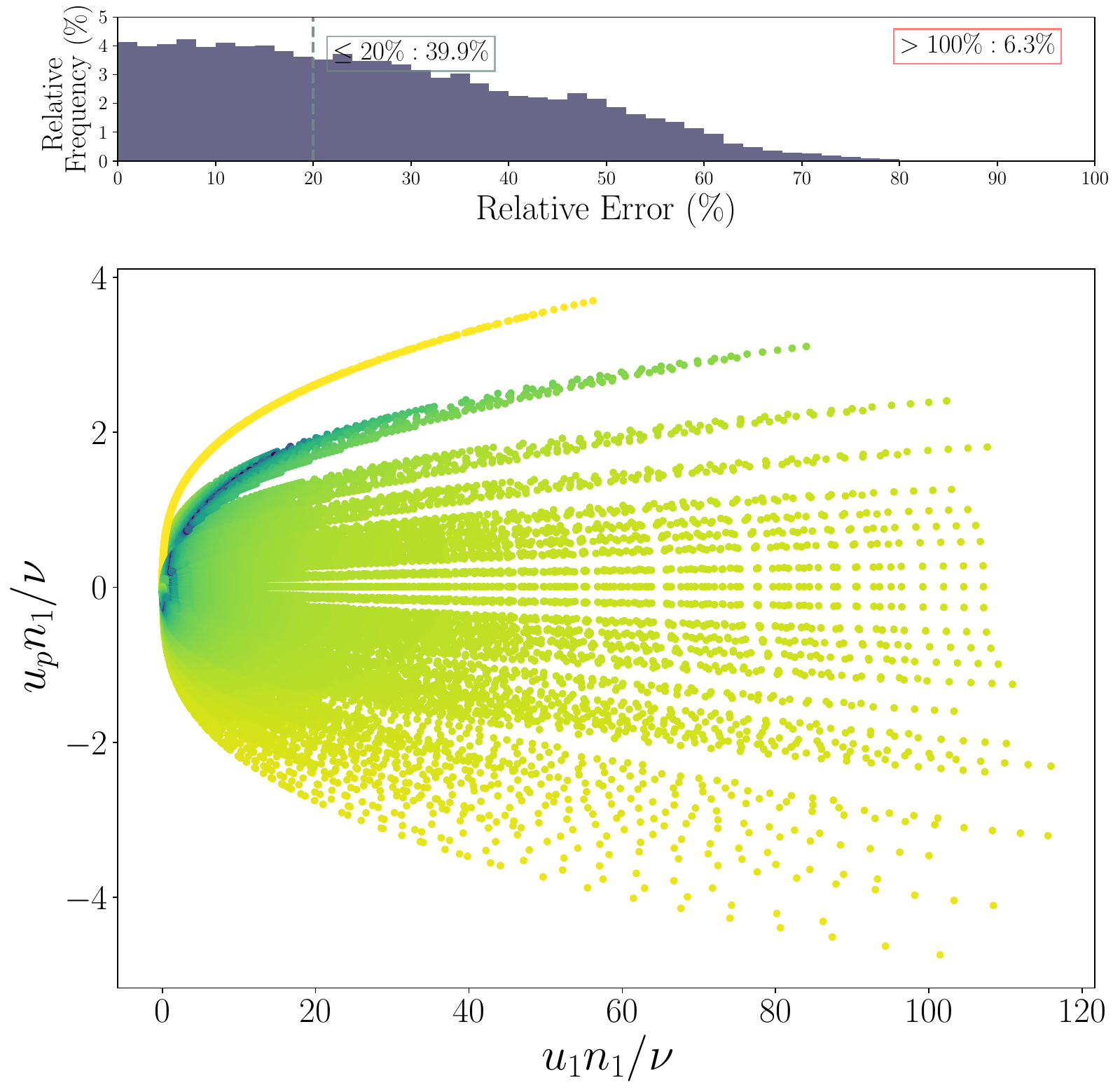}
        \caption{EQWM relative error}
        \label{fig:sub:fs_all}
    \end{subfigure}
    \hfill
    \begin{subfigure}[b]{0.45\textwidth}
        \includegraphics[width=\linewidth]{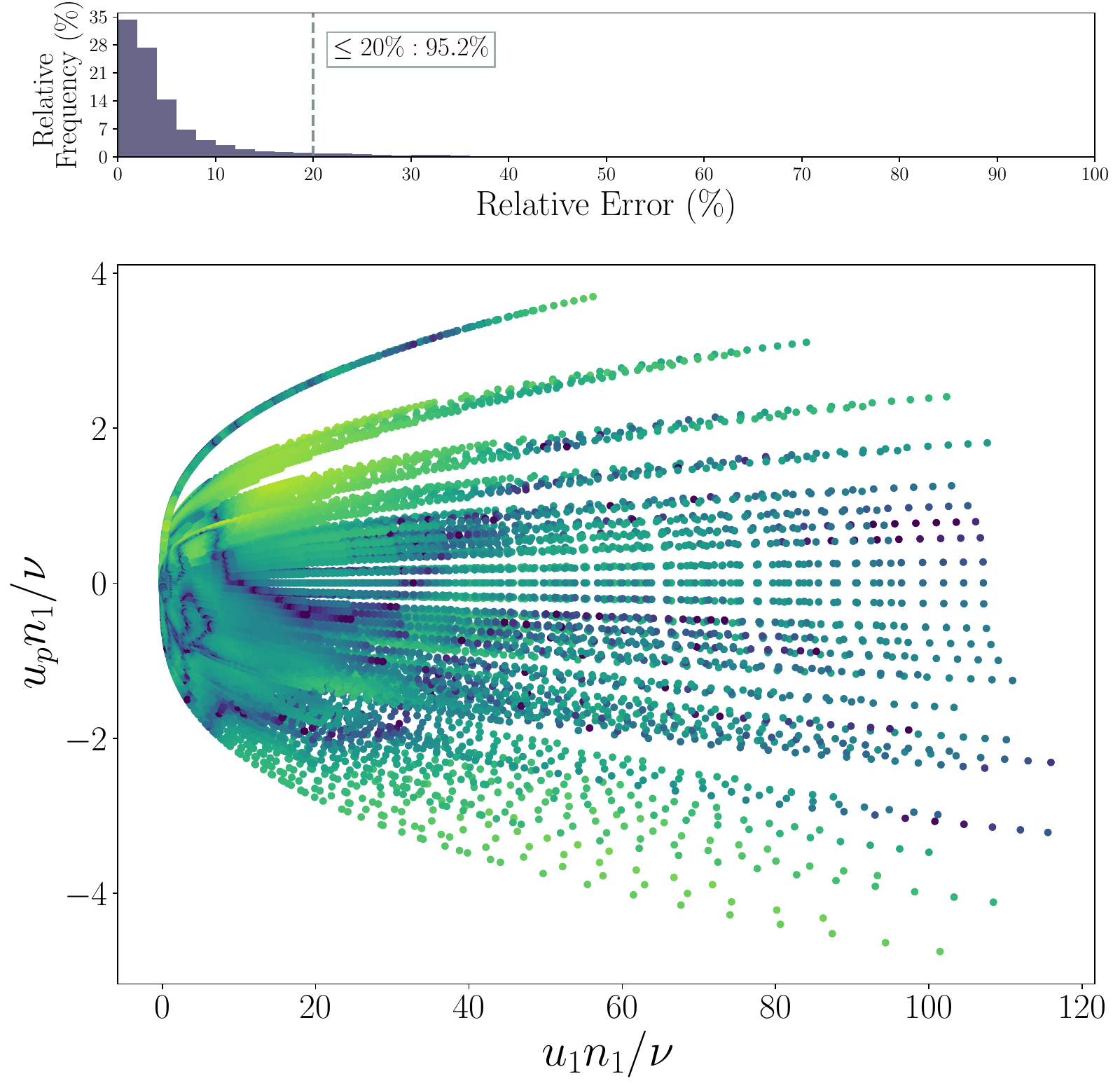}
        \caption{BFM relative error}
        \label{fig:sub:fs_bfm}
    \end{subfigure}
    \caption{ \textit{A priori} relative errors of (b) EQWM and (c)
      BFM for the Falkner--Skan laminar boundary layer, shown as a
      function of two BFM input variables.}
    \label{fig:fs}
\end{figure}

\paragraph{Falkner-Skan laminar boundary layer.}
Figure~\ref{fig:fs} shows the results for the laminar cases included in
the BFM training set. As expected, EQWM—originally designed for turbulent
flows—fails to generalize to laminar regimes. Its performance degrades
substantially, with more than 60\% of data points exhibiting prediction
errors greater than 20\%, and 6.3\% exceeding 100\% error. In contrast,
BFM performs robustly: 96.2\% of predictions fall within the 20\% error
threshold, and the vast majority lie below 5\%.

\begin{figure}
    \centering
    \begin{subfigure}[b]{0.6\textwidth}
        \includegraphics[width=\linewidth]{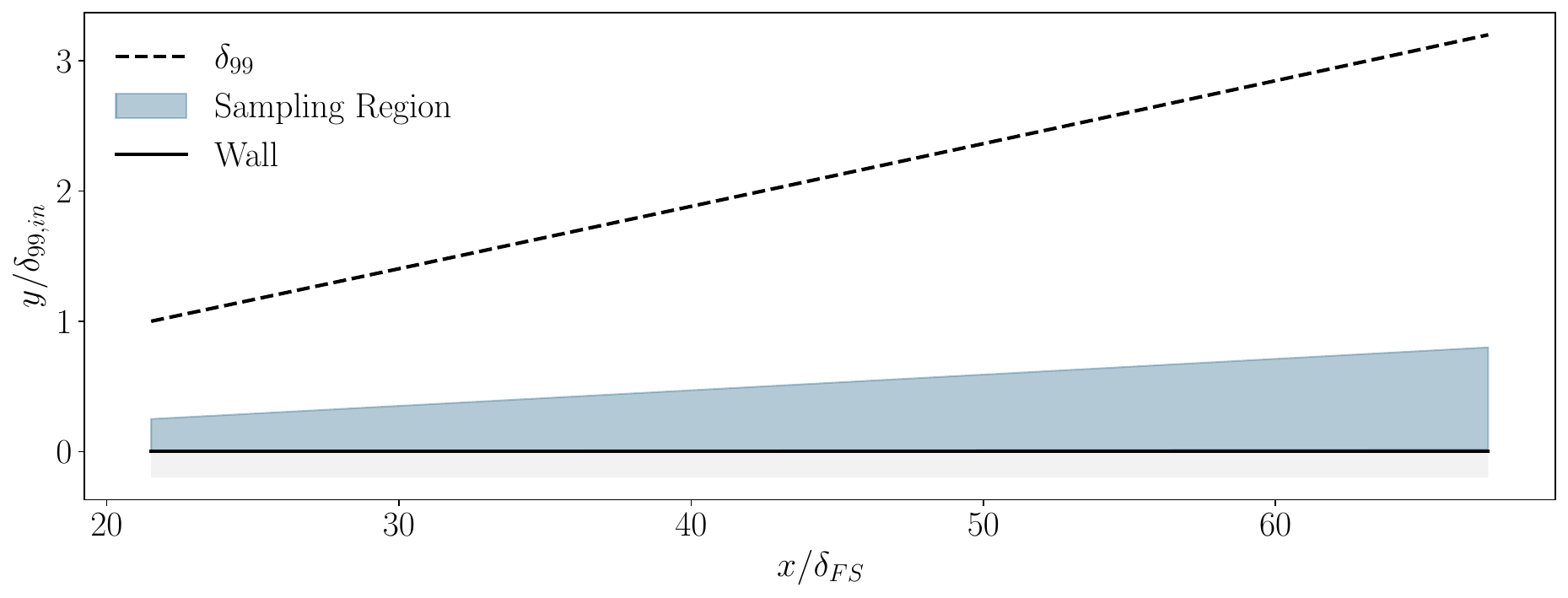}
        \caption{Sampling region for matching location}
        \label{fig:sub:FS_m_region}
    \end{subfigure}
    \centering
    \begin{subfigure}[b]{0.45\textwidth}
        \includegraphics[width=\linewidth]{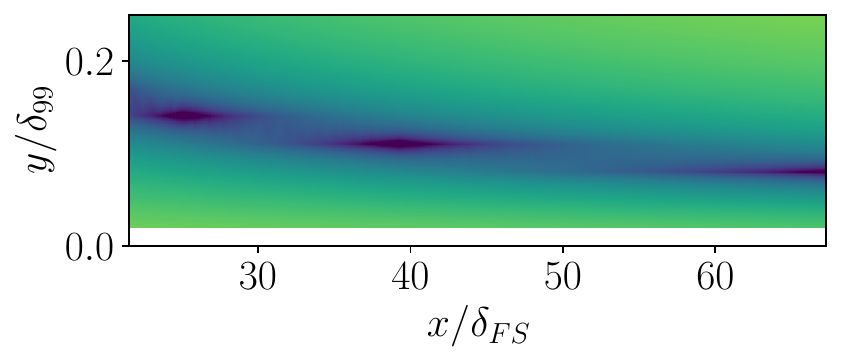}
        \caption{EQWM relative error}
        \label{fig:sub:FS_m_contour_eqwm}
    \end{subfigure}
    \hfill
    \begin{subfigure}[b]{0.45\textwidth}
        \includegraphics[width=\linewidth]{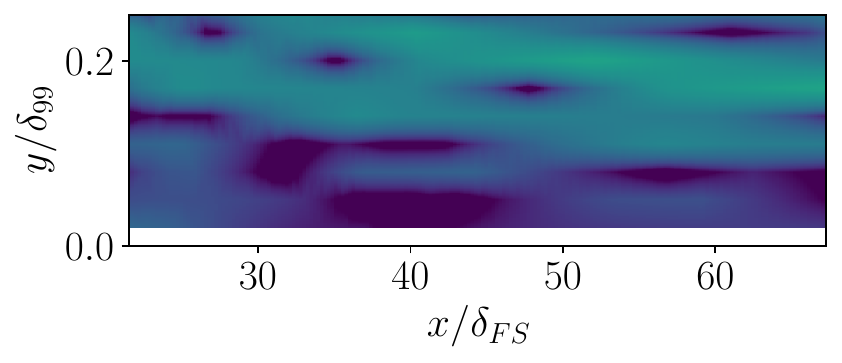}
        \caption{BFM relative error}
        \label{fig:sub:FS_m_contour_bfm}
    \end{subfigure}
    \caption{\textit{A priori} testing of an APG Falkner--Skan laminar
      boundary layer with edge velocity $U_e = C x^{-0.05}$.  (a)
      Sampling region (blue) used to select wall-model matching
      locations.  Here, $\delta_{99}$ denotes the boundary-layer
      thickness based on  0.99 value of the freestream, 
      $\delta_{99,\text{in}}$ its inlet value, and
      $\delta_{FS} = \sqrt{\nu/C}\, x^{(1-a)/2}$ the Falkner--Skan
      length scale.  (b) Spatial distribution of the EQWM relative
      error.  (c) Spatial distribution of the BFM relative error.  The
      colormaps in panels (b) and (c) are identical to those used in
      Figure~\ref{fig:fs}.  }
    \label{fig:FS_m_apg_spatial}
\end{figure}

Figure~\ref{fig:FS_m_apg_spatial} shows the spatial distribution
(streamwise and wall-normal) of the model error for one particular APG
Falkner--Skan laminar boundary layer, providing a spatial perspective
on the performance trends observed in Figure~\ref{fig:fs}.
Figure~\ref{fig:sub:FS_m_region} indicates the sampling region. When
the errors are visualized in physical space
(Fig.~\ref{fig:sub:FS_m_contour_eqwm}--\ref{fig:sub:FS_m_contour_bfm}),
BFM performs well across the entire domain. In contrast, EQWM exhibits
large errors both along the streamwise direction and across the
wall-normal extent, consistent with the poor performance reported
earlier.

\begin{figure}
    \centering
    \begin{subfigure}[b]{0.6\textwidth}
        \includegraphics[width=\linewidth]{Figures/Detailed_Breakdown/viridis_logscale.pdf}
    \end{subfigure}
    \centering
    \begin{subfigure}[b]{0.45\textwidth}
        \includegraphics[width=\linewidth]{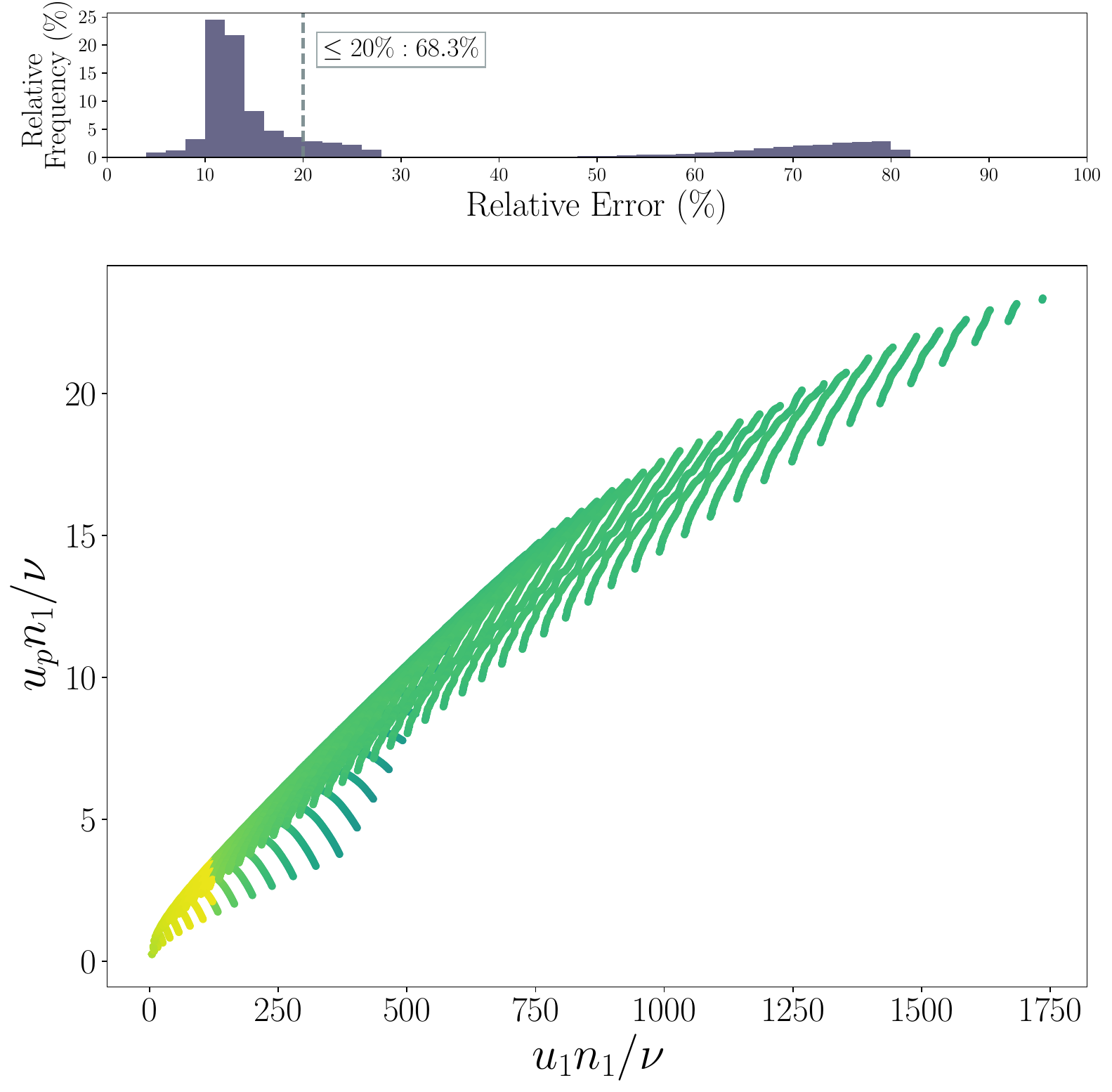}
        \caption{EQWM relative errors}
        \label{fig:sub:apg_log}
    \end{subfigure}
    \hfill
    \begin{subfigure}[b]{0.45\textwidth}
        \includegraphics[width=\linewidth]{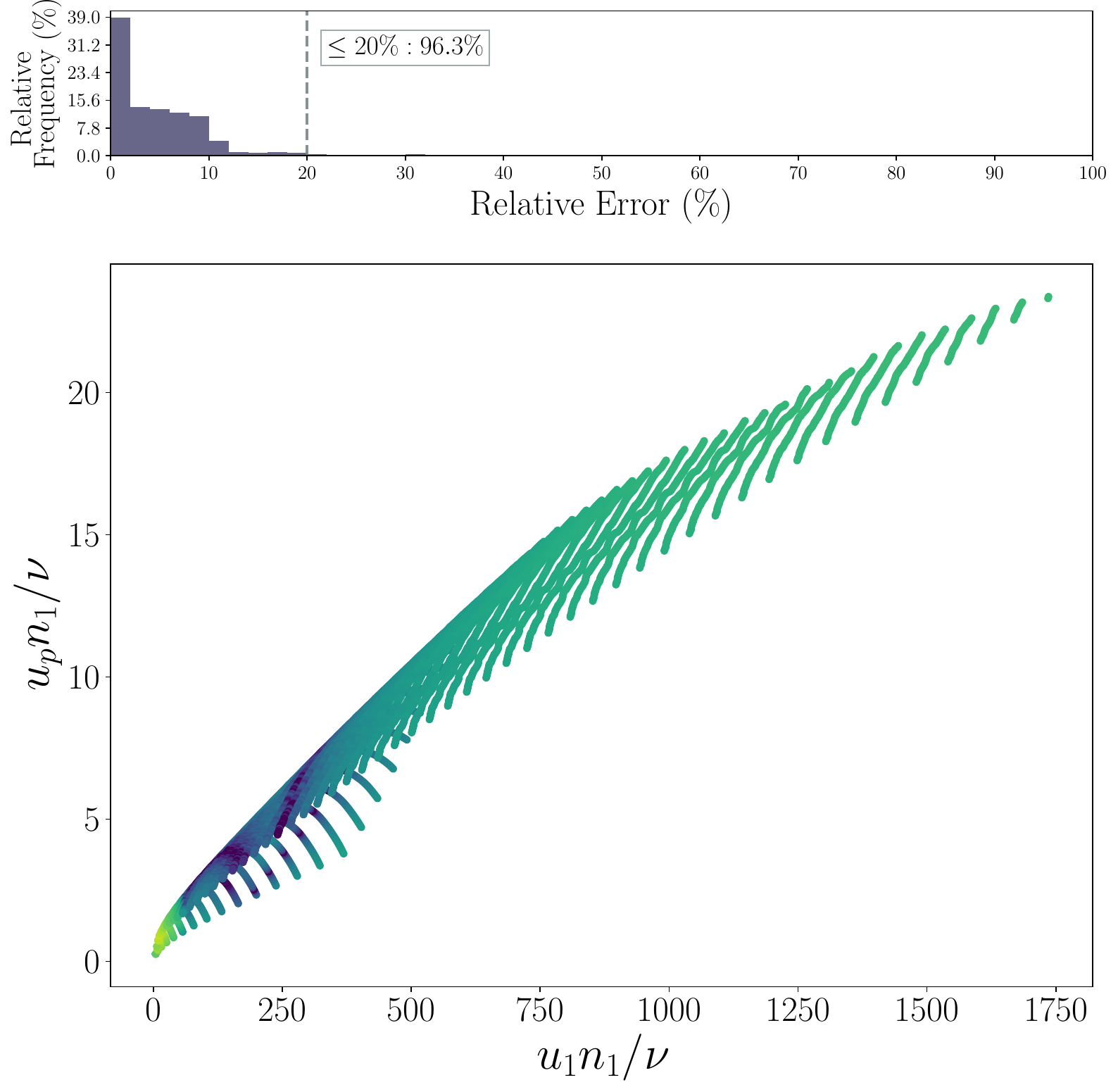}
        \caption{BFM relative errors}
        \label{fig:sub:apg_bfm}
    \end{subfigure}
    \caption{\textit{A priori} relative errors of (a) EQWM and (b) BFM
      for the APG TBL with constant $\beta\approx2$ from
      \citet{bobkeHistoryEffectsEquilibrium2017}.}
    \label{fig:apg}
\end{figure}

\paragraph{TBL with mild APG.}
We now examine an APG TBL case that was not included in the training
set. This configuration, introduced by
\citet{bobkeHistoryEffectsEquilibrium2017}, corresponds to a TBL with
a constant Clauser parameter of $\beta \approx 2$. The results are
shown in Figure~\ref{fig:apg}. Although
\citet{bobkeHistoryEffectsEquilibrium2017} reported that the
inner-layer velocity profile approximately follows a logarithmic law,
EQWM performs poorly: prediction errors exceed 10\% across most of the
profile and reach more than 60\% near the wall, particularly within
the buffer region. In contrast, BFM maintains high accuracy even
though the streamwise evolution of $\beta$ in this mild APG case
differs substantially from the training data (see
Figure~\ref{subfig:APG_beta}). The accurate predictions of BFM extend
across all wall-normal matching locations, including the buffer layer,
with errors remaining below 10\%. These results support the modeling
assumption that history effects have limited influence on the
wall-model predictions.

\begin{figure}
    \centering
    \begin{subfigure}[b]{0.6\textwidth}
        \includegraphics[width=\linewidth]{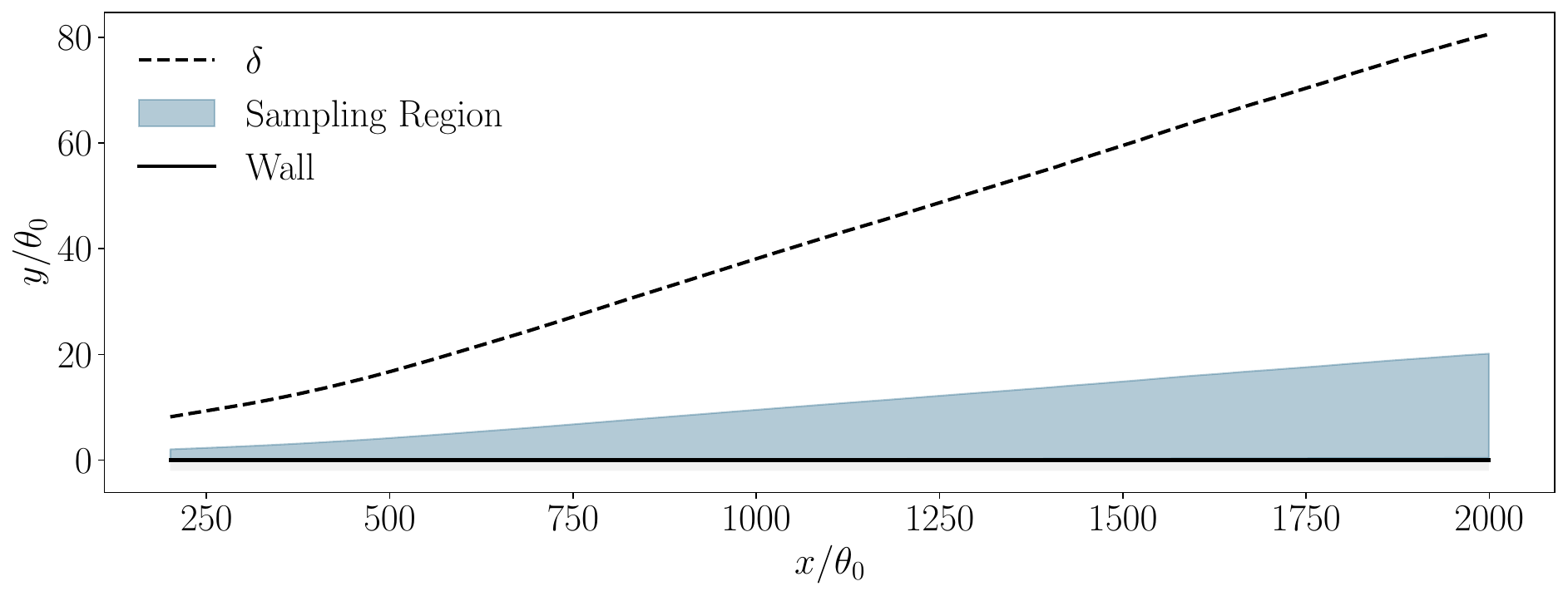}
        \caption{Sampling region for matching location}
        \label{fig:sub:apg_region}
    \end{subfigure}
    \centering
    \begin{subfigure}[b]{0.45\textwidth}
        \includegraphics[width=\linewidth]{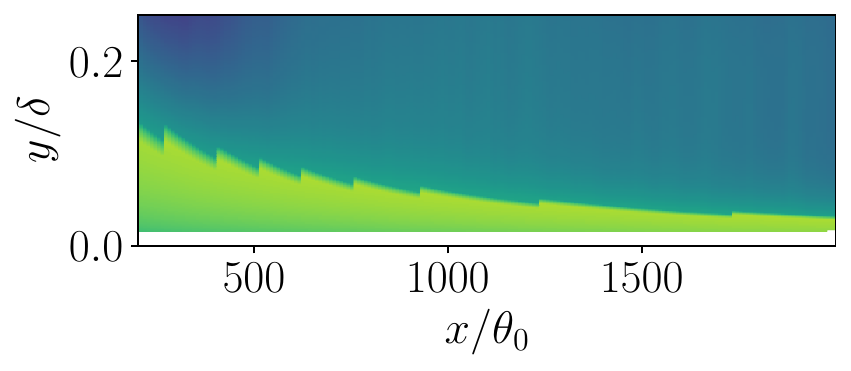}
        \caption{EQWM relative errors}
        \label{fig:sub:apg_contour_eqwm}
    \end{subfigure}
    \hfill
    \begin{subfigure}[b]{0.45\textwidth}
        \includegraphics[width=\linewidth]{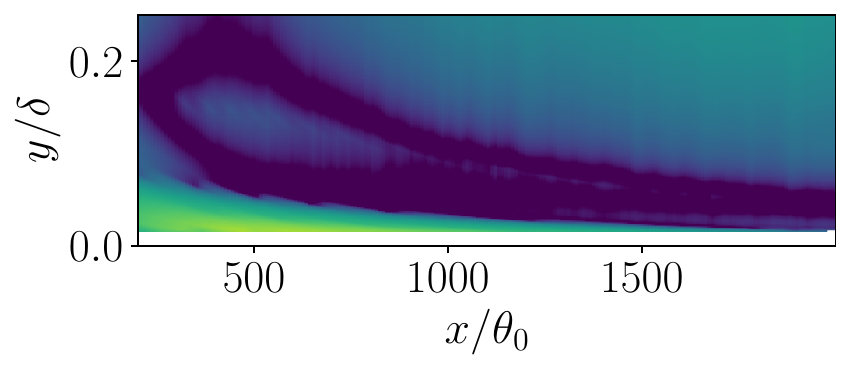}
        \caption{BFM relative errors}
        \label{fig:sub:apg_contour_bfm}
    \end{subfigure}
    \caption{\textit{A priori} testing of a TBL with mild APG.  (a)
      Sampling region (blue) used to select wall-model matching
      locations.  Here, $\theta_0$ denotes the inlet momentum
      thickness and $\delta$ the boundary-layer edge.  (b) Spatial
      distribution of the EQWM relative error.  (c) Spatial
      distribution of the BFM relative error.  The colormaps in panels
      (b) and (c) are identical to those used in
      Figure~\ref{fig:apg}.}
    \label{fig:apg_spatial}
\end{figure}

The spatial distribution of the model errors in
Figure~\ref{fig:apg_spatial} offers additional insight. First, it
confirms that the superior accuracy of BFM relative to EQWM is
primarily localized in the near-wall region. Second, the plot shows a
downstream improvement in BFM performance. This trend may be
attributed to larger errors in the upstream region, where the boundary
layer is thin---a condition that manifests as points clustered near
the origin in the model-input space.

\begin{figure}
    \centering
    \begin{subfigure}[b]{0.6\textwidth}
        \includegraphics[width=\linewidth]{Figures/Detailed_Breakdown/viridis_logscale.pdf}
    \end{subfigure}
    \centering
    \begin{subfigure}[b]{0.45\textwidth}
        \includegraphics[width=\linewidth]{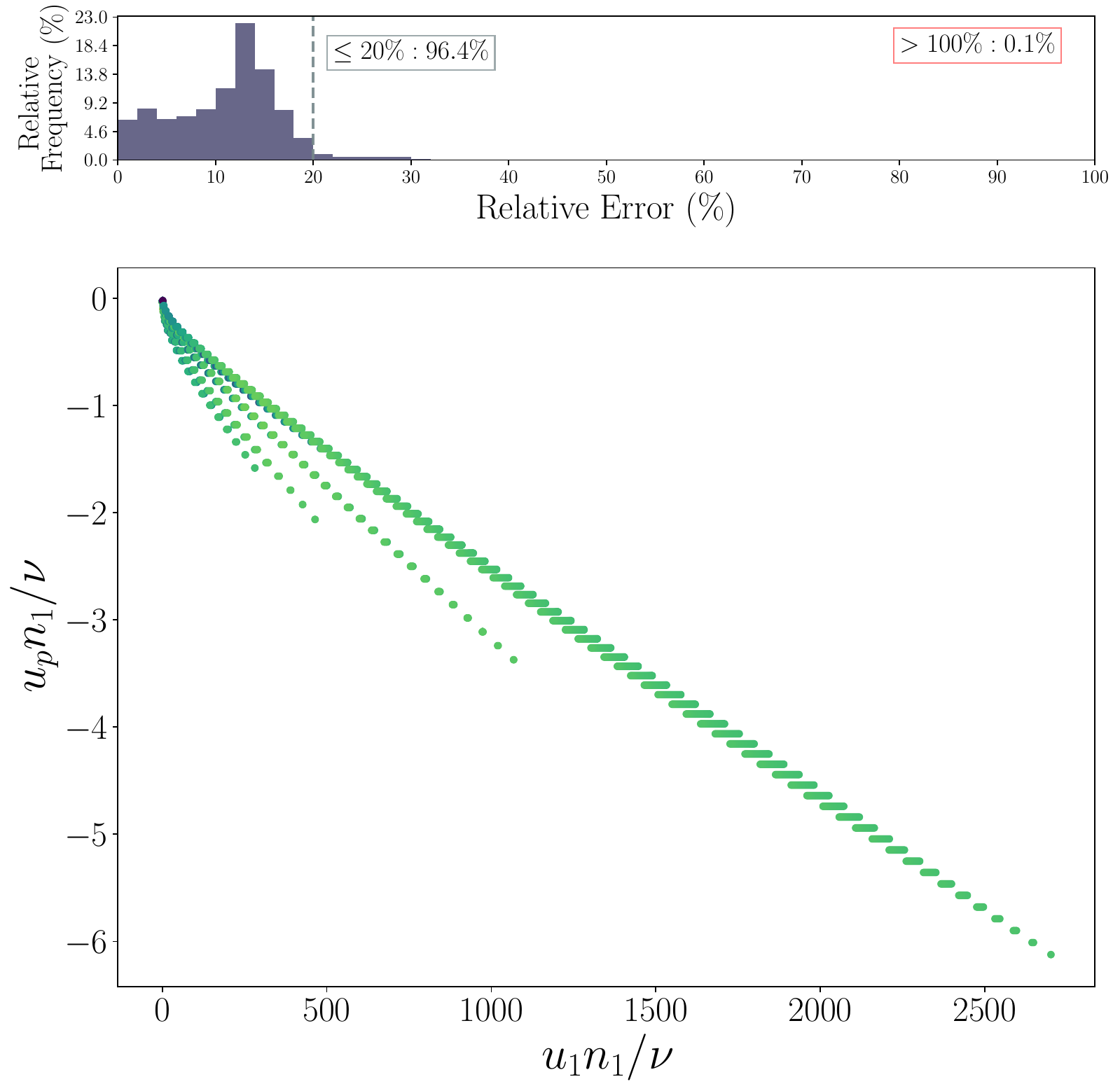}
        \caption{EQWM relative errors}
        \label{fig:sub:duct_log}
    \end{subfigure}
    \hfill
    \begin{subfigure}[b]{0.45\textwidth}
        \includegraphics[width=\linewidth]{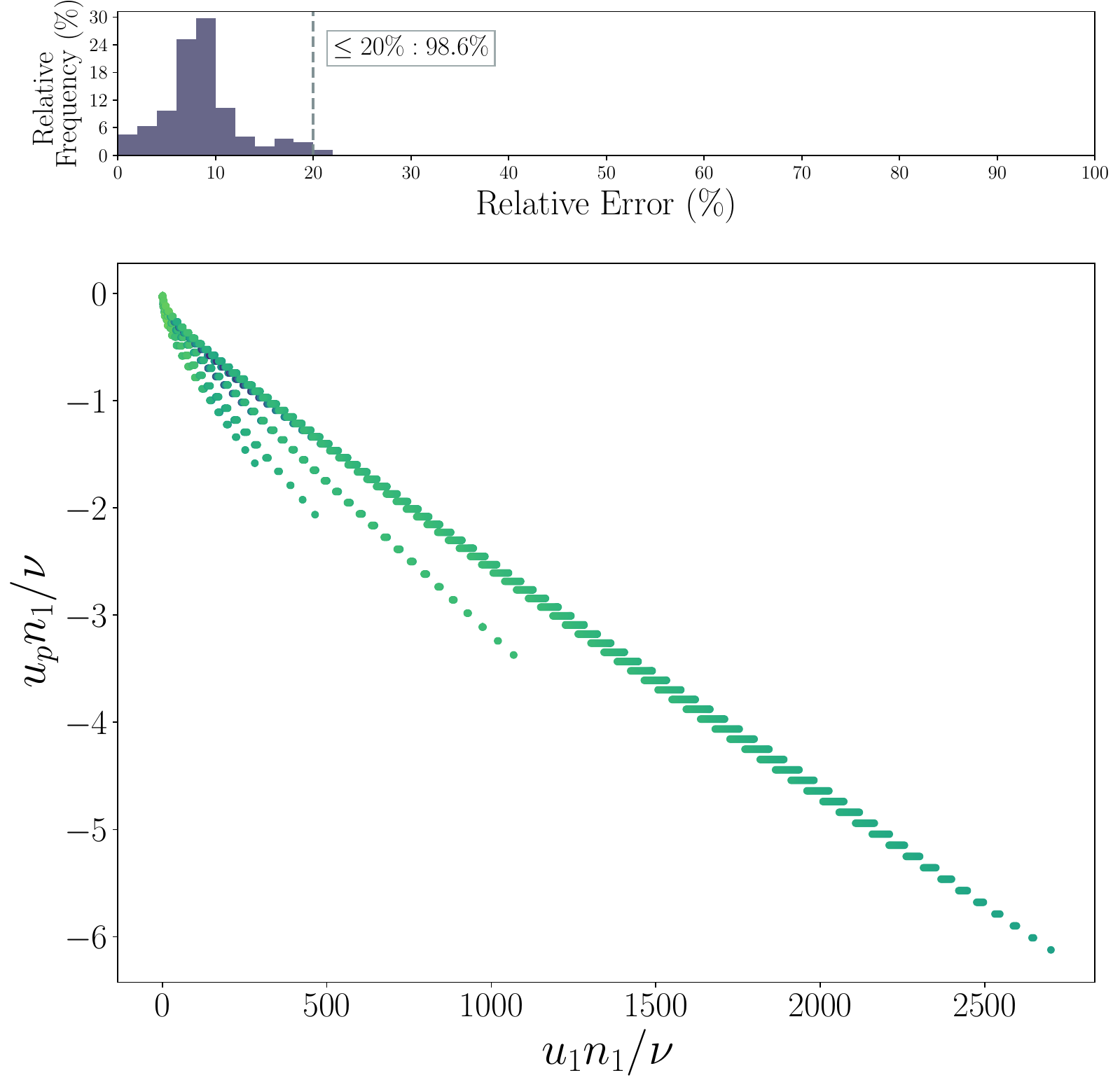}
        \caption{BFM relative errors}
        \label{fig:sub:duct_bfm}
    \end{subfigure}
    \caption{\textit{A priori} relative error of (a) EQWM and (b) BFM
      for the turbulent flow in a square duct from
      \citet{pirozzoli_turbulence_2018}.}
    \label{fig:duct}
\end{figure}

\paragraph{Square duct.} 
This case test the ability of the model to predict previously unseen
secondary flows. We consider the DNS data of
\citet{pirozzoli_turbulence_2018} at $Re_{\tau} = 220, 500, 1000$, and
$2000$. Whereas prior studies~\cite{modesti_role_2018,
  pirozzoli_turbulence_2018} examined only the streamwise velocity,
our objective here is to predict the \emph{total} wall-shear-stress
magnitude based on the magnitude of the wall-parallel velocity
vector. This task is challenging because BFM was not trained on
secondary flows or on cases exhibiting mean-flow
three-dimensionality. Nonetheless, BFM performs accurately as shown in Figure~\ref{fig:duct}. EQWM also
yields small errors, which is expected given that the cross-flow
components are weak relative to the dominant streamwise
motion. Overall, both models perform well, showing only slightly
larger errors than in a comparable turbulent channel flow without
secondary motions. 

\begin{figure}
    \centering
    \begin{subfigure}[b]{0.75\textwidth}
        \includegraphics[width=\linewidth]{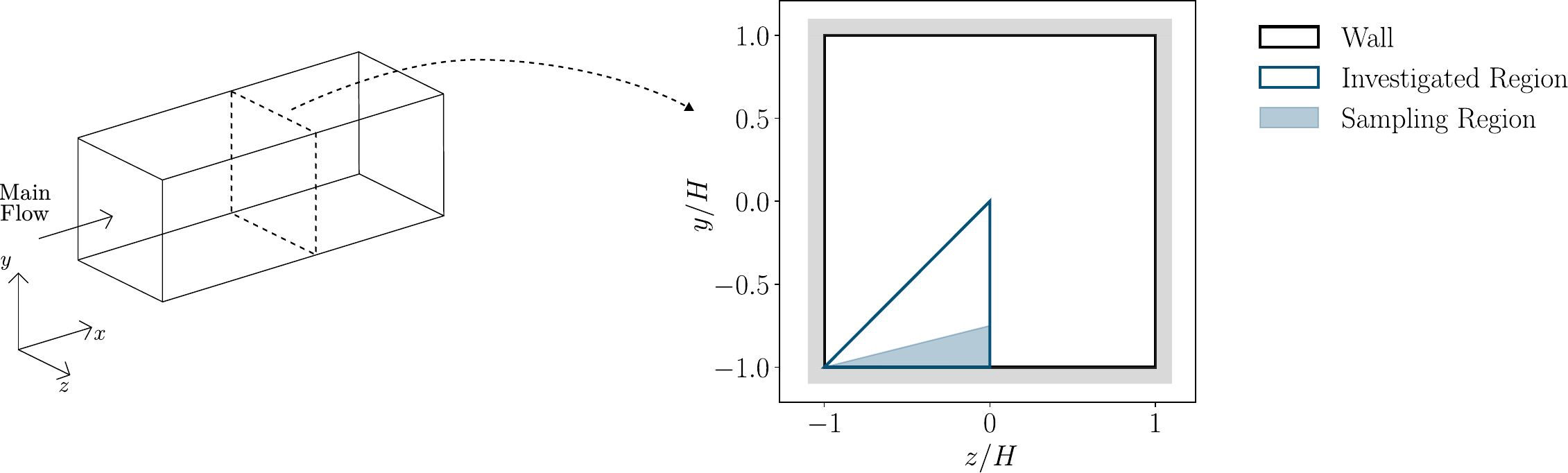}
        \caption{Schematic of the duct flow simulation and the
          sampling region for matching location}
        \label{fig:sub:duct_region}
    \end{subfigure}
    \centering
    \begin{subfigure}[b]{0.45\textwidth}
        \includegraphics[width=\linewidth]{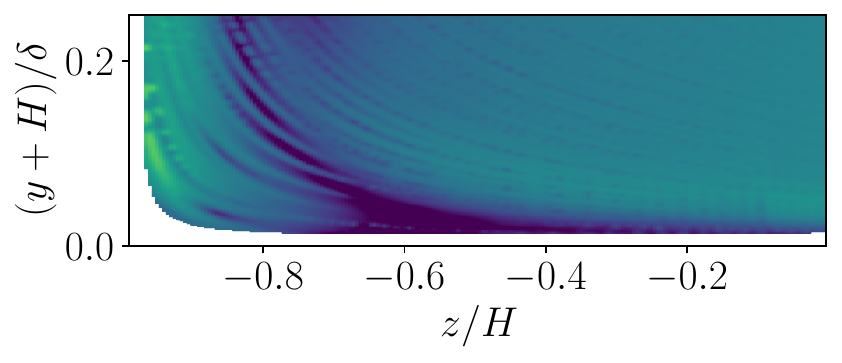}
        \caption{EQWM relative errors}
        \label{fig:sub:duct_contour_eqwm}
    \end{subfigure}
    \hfill
    \begin{subfigure}[b]{0.45\textwidth}
        \includegraphics[width=\linewidth]{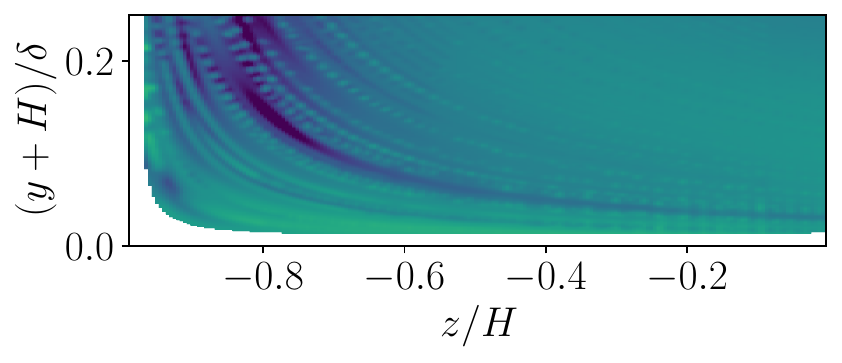}
        \caption{BFM relative errors}
        \label{fig:sub:duct_contour_bfm}
    \end{subfigure}
    \caption{\textit{A priori} testing of a square duct flow. (a)
      Sampling region (blue) used to select wall-model matching
      locations. Here, $H$ is the half duct height and $\delta$ is the
      boundary layer edge defined as the top of the bisector of the
      corner region. (b) Spatial distribution of the EQWM relative
      error. (c) Spatial distribution of the BFM relative error. The
      colormaps in panels (b) and (c) are identical to those used in
      Figure~\ref{fig:duct}.}
    \label{fig:duct_spatial}
\end{figure}

The spatial error distribution, shown in Figure~\ref{fig:duct_spatial},
reveals a clear trend for both models: prediction errors increase in
proximity to the corners of the domain. This result is consistent with
physical intuition, as the models, which are derived from 2D mean-flow
assumptions, are expected to degrade where the 3D secondary motions
are most intense. Despite this localized deterioration, the overall
errors remain small. Furthermore, \citet{pirozzoli_turbulence_2018}
demonstrated that the effect of this secondary motion diminishes as
the Reynolds number increases. Since the current cases are at moderate
Reynolds numbers, we hypothesize that the errors induced by the
secondary flow would be mitigated even further at larger Reynolds
numbers.

\begin{figure}
    \centering
    \begin{subfigure}[b]{0.6\textwidth}
        \includegraphics[width=\linewidth]{Figures/Detailed_Breakdown/viridis_logscale.pdf}
    \end{subfigure}
    \centering
    \begin{subfigure}[b]{0.45\textwidth}
        \includegraphics[width=\linewidth]{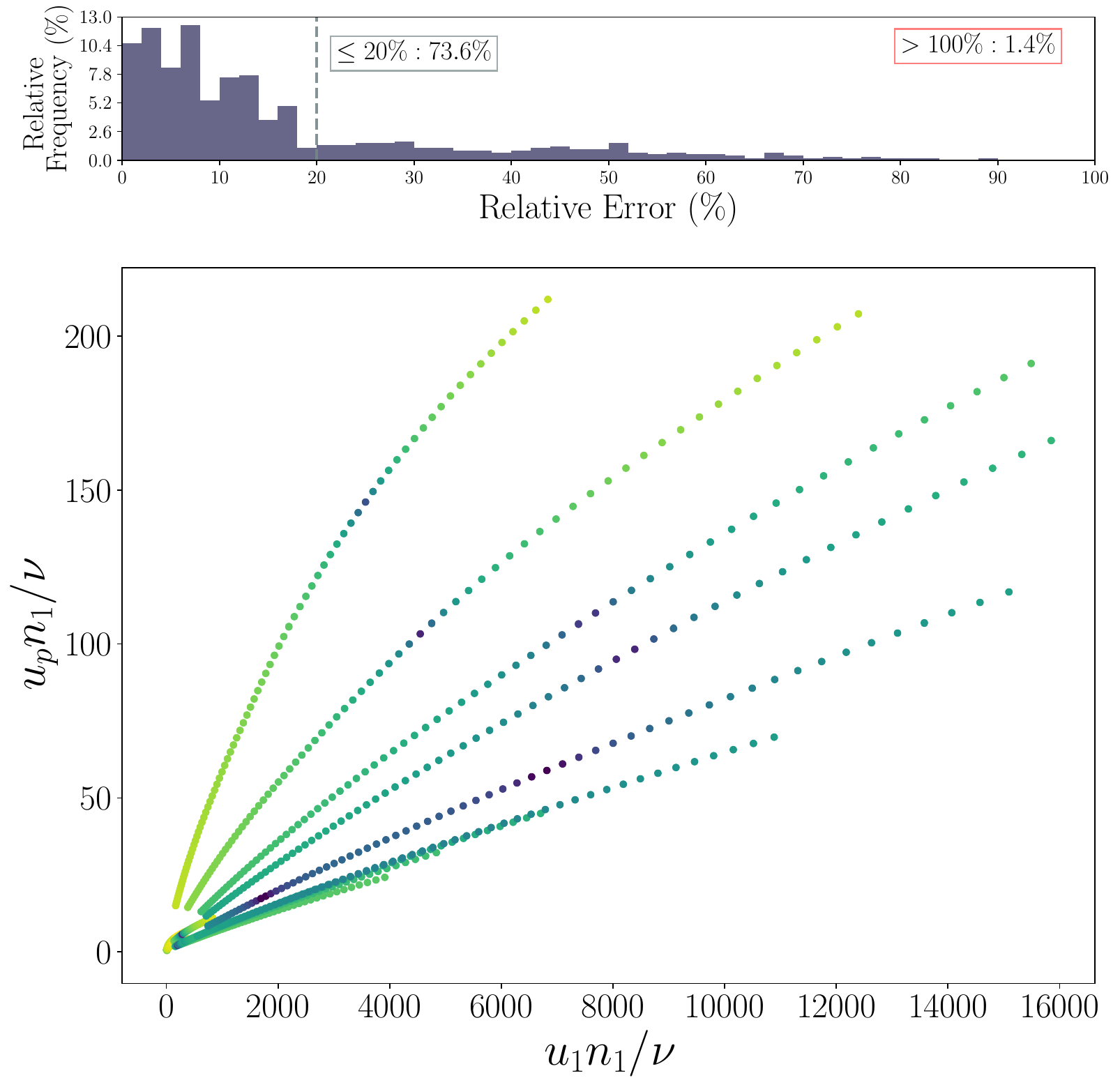}
        \caption{EQWM relative errors}
        \label{fig:sub:aa_10M_log}
    \end{subfigure}
    \hfill
    \begin{subfigure}[b]{0.45\textwidth}
        \includegraphics[width=\linewidth]{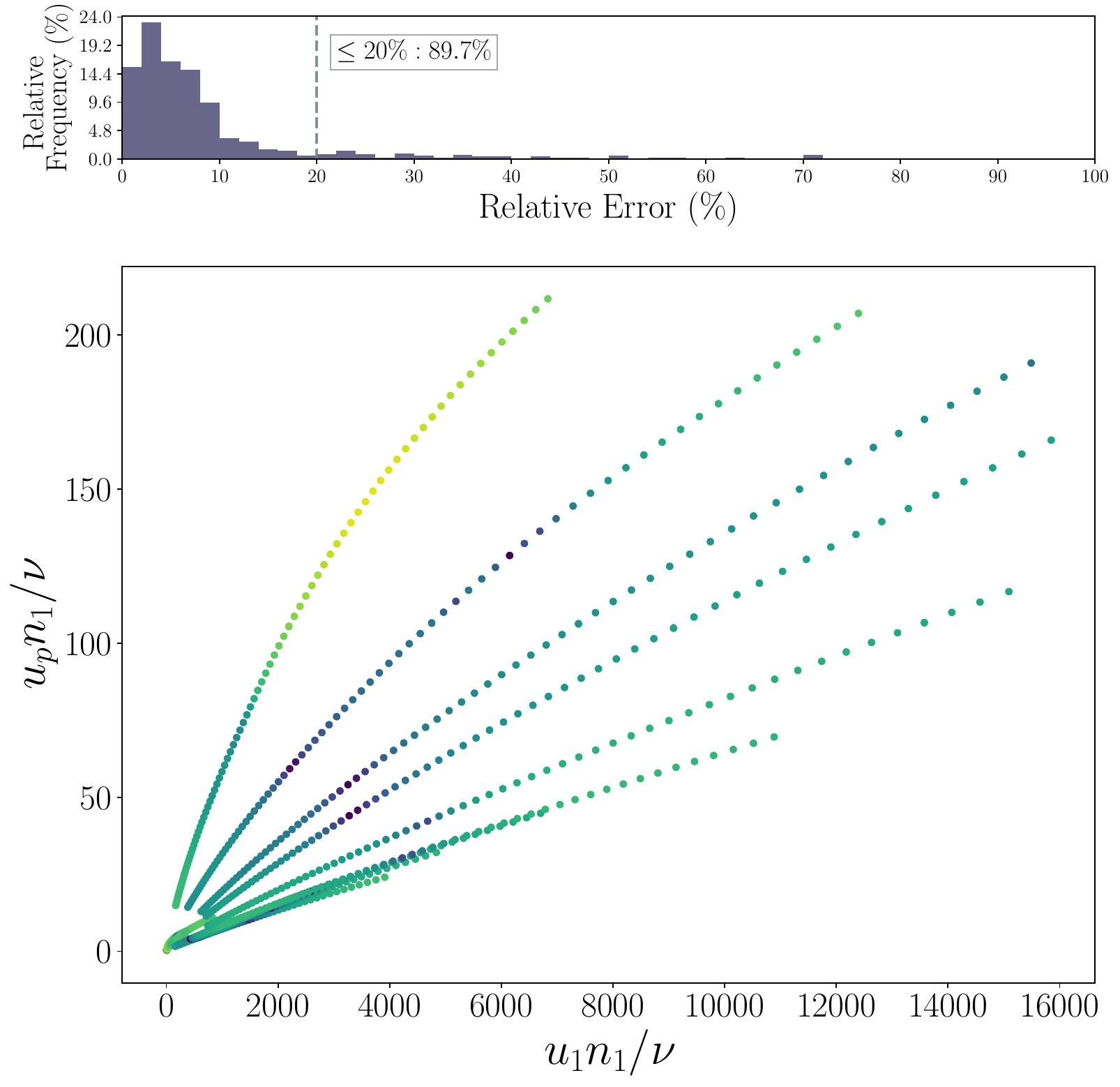}
        \caption{BFM relative errors}
        \label{fig:sub:aa_10M_bfm}
    \end{subfigure}
    \caption{\textit{A priori} relative errors of (a) EQWM and (b) BFM
      for the Aerospatiale A-airfoil near-stall from
      \citet{tamakiWallResolvedLargeEddySimulation2023}.}
    \label{fig:aair_10M}
\end{figure}

\paragraph{Aerospatiale A-airfoil near-stall.}
We further evaluate the models using a realistic, near-stall,
high--Reynolds-number airfoil studied by
\citet{tamakiWallResolvedLargeEddySimulation2023}. The test case
corresponds to an airfoil at a chord-based Reynolds number of $Re_c
\sim 1 \times 10^7$ and an angle of attack of $13.3^{\circ}$, with
solution data available at ten streamwise stations along the suction
side. Because such flows lie far outside the low- to moderate-$Re$
canonical regimes typically used for model development, they
constitute a stringent test of model generalizability. Both BFM and
EQWM achieve reasonable accuracy across all stations as shown in Figure~\ref{fig:aair_10M}. However, BFM
shows a modest but clear advantage: a larger fraction of its
predictions fall below the 20\% error threshold, and none exceed 100\%
error, whereas EQWM exhibits a few outliers.

\begin{figure}
    \centering
    \includegraphics[width=0.8\linewidth]{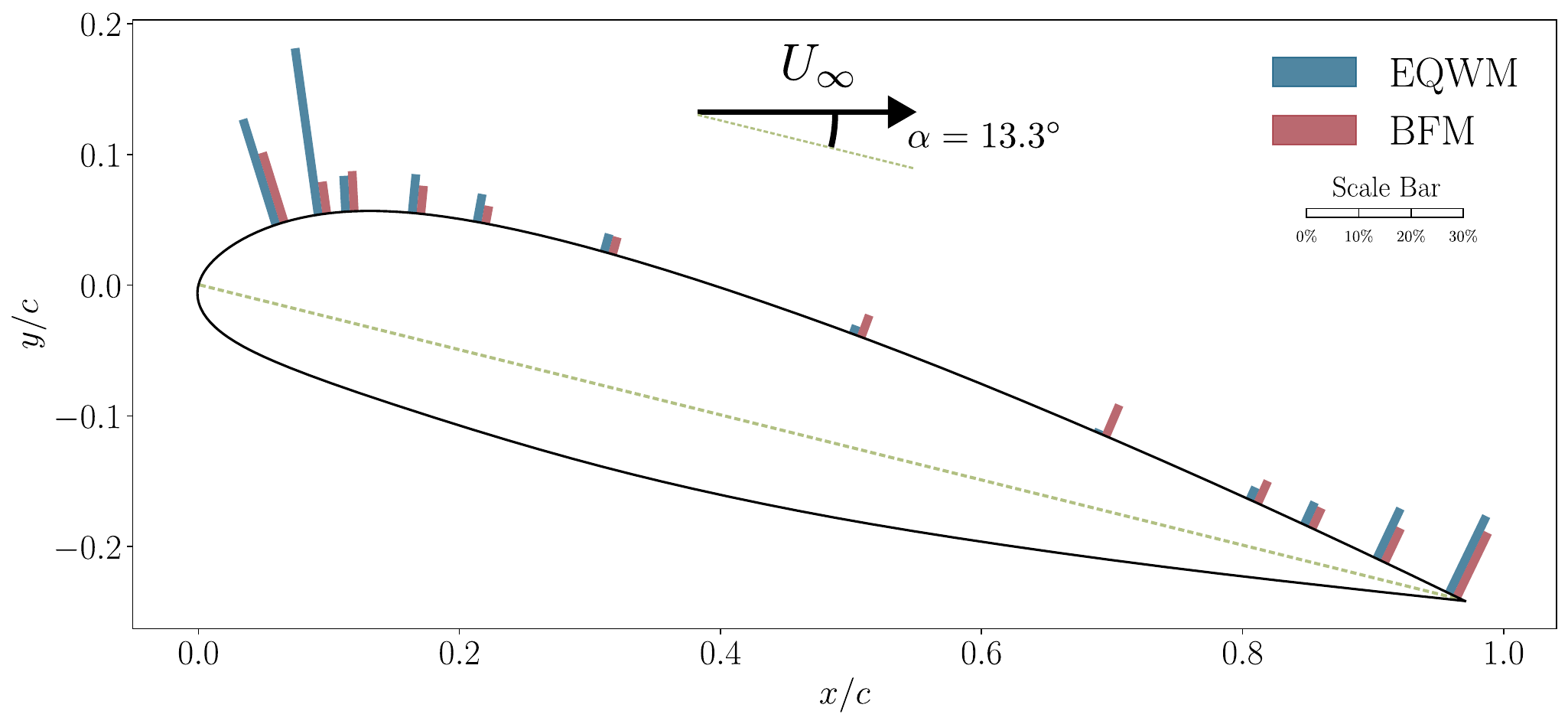}
    \caption{\textit{A priori} testing of the Aerospatiale A-airfoil
      at near-stall angle of attack. A comparison of mean errors at
      stations distributed at different locations of the suction
      side. Here, $c$ is the chord length, $U_\infty$ is the
      freestream velocity, and the dashed green line is the chord
      line. The length of bars is proportional to the model
      error as shown by the scale bar.}
    \label{fig:aairfoil_bar}
\end{figure}

Figure~\ref{fig:aairfoil_bar} shows the mean relative error at each
station and highlights the advantage of BFM. For EQWM, although the
errors are small over the mid-chord region, they increase sharply near
both the leading and trailing edges. In contrast, the errors from BFM
remain lower across the chord. This behavior reflects the ability of
BFM to handle the transition occurring at the leading edge---where
EQWM breaks down---and to better predict the flow approaching
separation near the trailing edge. While the errors of both models
rise in the vicinity of separation, BFM remains more accurate,
underscoring the benefits of its diverse training set.

\begin{figure}
    \centering
    \includegraphics[width=0.5\linewidth]{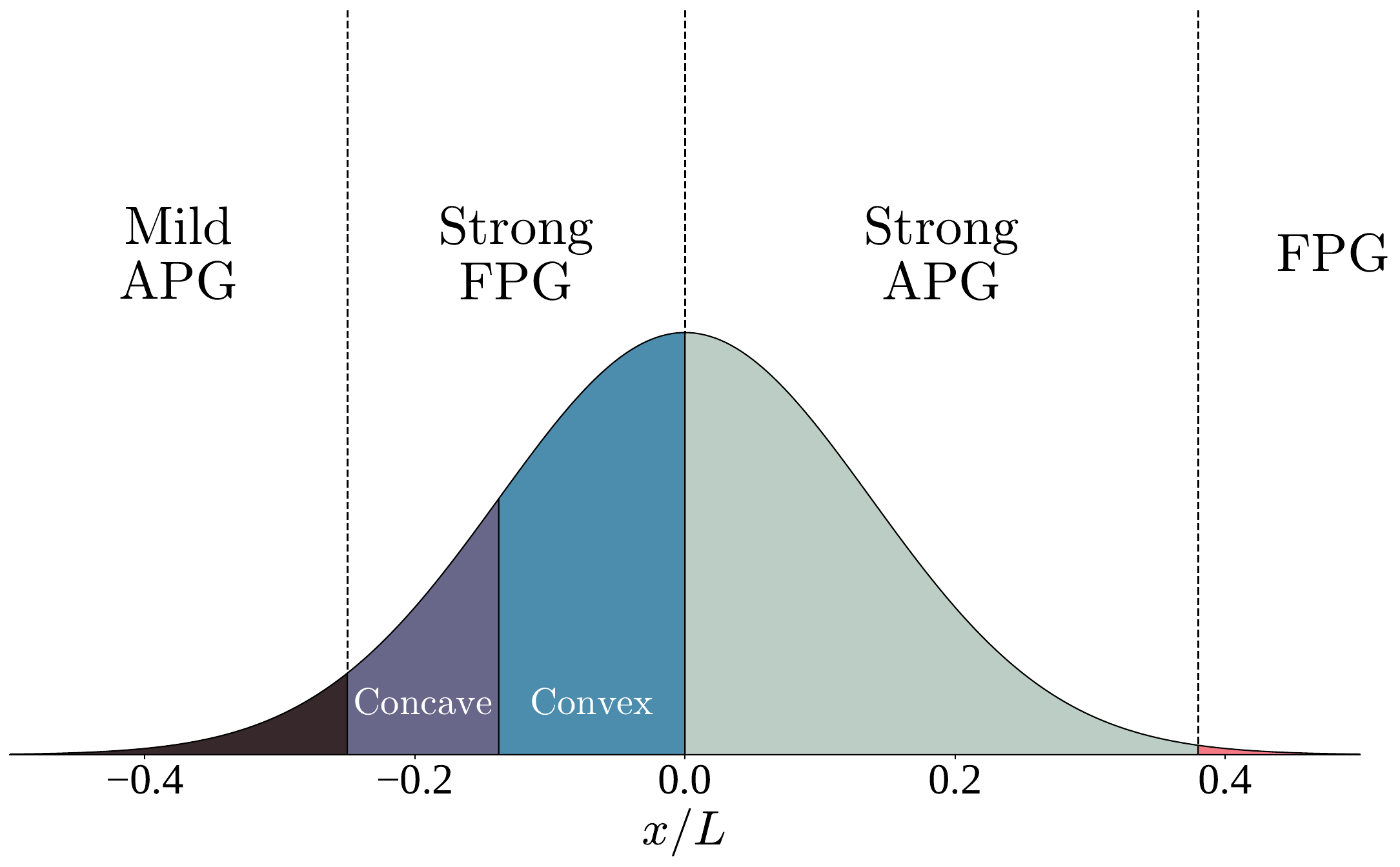}
    \caption{Flow regimes for the Gaussian bump with $Re_L=1M$ as a
      function of streamwise coordinates ($x/L$). From left to right:
      $x<-0.25$ (brown): mild APG; $-0.25<x<-0.138$ (purple): strong
      FPG with concave surface; $-0.138<x<0$ (blue): strong FPG with
      convex surface; $0<x<0.38$ (green): strong APG; $x>0.38$
      (orange): FPG recovery zone.  }
    \label{fig:GP1M}
\end{figure}

\paragraph{2D Gaussian bump ($Re_L\approx 1M$).}
We present results for a Gaussian bump case with a bump-length-based
Reynolds number of $Re_L \approx 1 \times 10^6$, which was not
included in our training data. This case differs fundamentally from
the training case at $Re_L = 2 \times 10^6$ due to flow
relaminarization near the apex, driven by low Reynolds number
effects. As shown in Figure~\ref{fig:GP1M}, the flow initially
encounters a mild APG near the root of the bump, followed by a strong
FPG. Near the apex, the combination of the convex surface and the FPG
induces relaminarization. Downstream of the apex, the APG intensifies,
triggering re-transition and pushing the flow to the verge of
separation at $x/L \approx 0.2$. Finally, the flow experiences another
FPG and approaches a ZPG state near the outflow.

In addition to these changes in pressure gradient behavior, the
curvature effects in this case also differ from those in the training
set. While the geometric curvature remains constant, the relevant
dimensionless parameter, $\delta/R$, varies significantly---primarily
because the boundary layer becomes thinner as the Reynolds number
increases at a fixed spatial location. Consequently, this case
introduces new physical regimes and parameter ranges, providing a
stringent out-of-sample test of the generalization capability of BFM.
We focus our attention on the two most challenging regions: the strong
FPG region with a convex surface and the strong APG region, as
indicated in Figure~\ref{fig:GP1M}. In the FPG region, partial
relaminarization degrades the performance of EQWM, whereas BFM remains
robust. In the APG region, BFM again significantly outperforms
EQWM. However, even in this region, the BFM model shows clear room for
improvement due to flow physics not represented in the training
set. Specifically, the APG region begins in a partially laminar state
and gradually transitions back to turbulence, which distinguishes it
from a purely transitional flow where the pre‑transition state is
fully laminar.
\begin{figure}
    \centering
    \begin{subfigure}[b]{0.61\textwidth}
        \includegraphics[width=\linewidth]{Figures/Detailed_Breakdown/viridis_logscale.pdf}
    \end{subfigure}
    \centering
    \begin{subfigure}[b]{0.45\textwidth}
        \includegraphics[width=\linewidth]{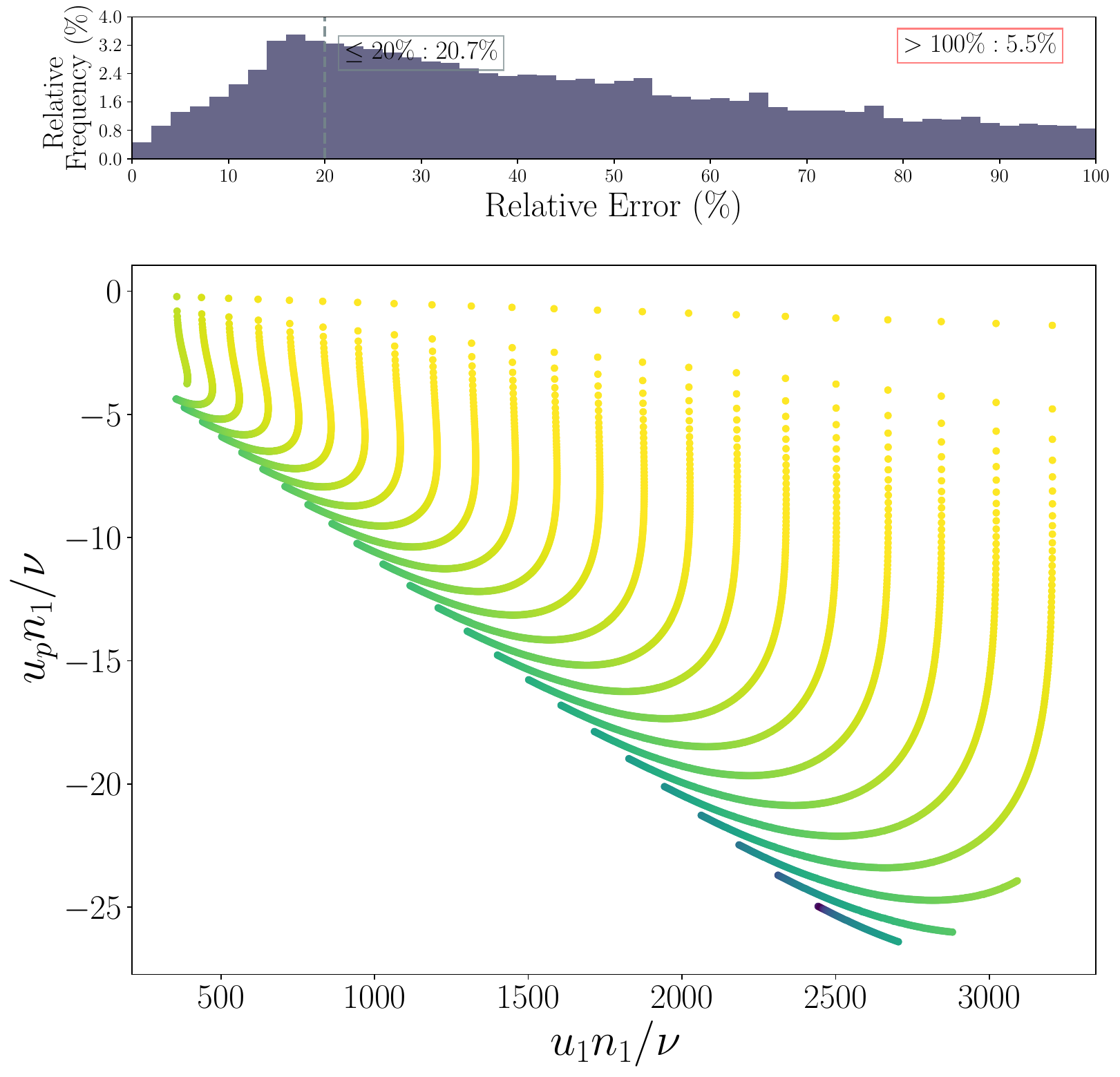}
        \caption{EQWM relative error for strong FPG (convex)}
        \label{fig:sub:gp_1M_log}
    \end{subfigure}
    \hfill
    \begin{subfigure}[b]{0.45\textwidth}
        \includegraphics[width=\linewidth]{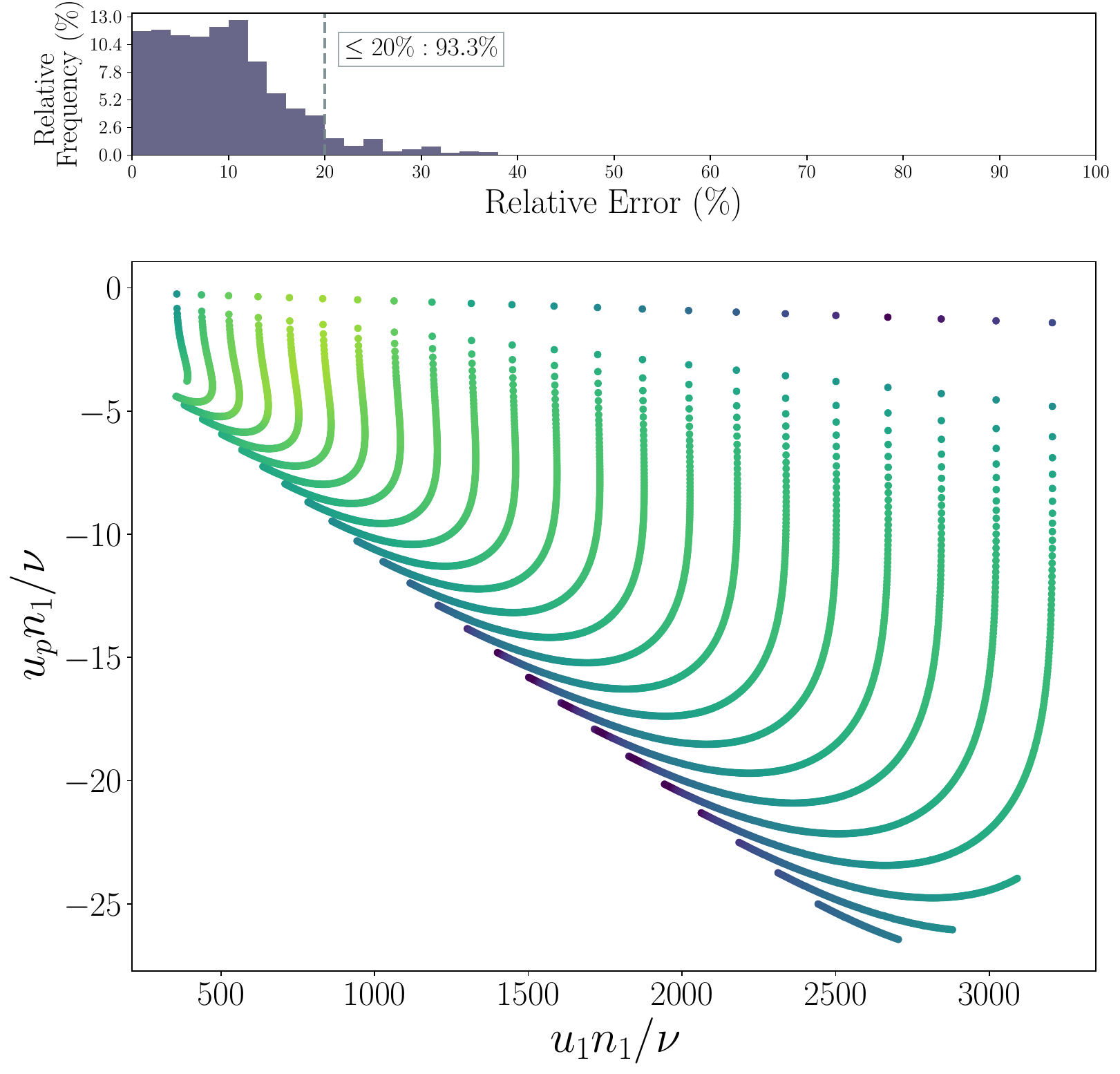}
        \caption{BFM relative error for strong FPG (convex)}
        \label{fig:sub:gp_1M_bfm}
    \end{subfigure}
    \begin{subfigure}[b]{0.45\textwidth}
        \includegraphics[width=\linewidth]{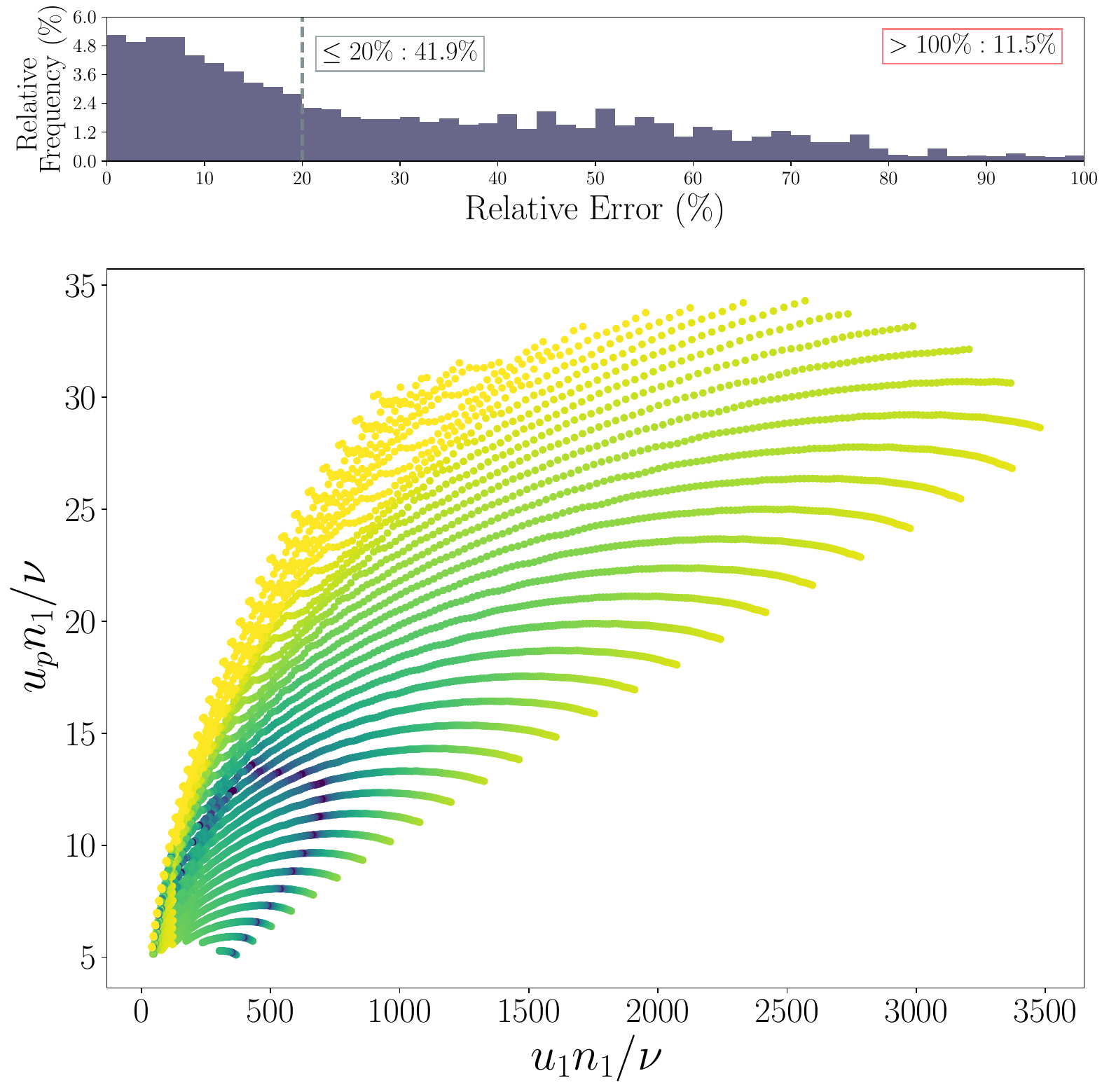}
        \caption{EQWM relative error for strong APG}
        \label{fig:sub:gp_1M_log_apg}
    \end{subfigure}
    \hfill
    \begin{subfigure}[b]{0.45\textwidth}
        \includegraphics[width=\linewidth]{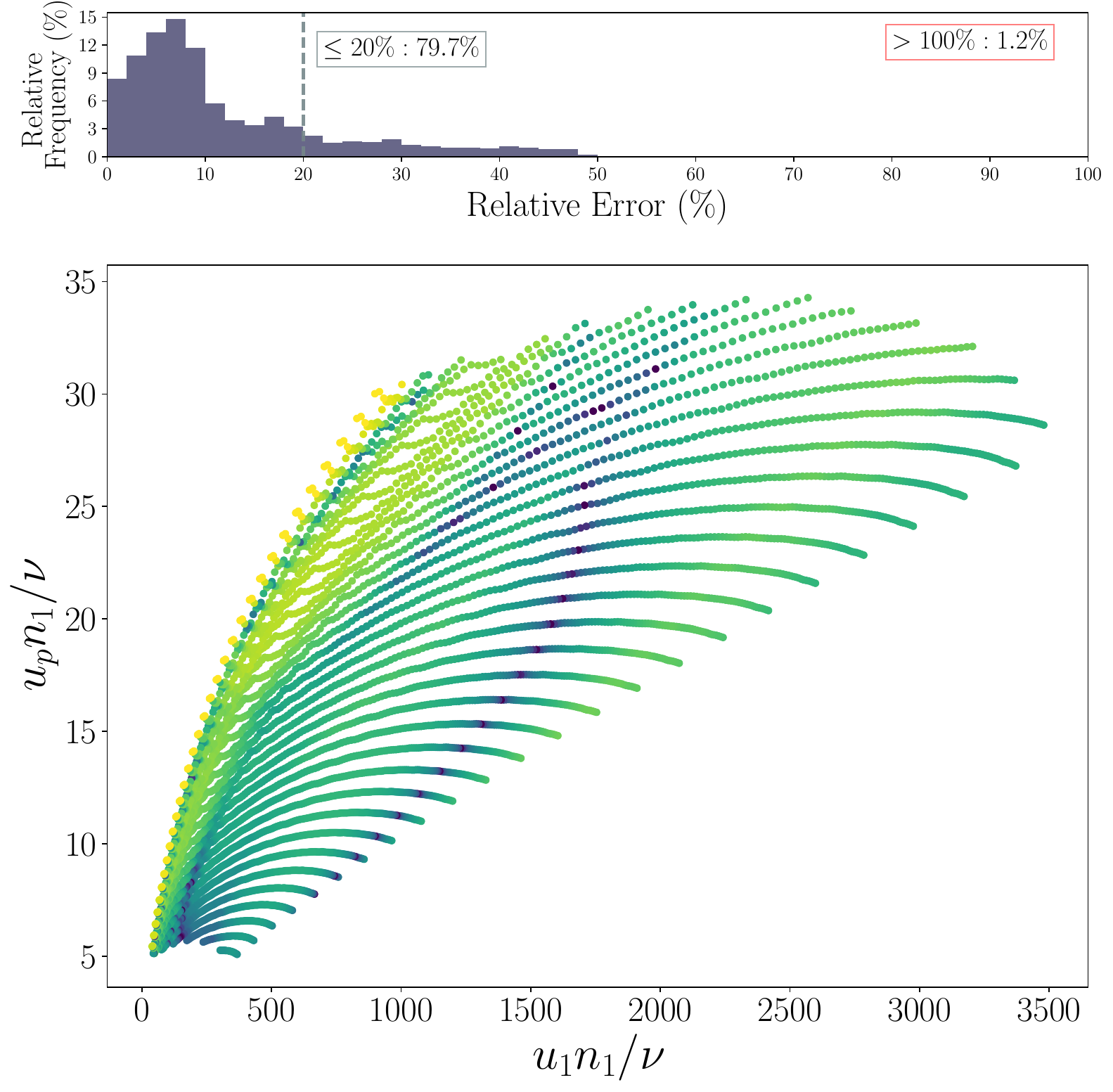}
        \caption{BFM for strong APG}
        \label{fig:sub:gp_1M_wm_apg}
    \end{subfigure}
    \caption{\textit{A priori} relative errors for (a) EQWM and (b)
      BFM in the strong FPG region with convex curvature of a 2D
      Gaussian bump at $Re_L = 1\,\text{M}$, and for (c) EQWM and (d)
      BFM in the strong APG region. The two regions are delineated in
      Figure~\ref{fig:GP1M}.}
    \label{fig:gp_1M}
\end{figure}

\begin{figure}
    \centering
    \begin{subfigure}[b]{0.6\textwidth}
        \includegraphics[width=\linewidth]{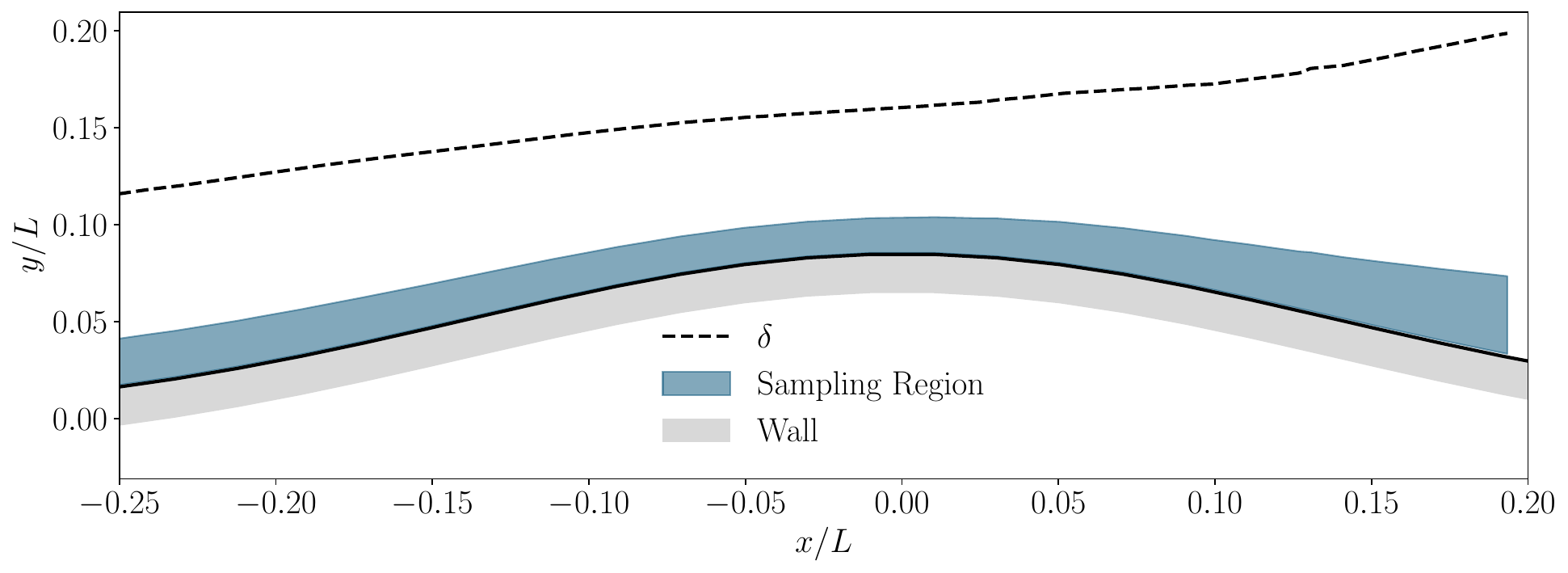}
        \caption{Sampling region for the matching location}
        \label{fig:sub:gp_1M_region}
    \end{subfigure}
    \centering
    \begin{subfigure}[b]{0.45\textwidth}
        \includegraphics[width=\linewidth]{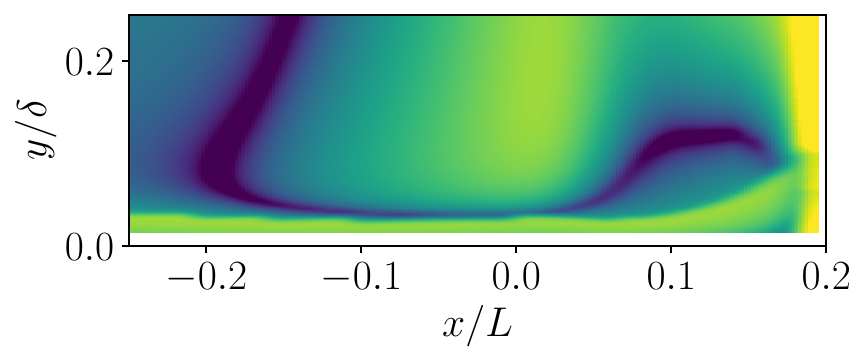}
        \caption{EQWM relative error}
        \label{fig:sub:gp_1M_contour_eqwm}
    \end{subfigure}
    \hfill
    \begin{subfigure}[b]{0.45\textwidth}
        \includegraphics[width=\linewidth]{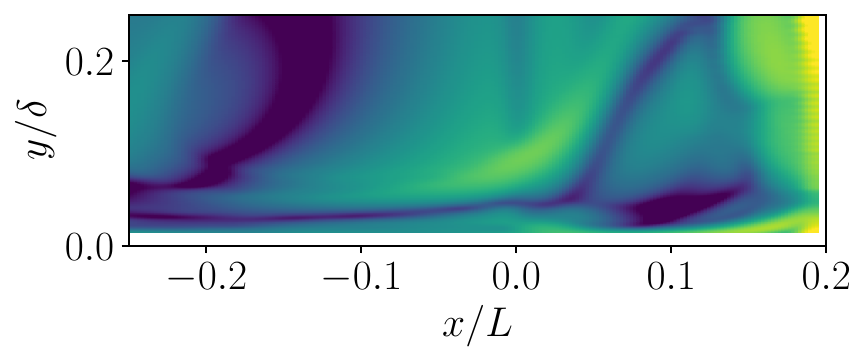}
        \caption{BFM relative error}
        \label{fig:sub:gp_1M_contour_log}
    \end{subfigure}
    \caption{\textit{A priori} testing of a 2D Gaussian bump at
      $Re_L=1M$. (a) Sampling region (blue) used to select wall-model
      matching locations. Here, $L$ is the bump length and $\delta$ is
      the boundary layer edge. (b) Spatial distribution of the EQWM
      relative error. (c) Spatial distribution of the BFM relative
      error.  The colormaps in panels (b) and (c) are identical to
      those used in Figure~\ref{fig:GP1M}.}
    \label{fig:gp_1M_spatial}
\end{figure}

The spatial error distributions in Figure~\ref{fig:gp_1M_spatial}
reinforce these observations. In the FPG region, the strong
performance of BFM originates from its high near-wall accuracy,
consistent with the behavior seen in earlier TBL cases. In the
subsequent APG region, the performance of both models deteriorates;
however, BFM remains noticeably more accurate. This outcome
underscores that while BFM generalizes more effectively through the
re-transition, the elevated errors also reflect the inherent
difficulty of this case, where the transition begins from a partially
laminar state rather than from a fully laminar boundary layer.

\paragraph{Underperforming cases.} 
There are three main cases (or five if counting subcases) in which
BFM, although outperforming EQWM, still exhibits large errors (i.e.,
relative errors above 20\%). These cases correspond to the curved TBL
studied by \citet{appelbaumSystematicDNSApproach2025}, the
transitional boundary layer reported by \citet{roach1990influence},
and the three-dimensional TBL induced by a spinning cylinder
investigated by \citet{driver1987experimental}.

\begin{figure}
    \centering
    \includegraphics[width=0.5\linewidth]{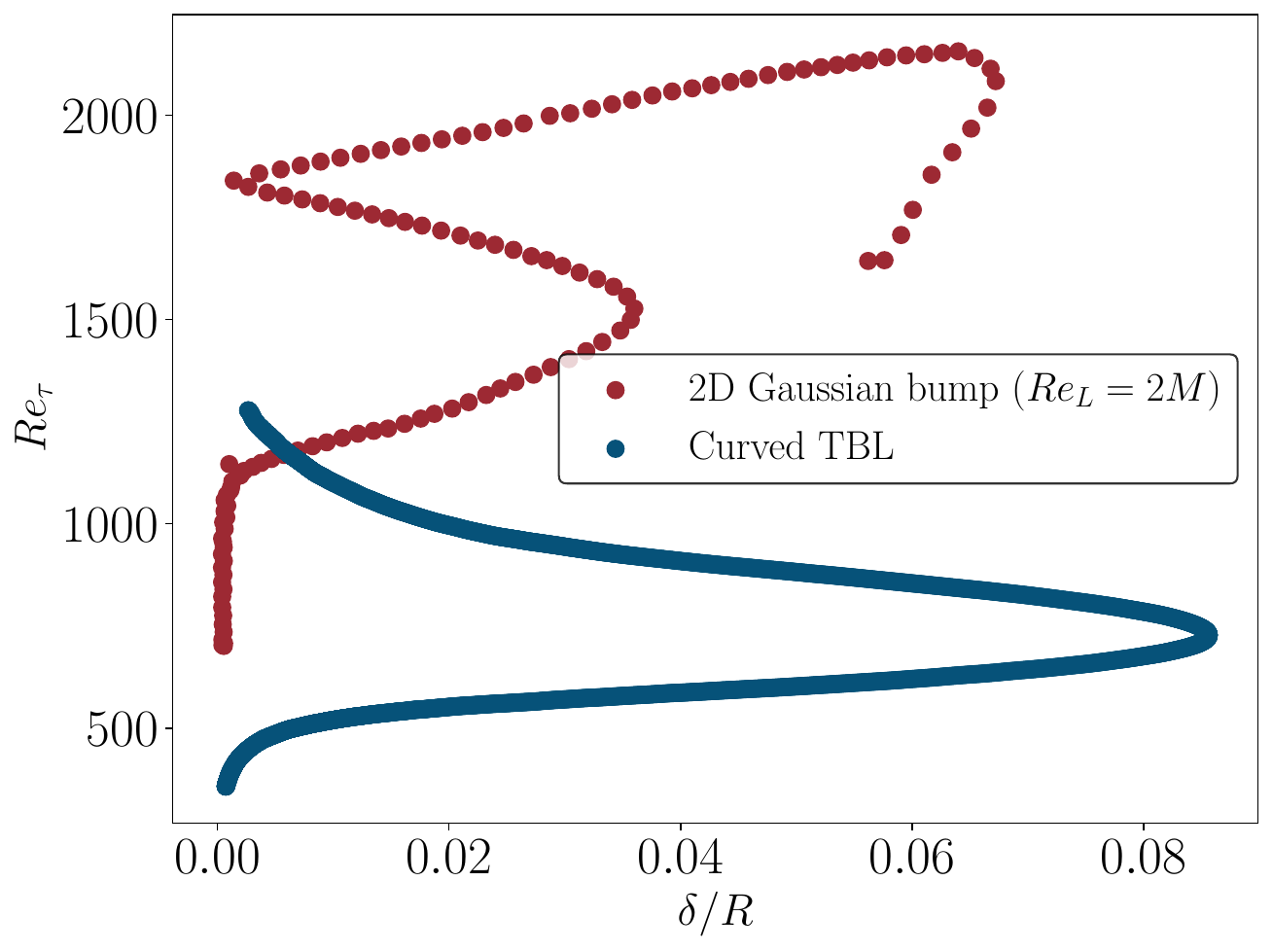}
    \caption{Friction Reynolds number ($Re_{\tau}$) versus the ratio
      of boundary layer thickness to curvature radius ($\delta/R$) for
      training (Gaussian bump with $Re_L=2M$) and testing case (curved
      TBL~\cite{appelbaumSystematicDNSApproach2025}).}
    \label{fig:curved_TBL_exp}
\end{figure}

In the first case (curved TBL), the observed errors are likely due to
the limited exposure of the model to additional curvature effects during
training. Figure~\ref{fig:curved_TBL_exp} illustrates this data gap by
showing the two-dimensional parameter space spanned by the friction
Reynolds number ($Re_{\tau}$) and the ratio of boundary-layer thickness
to curvature radius ($\delta/R$). The curved TBL case (blue) clearly
occupies a region that is only partially represented by the training
data incorporating curvature effects (red). This observation suggests
the need for a more systematic expansion of the training set to better
populate this portion of the parameter space.

For the second and third cases---both experimental---it is important
to first acknowledge that measurement uncertainties can be
significant. As reported by \citet{driver1987experimental}, the
friction coefficient alone may carry an uncertainty of up to
$\pm10\%$. Such errors may hinder our ability to evaluate the
performance of BFM, as the observed discrepancies could largely stem
from measurement noise rather than modeling deficiencies.  In the
transitional boundary-layer case, this interpretation is further
supported by the DNS results of \citet{lee_large-scale_2017}, which
exhibit much smaller errors for BFM. For the 3D TBL case, the relatively large
discrepancies likely arise from two factors: the lack of training data
with strong mean-flow three-dimensionality and the experimental
uncertainties themselves.

\subsection{\textit{A posteriori} testing}
\label{Sec:Apost}

We assess the BFM by conducting \emph{a posteriori} WMLES under two
complementary modalities: \emph{nudged a posteriori} tests, which
apply a forcing term to correct deviations of the mean velocity
profile from the reference DNS solution, and \emph{true a posteriori}
tests, which expose the model to the full coupled LES/WM error without
any external correction.  The cases considered include: (i) turbulent
channel flow; (ii) flat-plate turbulent boundary layers under adverse
and favorable pressure gradients; (iii) a three-dimensional Gaussian
bump to evaluate curvature-induced non-equilibrium effects; and (iv) a
realistic aircraft geometry at multiple angles of attack. Cases (i)
and (ii) are performed using nudged WMLES, whereas cases (iii) and
(iv) are conducted as standard (unnudged) WMLES.

\subsubsection{Nudged WMLES}

The nudging method used for \textit{a posteriori} testing is implemented
by adding a corrective forcing term to the LES governing equations. For
an incompressible flow, the nudged momentum equation takes the form
\begin{equation}
  \frac{\partial \mathbf{u}}{\partial t} + \mathbf{u} \cdot \nabla
  \mathbf{u} = -\frac{1}{\rho}\nabla p
  + \nabla \cdot \big( (\nu + \nu_t)\mathbf{S} \big)
  + \mathbf{F}_{\text{nudging}},
  \label{eq:ns_nudging}
\end{equation}
where $\mathbf{F}_{\text{nudging}}$ is the forcing term used to
correct the mean velocity profile. We denote $u$, $v$, and $w$ as the
streamwise ($x$), wall-normal ($y$), and spanwise ($z$) velocity
components, respectively. For a channel flow, the nudging term is
$\mathbf{F}_{\text{nudging}} = [- \alpha (u_{\text{DNS}} - \langle
  u^{\mathrm{obs}} \rangle_{xzt}),\, 0,\, 0]$, where $u_{\text{DNS}}$
is the DNS mean velocity profile and $\langle u \rangle_{xzt}$ is the
LES mean velocity profile averaged over the streamwise and spanwise
directions and a period of time.  For a turbulent boundary layer, the forcing
becomes $\mathbf{F}_{\text{nudging}} = [- \alpha (u_{\text{DNS}} -
  \langle u \rangle_{zt}),\, 0,\, 0]$, where the only difference is
that the LES velocity is no longer averaged in the streamwise
direction due to streamwise inhomogeneity.

\subsubsection{Flow solver}

The BFM is implemented in the GPU-accelerated flow solver
\texttt{charLES} (developed by Cadence, Inc.). The code is a
finite-volume, second-order-accurate, compressible solver that employs
a low-dissipation spatial discretization scheme, making it well suited
for WMLES. A Voronoi-diagram-based meshing strategy is used to
generate high-quality unstructured meshes. The solver has been
extensively validated across a wide range of configurations
\cite{goc2021large, goc2024wind, goc2025studies,
  bres2017unstructured}.

An exponential time filter, similar to that proposed by
\citet{yangIntegralWallModel2015}, is applied to smooth the model
inputs. The filter timescale is defined as $T_f = 100 /
\|\mathbf{S}_1\|$, where $\|\mathbf{S}_1\|$ is the magnitude of the
rate-of-strain tensor at the first matching location.  This value of
$T_f$, used in all simulations reported below, is empirically chosen
to suppress instantaneous oscillations in the pressure gradient that
would otherwise degrade the accuracy of the wall model inputs. Further
analysis of this choice is provided in
Appendix~\ref{Sec:App:TimeFilter}. Unless otherwise noted, the Vreman
SGS model~\cite{vreman2004eddy} is employed in all cases.

\subsubsection{Computational cost}

We assessed the computational cost of BFM on the MIT Satori cluster
using two GPU nodes, each equipped with four NVIDIA V100 32GB GPUs
(for a total of 8 GPUs). The cost of the wall model is a critical
factor to consider, as excessive overhead could offset the potential
benefits of using WMLES.  Our results show that BFM offers inference
costs comparable to EQWM. This efficiency is primarily enabled by a
custom CUDA kernel that leverages shared memory for storing model
weights. To quantify performance, we measured the total solver cost
for 100,000 time steps (excluding solver initialization) using the
Gaussian bump case described in \S\ref{Subsubsec:Case}. The normalized
solver speed was 0.0144 GPU-s/Mcv/step for EQWM and 0.0150
GPU-s/Mcv/step for BFM, where GPU-s, Mcv, and step denote GPU-seconds,
million control volumes, and number of time steps, respectively.  This
represents only a 4\% increase in total simulation time for BFM---an
overhead that is well justified by its improved accuracy compared to
EQWM.

\subsubsection{Results}
\label{Subsubsec:Case}

\begin{figure}
    \centering
    \includegraphics[width=.9\linewidth]{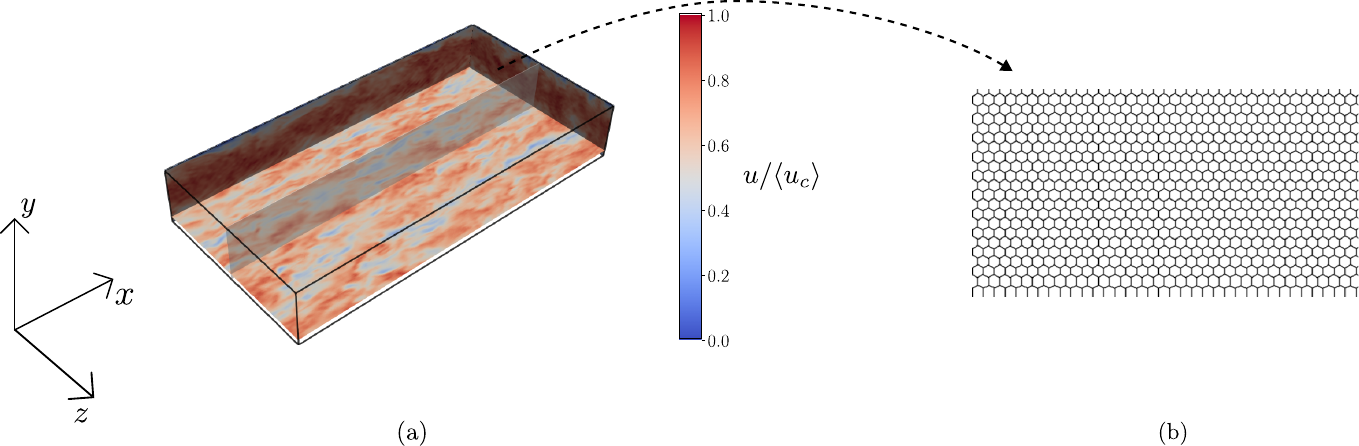}
    \caption{(a) Schematic of the turbulent channel flow
      simulation. The quantity represented is the instantaneous
      streamwise velocity from \textit{a posteriori} testing and
      normalized by mean centerline velocity at different planes.  (b)
      A plane cut visualizing the fine computational grid
      ($\Delta/h=0.05$).}
    \label{fig:TCH}
\end{figure}

\paragraph{Turbulent channel flow.}
A canonical turbulent channel flow was simulated at a friction
Reynolds number of $Re_\tau \approx 4,200$. The computational domain
has dimensions of $(L_x,L_y,L_z)=(4\pi h, 2h, 2\pi h)$, where $h$ is
the half-channel height. A schematic of the domain is shown in
Figure~\ref{fig:TCH}(a), with $x$, $y$, and $z$ representing the
streamwise, wall-normal, and spanwise directions, respectively. The
flow is periodic in the streamwise and spanwise directions. A pressure
gradient is applied in the streamwise direction to drive the flow.

Three cases are simulated using coarse ($\Delta/h=0.2$), medium
($\Delta/h=0.1$), and fine ($\Delta/h=0.05$) isotropic grid
resolutions, where $\Delta$ is the grid size. These resolutions are
representative of practical external aerodynamics simulations,
corresponding to approximately 5, 10 and 20 points per boundary layer,
respectively~\cite{goc2021large, lozano2022performance}. A sample of
the fine isotropic grid is shown in Figure~\ref{fig:TCH}(b).

\begin{figure}
    \centering
    \includegraphics[width=0.5\linewidth]{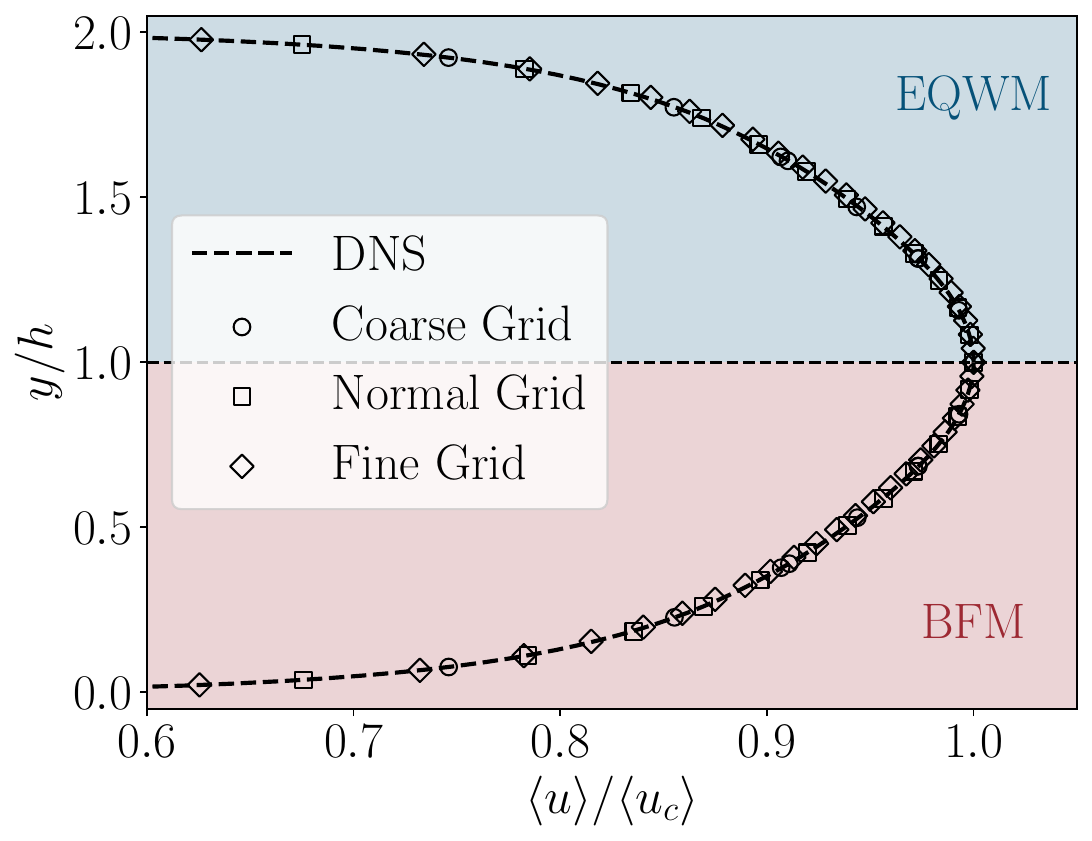}
    \caption{Mean streamwise velocity profile $\langle u\rangle$ in
      \textit{a posteriori} testing of a turbulent channel flow
      normalized by the mean centerline velocity $\langle u_c\rangle$
      obtained using the EQWM (top half) and BFM (bottom half)
      simulated on coarse, medium and fine grids.}
    \label{fig:CH_umean}
\end{figure}

The mean profiles from the nudged WMLES, shown in
Figure~\ref{fig:CH_umean}, demonstrate that the nudging scheme
successfully corrects the mean profiles to match the DNS data. The
predicted wall shear stresses for the three grid resolutions are
summarized in Table~\ref{tab:apost_CH}. Consistent with our \textit{a
  priori} analysis, both BFM and EQWM perform well for this case, with
BFM exhibiting a slightly worse performance.
\begin{table}
    \centering
    \begin{tabular}{ccccc}
    \toprule
    Grid Name& $\Delta/h$ & Grid points/$h$ &   $\varepsilon_\text{BFM}$\%              & $\varepsilon_\text{EQWM}$\%    \\
    \midrule
    Coarse & 0.20 & 5  &0.05\%      & 1.12\%     \\ 
    Medium & 0.10 & 10 &2.03\%      & 1.03\%     \\ 
    Fine & 0.05 & 20 &1.76\%      & 0.84\%     \\ 
    \bottomrule
    \end{tabular}
    \caption{\textit{A posteriori} testing of BFM and EQWM in WMLES of
      a turbulent channel flow at three grid resolutions, defined by
      $\Delta/h$, where $h$ is the channel half-height. Errors are
      reported as percentages relative to the reference DNS solution.}
    \label{tab:apost_CH}
\end{table}

\begin{figure}[!htbp]
    \centering
    \includegraphics[width=.9\linewidth]{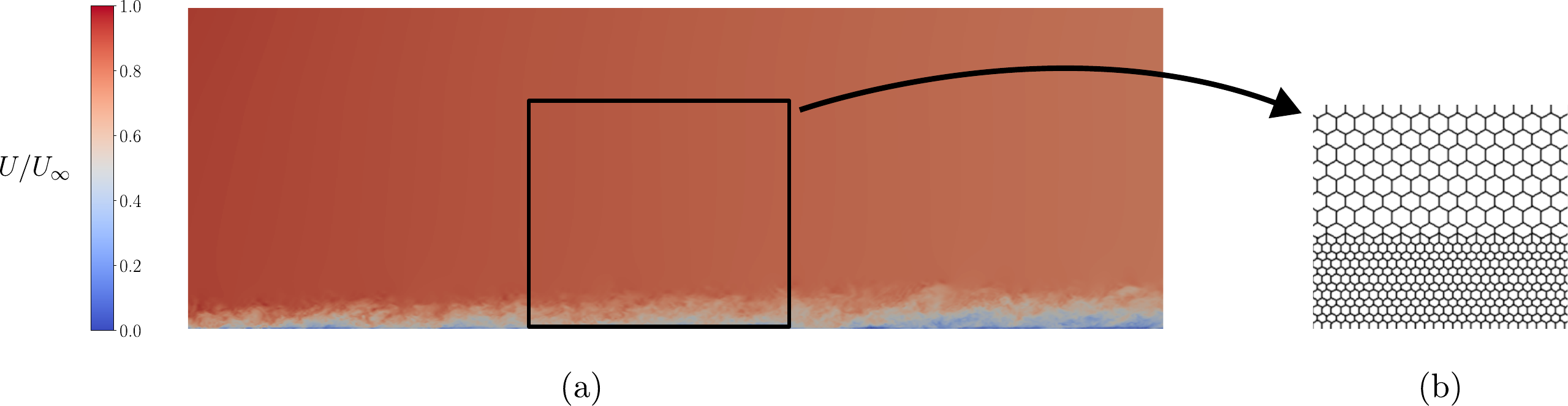}
    \caption{(a) The contour plot of the normalized instantaneous
      streamwise velocity at the center plane of the computational
      domain for \textit{a posteriori} testing of the APG case. (b)
      The grid in the black rectangular region shown in the left
      plot.}
    \label{fig:TBL_schematic}
\end{figure}

\paragraph{APG/FPG turbulent boundary layers.}
Turbulent boundary layers subjected to APG and FPG are evaluated
through \textit{a posteriori} testing. The geometry and boundary
conditions exactly match those of the corresponding top-ramp TBL cases
in the training database, featuring ramp angles of $5^{\circ}$ for the
APG case and $-4^{\circ}$ for the FPG case. The Reynolds number based
on momentum thickness ranges from $Re_{\theta} = 670$ to $6,000$ for
the APG case and from $Re_{\theta} = 670$ to $1,000$ for the FPG case,
while the Clauser parameter spans $\beta = 0$ to $2$ (APG) and $\beta
= 0$ to $-0.6$ (FPG).
WMLES is performed using grid resolutions corresponding to 5--40
points per boundary-layer thickness for the APG case and 5--15 points
for the FPG case. The mesh used in the APG simulation is shown in
Figure~\ref{fig:TBL_schematic}(a) and corresponds to the highlighted
region in Figure~\ref{fig:TBL_schematic}(b). The total number of
control volumes is 4.7 million for APG and 2.1 million for FPG.

\begin{figure}
    \centering
    \begin{subfigure}[b]{0.7\textwidth}
        \includegraphics[width=\linewidth]{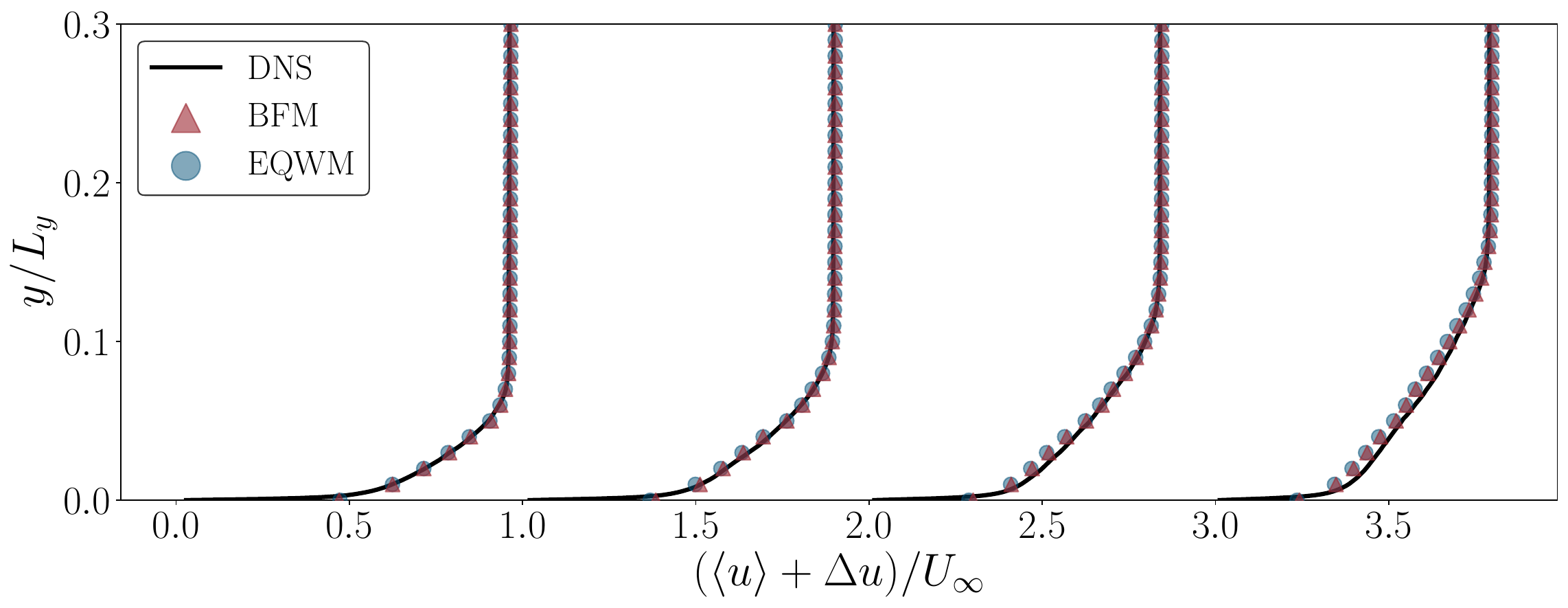}
        \caption{}
        \label{fig:sub:u}
    \end{subfigure}
    \hfill
    \begin{subfigure}[b]{0.5\textwidth}
        \includegraphics[width=\linewidth]{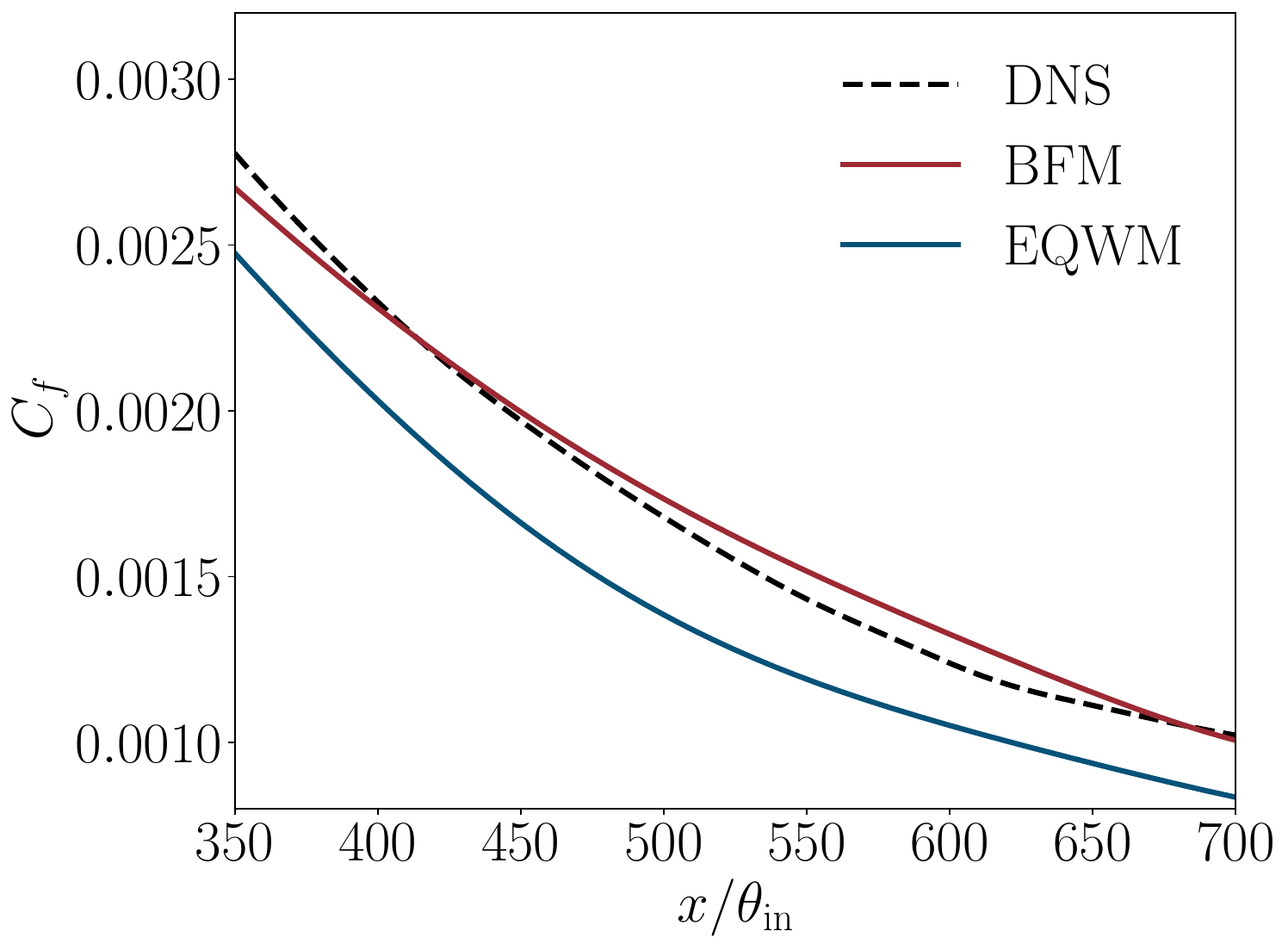}
        \caption{}
        \label{fig:sub:tauw_deg5_comp}
    \end{subfigure}
    \hfill
    \begin{subfigure}[b]{0.46\textwidth}
        \includegraphics[width=\linewidth]{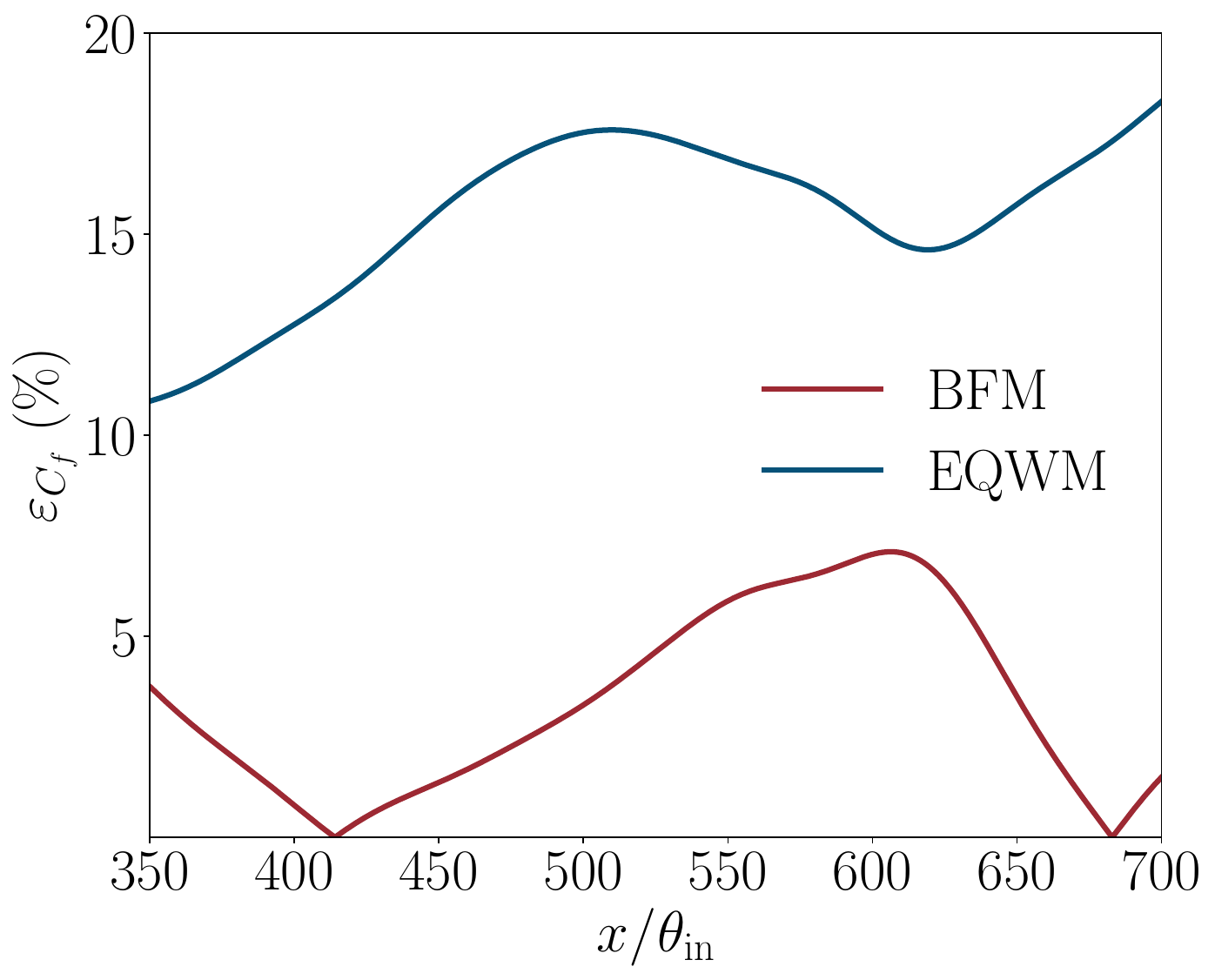}
        \caption{}
        \label{fig:sub:tauw_deg5_comp_err}
    \end{subfigure}
    \caption{\textit{A posteriori} testing of BFM and EQWM in the
      WMLES of an APG TBL. The configuration corresponds to a
      turbulent boundary layer developing under a top ramp inclined at
      $5^{\circ}$.  (a) Mean velocity profiles at four streamwise
      locations, $x/L_y = 3, 4, 5,$ and $6$. Profiles are vertically
      shifted by $\Delta u / U_{\infty} = x/L_y - 3$ for clarity.  (b)
      Streamwise distribution of the predicted skin-friction
      coefficient $C_f$.  (c) Relative error $\varepsilon_{C_f} =
      (C_{f,\text{pred}} - C_{f,\text{DNS}}) / C_{f,\text{DNS}}$.}
    \label{fig:tauw_deg5_apot}
\end{figure}

\begin{figure}
    \centering
    \begin{subfigure}[b]{0.7\textwidth}
        \includegraphics[width=\linewidth]{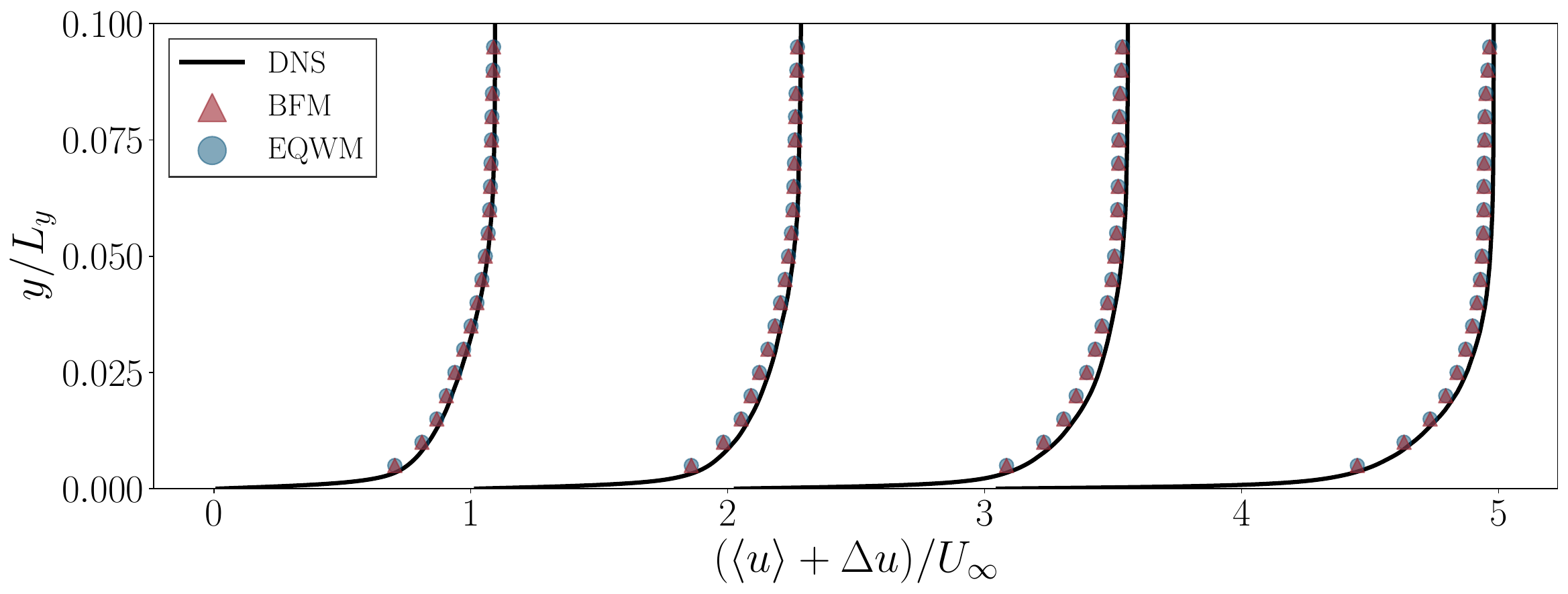}
        \caption{}
        \label{fig:sub:u-4}
    \end{subfigure}
    \hfill
    \begin{subfigure}[b]{0.5\textwidth}
        \includegraphics[width=\linewidth]{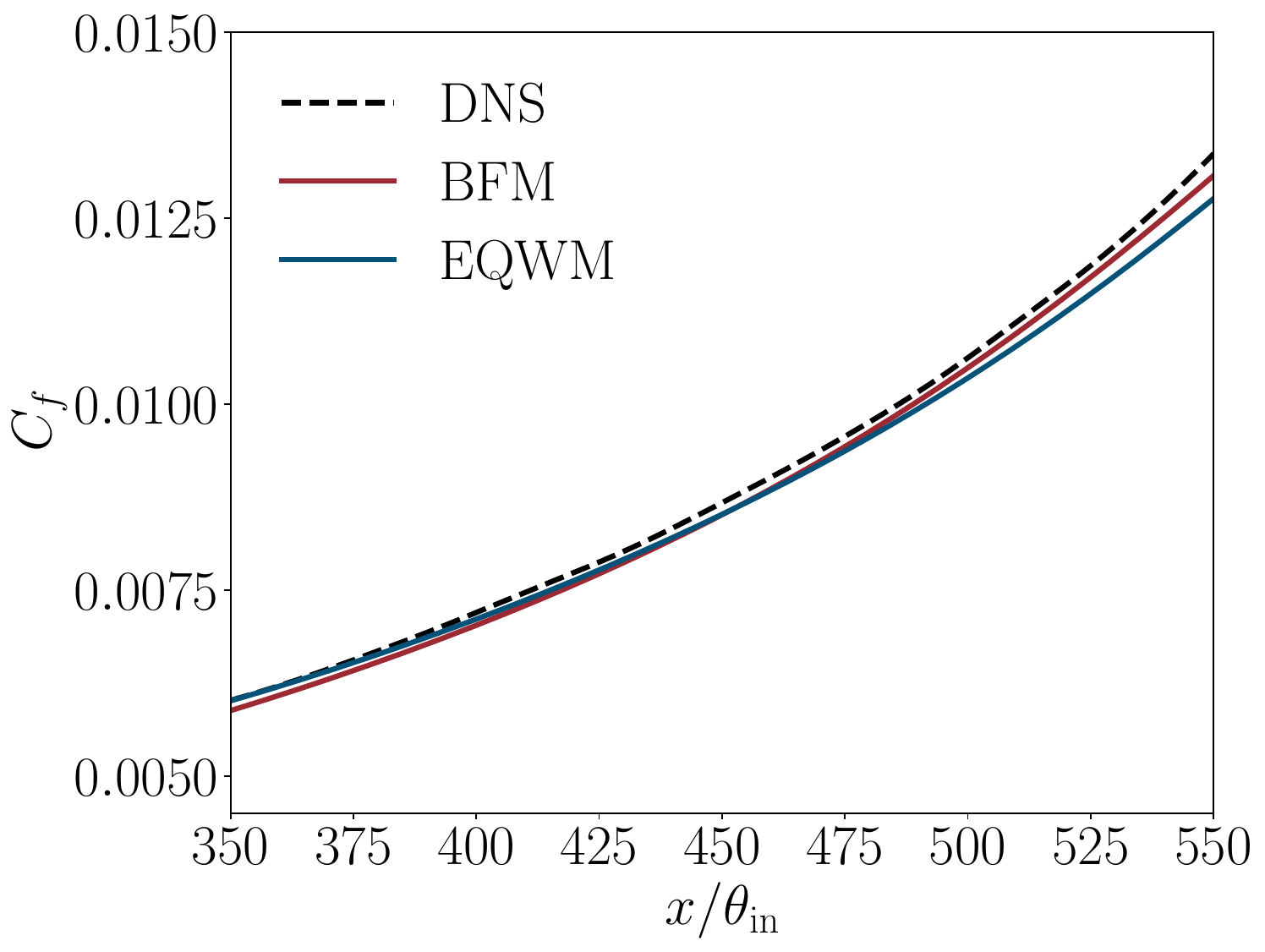}
        \caption{}
        \label{fig:sub:tauw_deg-4_comp}
    \end{subfigure}
    \hfill
    \begin{subfigure}[b]{0.46\textwidth}
        \includegraphics[width=\linewidth]{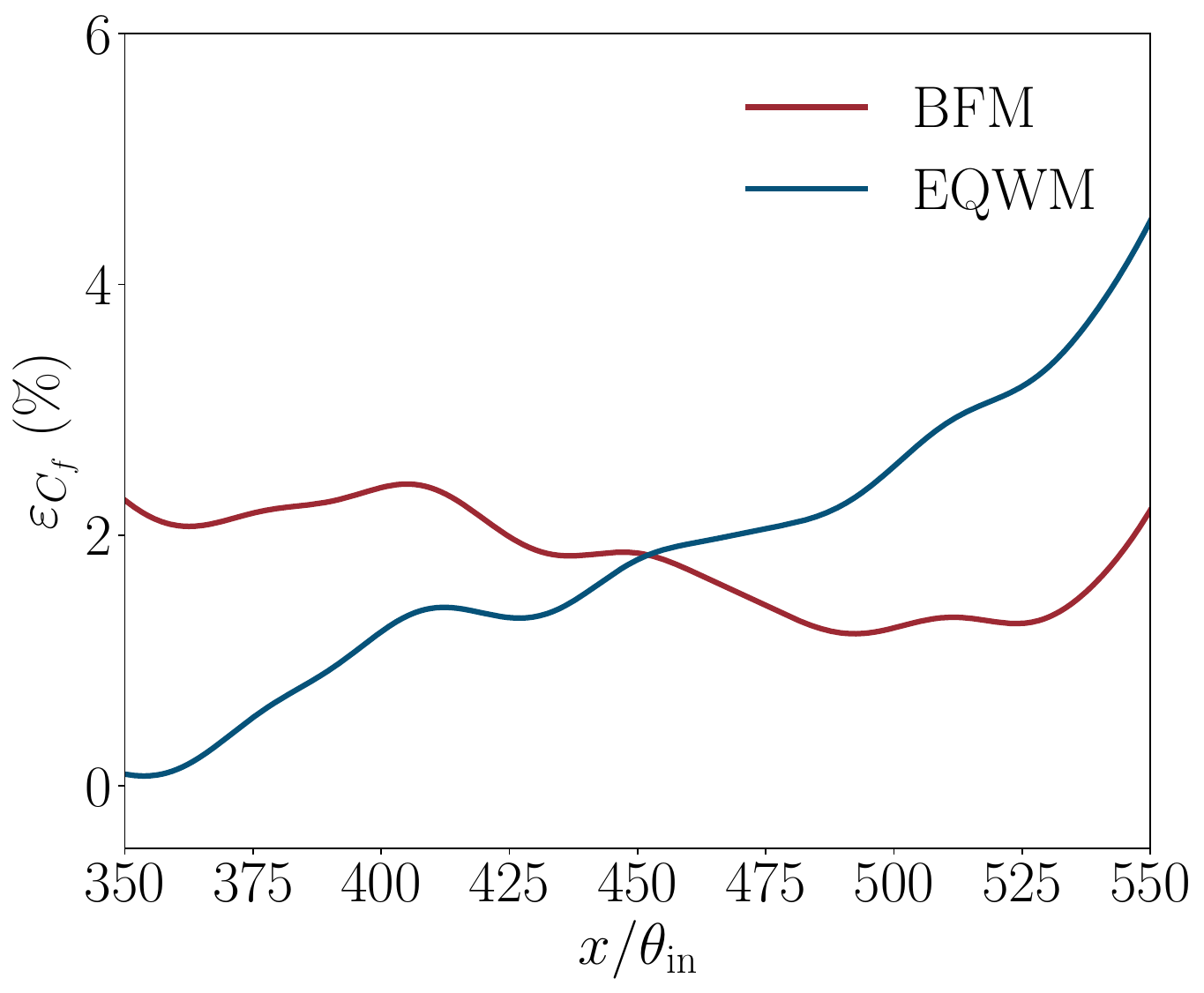}
        \caption{}
        \label{fig:sub:tauw_deg-4_comp_err}
    \end{subfigure}
    \caption{\textit{A posteriori} testing of BFM and EQWM in the
      WMLES of an FPG TBL. The configuration corresponds to a
      turbulent boundary layer developing under a top ramp inclined at
      $-4^{\circ}$.  (a) Mean velocity profiles at four streamwise
      locations, $x/L_y = 3, 4, 5,$ and $6$. Profiles are vertically
      shifted by $\Delta u / U_{\infty} = x/L_y - 3$ for clarity.  (b)
      Streamwise distribution of the predicted skin-friction
      coefficient $C_f$.  (c) Relative error $\varepsilon_{C_f} =
      |C_{f,\text{pred}} - C_{f,\text{DNS}}|/C_{f,\text{DNS}}$.}
    \label{fig:tauw_deg-4_apot}
\end{figure}

We focus on the downstream region where the flow experiences adverse
or favorable pressure-gradient effects. In both cases, the nudged
WMLES accurately reproduces the mean velocity profiles at all
streamwise stations, independent of the wall model used
(Figures~\ref{fig:sub:u} and~\ref{fig:sub:u-4}). For the APG case
(Figures~\ref{fig:sub:tauw_deg5_comp}
and~\ref{fig:sub:tauw_deg5_comp_err}), WMLES with BFM maintains a
relative error between 0\% and 7\% in predicting the friction
coefficient across the entire domain, whereas EQWM exhibits noticeably
larger errors, ranging from 10\% to 18\%. For the FPG case
(Figure~\ref{fig:tauw_deg-4_apot}), both wall models provide accurate
predictions of the wall-shear stress; EQWM errors fall between 0\% and
4\%, while BFM consistently achieves errors near 2\% throughout the
region considered.

\paragraph{3D Gaussian bump.}
The Gaussian bump considered here corresponds to the three-dimensional
configuration tested experimentally by
\citet{gray2022experimental}. The flow in this geometry is highly
three-dimensional---a feature not explicitly included in our training
data, which only comprised a spanwise-periodic, 2-D Gaussian
bump. Unlike the previous tests, this case is conducted as a
\textit{true a posteriori} evaluation, with no nudging applied to
correct deviations in the mean profiles caused by suboptimal SGS model
performance or deficient inflow conditions. As a result, the errors
reported here include both internal and external wall modeling errors.

\begin{figure}
    \centering
    \begin{subfigure}[b]{0.45\textwidth}
        \includegraphics[width=\linewidth]{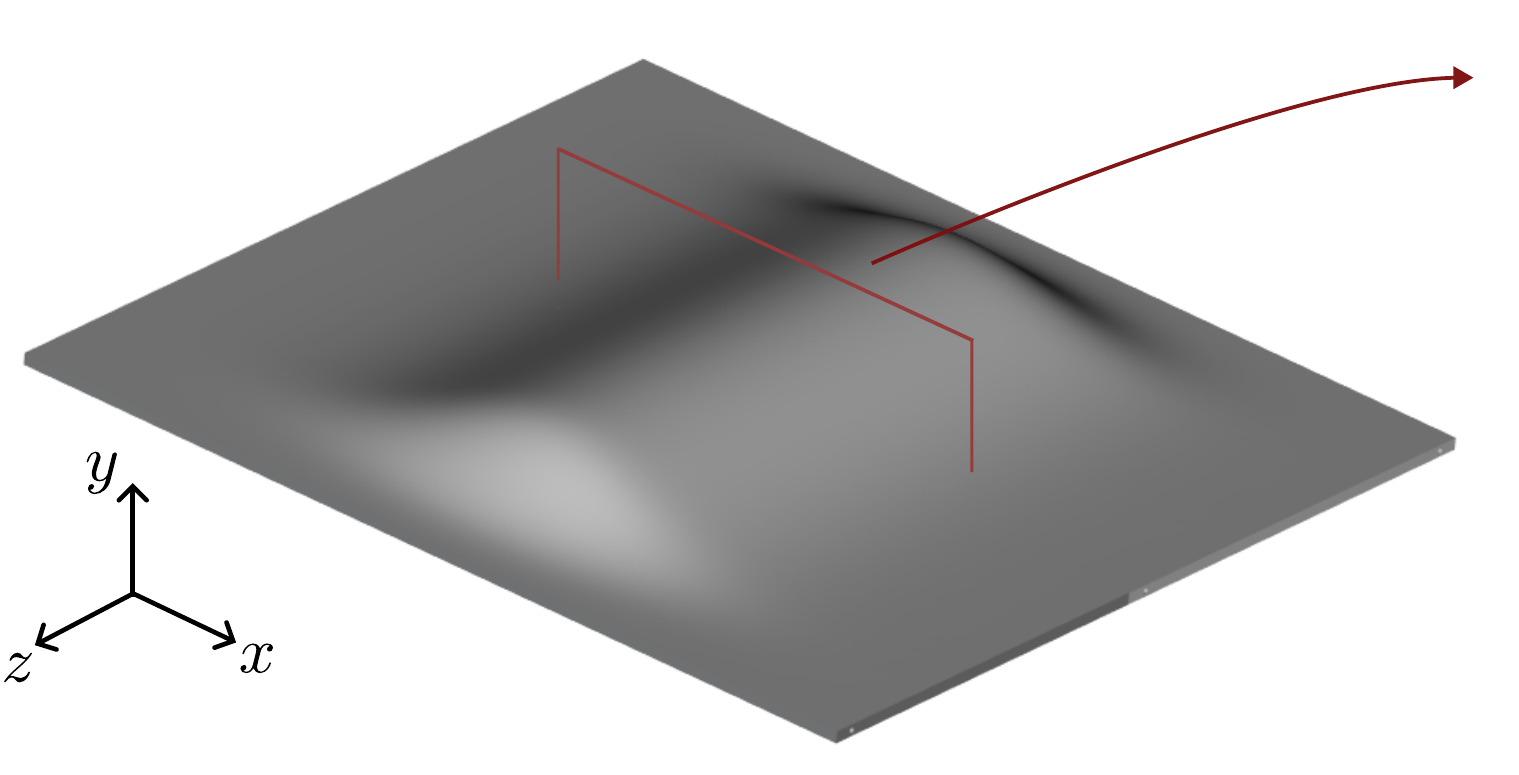}
        \caption{}
        \label{subfig:GP_schematic}
    \end{subfigure}
    \hfill
    \begin{subfigure}[b]{0.48\textwidth}
        \includegraphics[width=\linewidth]{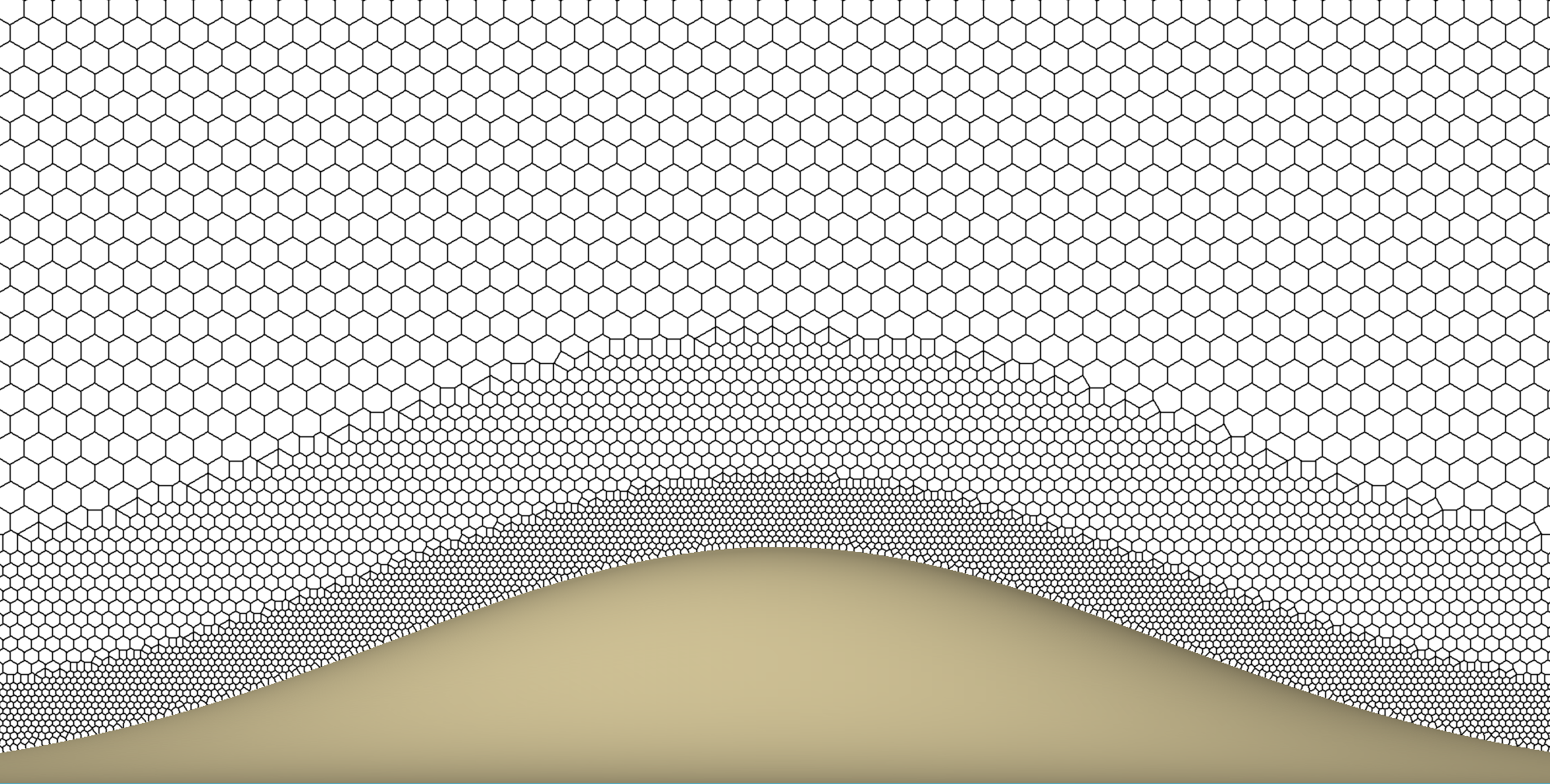}
        \caption{}
        \label{subfig:GP_grid}
    \end{subfigure}
    \caption{(a) The geometry of the three-dimensional Gaussian bump
      case used by \citet{gray2022experimental}. (b) A cut of the
      Voronoi grid used for \textit{a posteriori} test.}
    \label{fig:GP_schematic}
\end{figure}

The geometry, shown in Figure~\ref{subfig:GP_schematic}, is defined as
$y(x, z) = h \cdot \left( \left(1 + \operatorname{erf} \left(
\left(L/2 - 2z_0 - |z|\right) / z_0 \right) \right) / 2 \right) \cdot
\exp\left( -\left(x / x_0\right)^2 \right)$, where $L =
0.9144~\mathrm{m}$ is the bump length, $x_0 / L \approx 0.195$, $z_0 /
L = 0.06$, $h / L = 0.085$, and $\operatorname{erf}$ denotes the error
function, which provides tapering in the spanwise direction.  A
two-dimensional cross-section of the computational grid is shown in
Figure~\ref{subfig:GP_grid}, which employs three levels of isotropic
refinement. The simulation comprises 28.96 million control volumes,
providing approximately 10 points per boundary layer thickness at the
bump apex. The Reynolds number based on the bump length is $Re_L =
3.41 \times 10^6$.  The computational domain is a rectangular box
extending from $-L$ to $1.5L$ in the streamwise direction (with the
bump apex located at $x = 0$), from $-0.5L$ to $0.5L$ in the spanwise
direction, and from $0$ to $0.5L$ in the wall-normal
direction. Boundary conditions are defined as follows: a constant,
uniform inflow is imposed at the inlet; a non-reflecting
characteristic boundary condition with constant pressure is applied at
the outlet; and free-slip conditions are enforced at the lateral and
top boundaries.

\begin{figure}
    \centering
    \includegraphics[width=\linewidth]{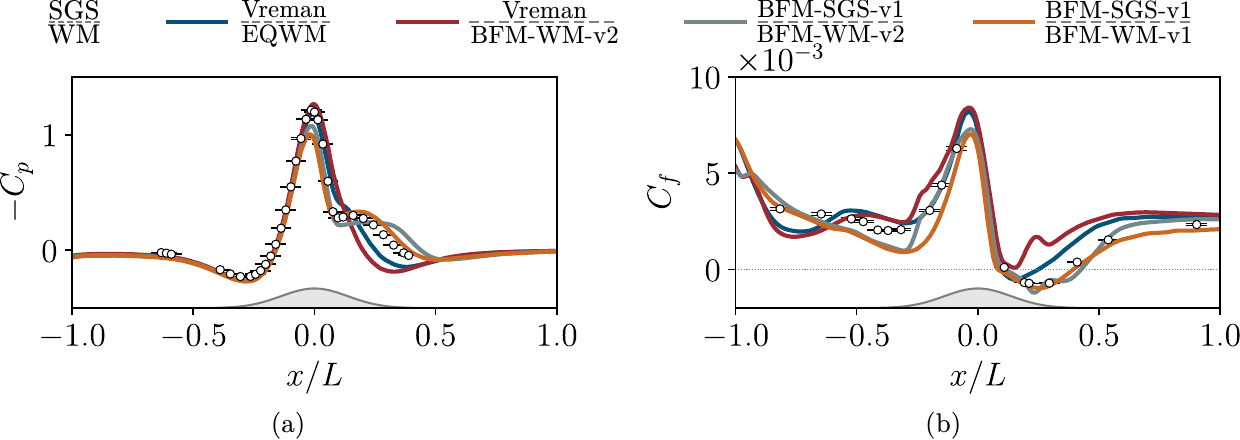}
    \caption{\textit{A posteriori} testing of BFM and EQWM in the
      WMLES of a Gaussian bump.  (a) Pressure coefficient and (b)
      friction coefficient at the mid-span plane ($z/L = 0$). The blue
      and red lines correspond to EQWM and BFM coupled with the Vreman
      SGS model, respectively. The green line shows results from BFM
      coupled with the BFM-SGS-v1 model of
      \citet{arranzBuildingblockflowComputationalModel2024}. The
      orange line uses the same BFM-SGS-v1 model combined with the
      earlier BFM wall model (BFM-WM-v1) from
      \citet{arranzBuildingblockflowComputationalModel2024}.
      Experimental measurements from \citet{gray2022experimental} are
      shown as black circles with error bars.}
    \label{fig:GP_Results}
\end{figure}

\begin{figure}
    \centering
    \includegraphics[width=0.5\linewidth]{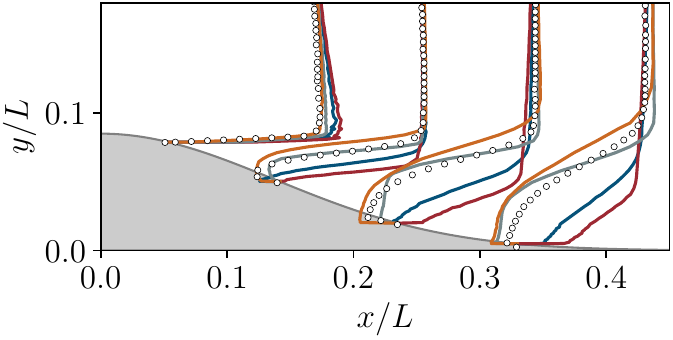}
    \caption{Mean streamwise velocity profile at four stations
      downstream of the apex for \textit{a posteriori} testing of the
      Gaussian bump. The colors of the line are the same as
      Figure~\ref{fig:GP_Results}.}
    \label{fig:GP_U}
\end{figure}

As shown in Figure~\ref{fig:GP_Results}, both BFM and EQWM perform
poorly when coupled with the Vreman SGS model. The EQWM yields a very
small separation bubble, while BFM predicts no separation at all. This
poor performance---further evidenced by the highly inaccurate mean
velocity profiles in Figure~\ref{fig:GP_U}---highlights the critical
importance of selecting an SGS model that minimizes external wall
modeling errors. The combined effect of external and internal wall
modeling errors leads to significantly degraded overall predictions.

Figure~\ref{fig:GP_Results} also includes results obtained by coupling
the current BFM (i.e., BFM-WM-v2) with BFM-SGS-v1 developed by
\citet{arranzBuildingblockflowComputationalModel2024}. The comparison
between SGS models (indicated in red and green) reveals a substantial
improvement in the predictions of both pressure and friction
coefficients when using BFM-SGS-v1.  This improvement is primarily due
to the enhanced accuracy of the mean velocity profiles, as shown in
Figure~\ref{fig:GP_U}, when the improved SGS model is employed. These
results emphasize that reducing the external modeling error
significantly enhances the performance of the wall model. In
Appendix~\ref{subsec:SGS}, we also present results for BFM coupled
with the dynamic Smagorinsky (DSM) SGS model, further illustrating the
sensitivity of the \textit{a posteriori} performance to the choice of
SGS model.

Finally, for completeness, we compare the current BFM to its previous
version BFM-WM-v1 coupled with
BFM-SGS-v1~\cite{arranzBuildingblockflowComputationalModel2024}. The
BFM-SGS-v1 model was specifically designed to be used in tandem with
BFM-WM-v1 to minimize external modeling errors. In principle, this
configuration should yield the lowest external error.  However, as
shown in the $C_f$ predictions in Figure~\ref{fig:GP_Results}, the
present BFM delivers improved accuracy on the upstream side of the
bump. This improvement is attributed to the enhanced ability of the
current version to reduce internal wall modeling errors compared to
its predecessor.  In conclusion, when coupled with an appropriate SGS
model, the current data-driven wall model (BFM-WM-v2) provides
superior performance. 

\begin{figure}
    \centering
    \includegraphics[width=0.9\linewidth]{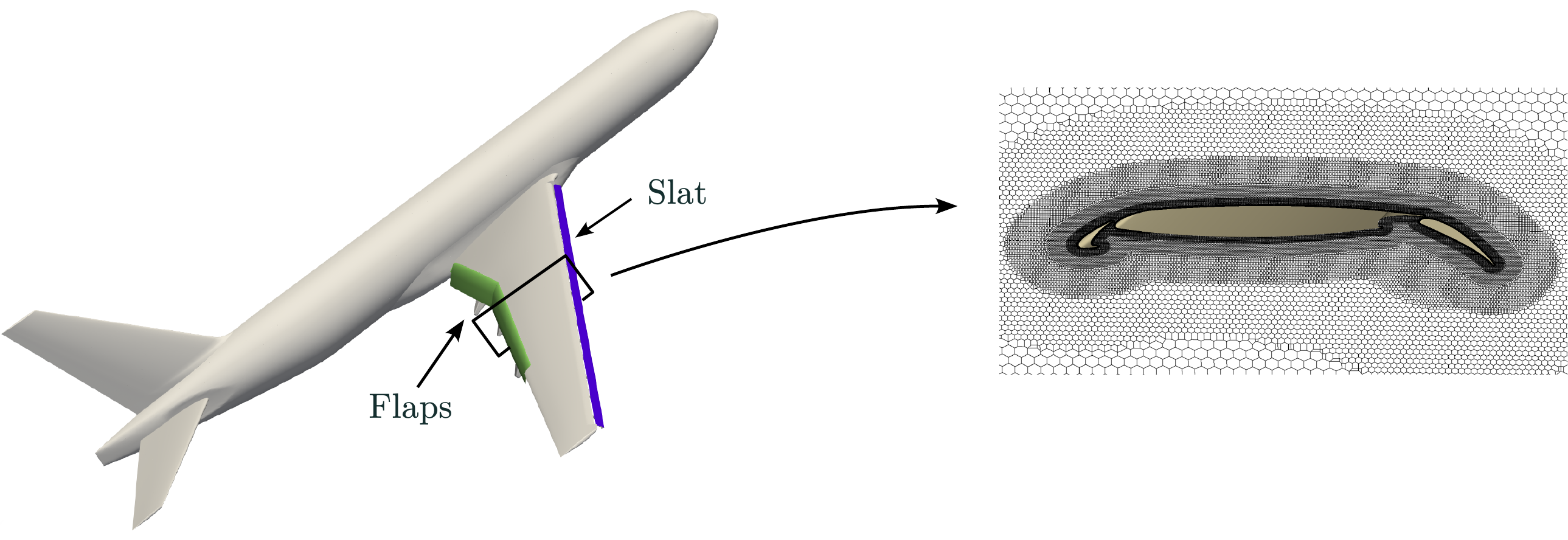}
    \caption{Geometry of the CRM-HL test case 2.3. The leading-edge
      slat (purple) and trailing-edge flaps (green) are deployed,
      while the nacelle and pylons are omitted compared to the
      complete CRM-HL configuration.}
    \label{fig:CRM_Geo}
\end{figure}

\paragraph{NASA High-Lift Common Research Model.}
We evaluate the performance of the current BFM on the NASA High-Lift
Common Research Model (CRM-HL), which served as the primary test case
for the 5th High-Lift Prediction Workshop
(HLPW5)~\cite{clark2025hlpw}. This configuration is widely recognized
as a highly challenging benchmark due to its complex geometry and the
intricate flow physics encountered near the maximum lift condition. In
particular, it poses significant challenges for WMLES, as previously
noted by \citet{kiris2025high}.  The specific configuration considered
here includes both a leading-edge slat and a trailing-edge flap,
corresponding to case 2.3 of HLPW5. The geometry of the configuration
is shown in Figure~\ref{fig:CRM_Geo}.

Following the computational setup of \citet{goc2024wind}, the
simulation is conducted at a freestream Mach number of 0.2 and a
Reynolds number of $5.49 \times 10^6$ based on the mean aerodynamic
chord. The geometry consists of a semi-span aircraft model placed
within a large hemispherical domain with a radius 1,000 times the mean
aerodynamic chord. A free-slip, no-penetration boundary condition is
imposed on the symmetry plane. The inlet, defined as the front half of
the hemisphere, features a uniform plug flow, while the outlet on the
rear half employs a non-reflecting boundary condition with the
freestream pressure specified. The simulation is performed with 35
million control volumes, with the minimum grid size
$\Delta_{\text{min}} \approx 0.002$ times the mean aerodynamic chord.

The quantities of interest are the integrated aerodynamic
forces---specifically the lift ($C_L$), drag ($C_D$), and pitching
moment ($C_M$) coefficients defined as
\begin{equation}
    C_L = \frac{L}{\frac{1}{2}\rho_{\infty}U_{\infty}^2S}, \quad
    C_D = \frac{D}{\frac{1}{2}\rho_{\infty}U_{\infty}^2S}, \quad
    C_M = \frac{M}{\frac{1}{2}\rho_{\infty}U_{\infty}^2Sc}, 
\end{equation}
where $\rho_{\infty}$ and $U_{\infty}$ are the freestream density and
velocity, $L$ is the lift force, $D$ is the drag force, $M$ is the
pitching moment, $S$ is the wing area and $c$ is the mean chord
length.

\begin{figure}
    \centering
    \includegraphics[width=1.0\linewidth]{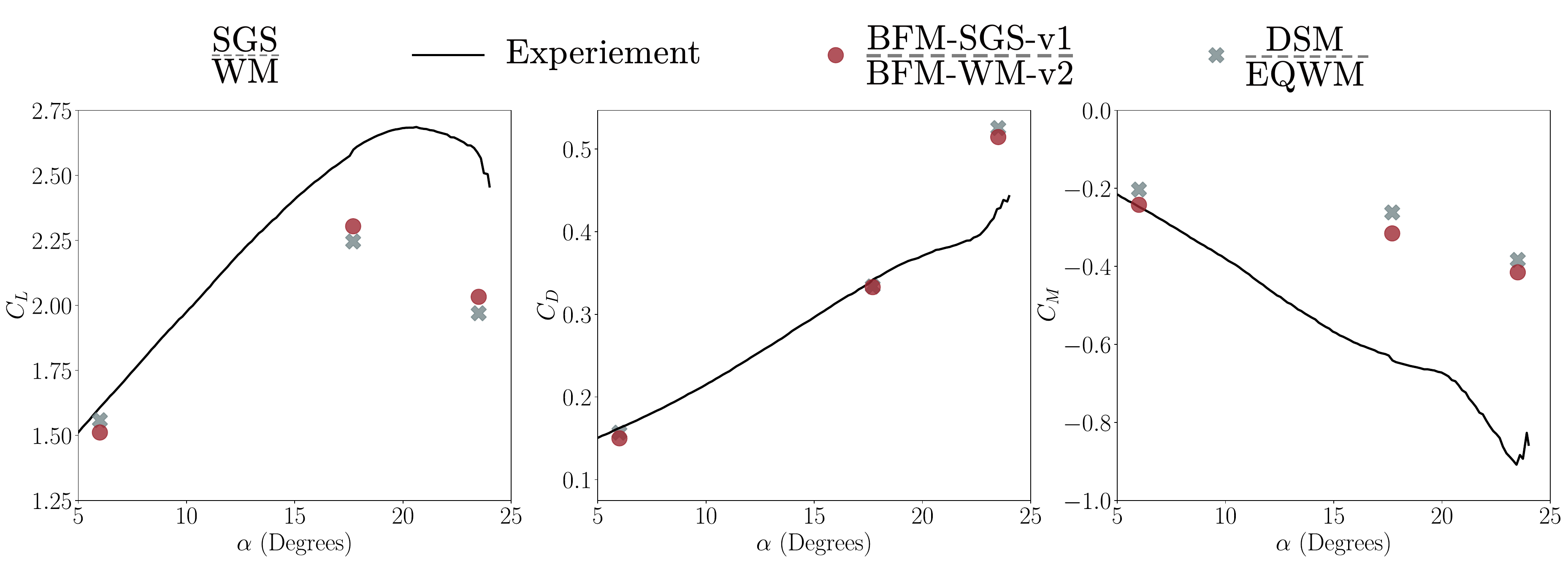}
    \caption{\textit{A posteriori} testing of the NASA High-Lift
      Common Research Model test case 2.3 using BFM and EQWM. (a) Lift
      ($C_L$); (b) Drag ($C_D$); (c) pitching moment ($C_M$)
      coefficients as a function of angle of attack $\alpha$.}
    \label{fig:CRM_F}
\end{figure}

The results are shown in Figure~\ref{fig:CRM_F}. The BFM model is
evaluated at three angles of attack ($\alpha = 6^{\circ}$,
$17.7^{\circ}$, and $23.5^{\circ}$). As expected, all models perform
well in the linear regime at $\alpha = 6^{\circ}$. However, under
near-stall and post-stall conditions, predictions from both BFM and
EQWM deviate significantly from the reference data. Although not shown
here, these discrepancies are primarily due to underprediction of flap
separation and inaccurate modeling of leading-edge
transition—phenomena that are well known to be challenging for
conventional approaches \cite{kiris2025high}. While the BFM approach
offers a avenue for improving predictions in such complex flow
regimes, the present case is likely dominated by wall-modeling errors
external to the BFM itself. Future tests will be conducted using BFM
combined with BFM-SGS-v2.

\section{Discussion}
\label{Sec:Discussion}

\subsection{Model Limitations}

We outline the primary limitations of BFM-WM-v2 along with potential
directions for improvement:
\begin{itemize}
\item \emph{Dependence on the building-block-flow assumption:} The
  core assumption underlying BFM is that a finite set of canonical
  flows (referred to as building-block flows) can encapsulate the
  essential near-wall physics necessary for generalizable wall
  modeling. While this approach provides a structured foundation for
  data-driven modeling, its validity may be limited in cases involving
  separation patterns or flow regimes not present in the training
  data. Two scenarios of this limitation are presented in the
  following points.
\item \emph{Transitional flow regimes:} Although the model performs
  well for laminar and transitional regimes in controlled \textit{a
    priori} settings, it was not trained on a comprehensive set of
  flows capturing laminar-to-turbulent transition
  dynamics. Incorporating such cases is nontrivial, as it may
  introduce numerical instabilities, trigger premature transition, and
  hinder the model to revert to laminar states.
\item \emph{Flows with mean-flow three-dimensionality:} The current
  model predicts only the magnitude of the wall-shear stress, assuming
  alignment between the stress vector $\boldsymbol{\tau}_w$ and the
  velocity vector at the first off-wall grid point,
  $\boldsymbol{u}_1$. This assumption is justified for high Reynolds
  numbers and typical WMLES resolutions, where the two vectors remain
  aligned on average~\cite{lozano2020non}. However, in boundary layers
  with strong mean-flow three-dimensionality, this assumption breaks
  down~\cite{spalart1989theoretical}, leading to misalignment and
  inaccurate predictions. Addressing this limitation would require
  both restructured model inputs and the inclusion of richer
  three-dimensional datasets, which are currently sparse and difficult
  to balance relative to two-dimensional cases.
\item \emph{Local-state sufficiency assumption:} The model relies
  exclusively on local flow quantities as input, under the premise
  that the near-wall region attains a quasi-equilibrium
  state~\cite{baxerres2024evidence}. While this design choice enhances
  generalizability to complex geometries, it limits the ability of the
  model to capture long-range interactions or unsteady effects. Prior
  work on turbulence forecasting has demonstrated that incorporating
  temporal history can substantially improve predictive
  accuracy~\cite{Lozano2020_cau, Srinivasan2019, Nakamura2021}.

\item \emph{Training on ensemble-averaged inputs:} BFM is trained
  using ensemble-averaged quantities from DNS, rather than
  instantaneous flow fields. This design choice simplifies training,
  avoids underfitting with high-dimensional fluctuating
  data~\cite{dupuyDatadrivenWallModeling2023}, and facilitates broader
  dataset inclusion. During inference, a temporal averaging filter is
  applied to approximate these
  conditions~\cite{yangIntegralWallModel2015,
    lozano-duranMachineLearningBuildingblockflow2023}. However, this
  restricts the model from accurately predicting instantaneous
  wall-shear stress and may lead to degraded performance in
  statistically unsteady regimes.

 \item \emph{Sensitivity to SGS model:} A key limitation identified in
   our study is the strong dependence of wall-model performance on the
   choice of SGS model. Errors introduced by underperforming SGS
   models (i.e., external wall-model errors) were found to be the
   dominant source of inaccuracy. This challenge was first discussed
   in detail during the development of
   BFM-WM-v1~\cite{lozano-duranMachineLearningBuildingblockflow2023},
   where the need for a unified wall/SGS modeling framework was
   highlighted. Our group has advanced this direction through the
   development of BFM-v1, a consistent framework that includes both an
   SGS model (BFM-SGS-v1) and a wall model (BFM-WM-v1) designed to
   operate in
   tandem~\cite{arranzBuildingblockflowComputationalModel2024}. Building
   on this foundation, \citet{lingNumericallyConsistentDataDriven2025}
   proposed a strategy for developing numerically consistent SGS
   models by enforcing compatibility with a pre-trained wall
   model. The method generates a training dataset via nudged WMLES,
   ensuring that the SGS model learns from flow fields already aligned
   with the wall model.  Importantly, the procedure makes no
   assumptions about the internal structure of the wall model,
   allowing the resulting SGS model to be integrated seamlessly. Once
   deployed, such an SGS model is expected to substantially improve
   \textit{a posteriori} performance when coupled with the original
   wall model.
  
\end{itemize}

\subsection{Outlook for future versions}

We conclude by outlining the main directions for improving future
versions of the Building‑Block Flow Model wall model. The current
approach primarily relies on a database of pressure‑gradient turbulent
boundary layers. While this foundation has proven effective, extending
the database and refining the formulation are essential for broader
applicability and improved accuracy. Below we discuss the most
relevant avenues for future development.
\begin{itemize}

  \item \emph{Turbulence with mean‑flow three‑dimensionality.}  A
    critical next step involves extending the model and database to
    include flows with mean three-dimensionality. Current datasets
    capture only quasi-two-dimensional physics, while boundary layers
    subjected to lateral pressure gradients exhibit complex
    shear-stress misalignment effects~\cite{lozano2020non}. Existing
    databases of 3D channel flows with imposed spanwise pressure
    gradients~\cite{lozano2020non} provide a useful starting point,
    but an ideal solution would be a new parametric dataset—analogous
    to the PG TBL series—spanning varying spanwise pressure gradients.
    Capturing these effects requires two key model modifications: (i)
    including the wall-parallel velocity-vector orientation as part of
    the model input, and (ii) extending the model output to predict
    both components of the wall-shear-stress vector instead of only
    its magnitude. With these changes, retraining on suitably enriched
    datasets would allow the model to generalize to flows with
    inherent three‑dimensionality.

  \item \emph{Wall-curvature effects.}  Another major extension
    concerns the inclusion of wall-curvature effects through a richer
    collection of training cases. The current version was trained only
    on a single Gaussian-bump configuration. New databases are being
    generated to include TBLs under adverse and favorable pressure
    gradients over curved surfaces. Such an expansion is crucial
    because, in practical flows, curvature and pressure‑gradient
    effects often appear simultaneously. Although parametric studies
    exist for curved turbulent channels~\cite{soldati2025reynolds,
      brethouwer2022turbulent}, no equivalent systematic dataset
    exists for TBLs combining both effects. Creating such a dataset
    remains a primary objective for future work.

  \item \emph{Laminar-to-turbulent transition.}  Accurately modeling
    transitional flows requires mechanisms for automatically switching
    between laminar and turbulent regimes. Developing reliable
    flow-state sensors for this purpose is an ongoing
    challenge. Existing sensor-based
    approaches~\cite{bodart2012sensor, xu2024wall} provide promising
    foundations, but more advanced or data-driven detection techniques
    will likely be needed to ensure robust performance across a wide
    range of Reynolds numbers and flow configurations.

  \item \emph{Compressible-flow extension.}  The building‑block
    modeling approach can be naturally extended to compressible
    regimes. Our group has successfully applied similar methodologies
    to high-speed flows~\cite{mabuilding}. Adapting the current model
    to compressibility will require modifications to account for
    variable density, temperature, and acoustic effects. The main
    limitation, however, is the scarcity of high-fidelity training and
    validation datasets for compressible wall-bounded turbulence,
    which remains a bottleneck for data-driven approaches.

  \item \emph{Wall-roughness effects.}  Incorporating the influence of
    wall roughness represents another important direction. Extending
    the model to account for subgrid-scale roughness elements would
    enable prediction of friction and heat-transfer modifications
    induced by surface texture, a critical capability for realistic
    aerodynamic and environmental applications. We have applied the
    building-block flow model approach to develop wall models for
    rough surfaces~\cite{mabuilding,
      maMachinelearningWallmodelLargeeddy2025}, although these studies
    assume isotropic roughness under equilibrium conditions.

  \item \emph{Numerical consistency with the SGS model.} As discussed
    above, the current BFM exhibits sensitivity to the choice of SGS
    model.  Developing an SGS model that is both numerically and
    physically consistent with BFM would help mitigate the external
    modeling errors identified in the \textit{a posteriori} tests. A
    unified wall/SGS framework is expected to deliver more stable and
    accurate WMLES predictions across a broad range of flow
    conditions.

  \item \emph{Continual learning and adaptive grid refinement.}
    Beyond expanding the physical database, continual-learning
    strategies~\cite{wang2024comprehensive} offer pathway for targeted
    model improvement. This approach allows retraining on new,
    specialized datasets without degrading performance on previously
    learned cases. It is particularly valuable for proprietary or
    complex datasets that cannot be publicly released. Continual
    learning enables users to fine-tune the base model to specific
    flow regimes---e.g., pressure-gradient ranges or Mach
    numbers---while preserving its general predictive capability. This
    strategy bridges the gap between generalizability and
    case-specific optimization. 
  
\end{itemize}

\section{Conclusions}
\label{Sec:Conclusion}

We have introduced a general-purpose wall model for LES grounded in
the building-block flow modeling approach. The method leverages the
hypothesis that the essential near-wall physics in complex turbulent
flows can be locally represented using a finite set of canonical,
simplified flows---termed building-block flows. Rather than
constructing a model that memorizes case-specific behaviors, the goal
of BFM is to learn the underlying flow physics and achieve
generalizability across a wide spectrum of geometries and flow
regimes. The resulting wall model is designed to handle a variety of
physical scenarios, including laminar and turbulent flows, adverse and
favorable pressure gradients, wall curvature, and flow separation. By
focusing on extracting essential flow physical, the model traces a
patchway to depart from case-tuned equilibrium models.

The current version of the model, referred to as BFM-WM-v2, builds on
our prior work (BFM-v1), which established the feasibility of
constructing generalizable wall and SGS models using the
building-block flow
assumption~\cite{lozano-duranMachineLearningBuildingblockflow2023,
  arranzBuildingblockflowComputationalModel2024}. In BFM-v1, the wall
model (BFM-WM-v1) and SGS model (BFM-SGS-v1) were developed jointly
and demonstrated strong performance across a range of canonical and
complex geometries.  In this work, we develop the second-generation
wall model for BFM-v2 to address the limitations in predictive
accuracy observed in off-training-distribution flows. The improvements
in BFM-WM-v2 are driven by insights gained from BFM-v1 and include a
more comprehensive training database, an improved input--output
design, balanced training via a weighted loss, and a substantially
expanded test suite for model validation.

The model formulation uses a feedforward artificial neural network
trained to predict the magnitude of the wall-shear-stress vector from
localized near-wall flow features. All inputs and outputs are
expressed in dimensionless form, ensuring physical invariance under
units transformations. To identify the optimal input variables, we
employ the Information-Theoretic Buckingham-$\pi$ method, which
systematically selects the most informative dimensionless variables
based on the irreducible model error. This procedure yields a compact
and physically interpretable input set consisting of local velocities
and wall-normal distances, without requiring wall-normal gradients or
non-local flow information.  The training dataset contains over one
million samples extracted from 67 building-block cases, including
turbulent channels, pressure-gradient boundary layers, laminar
Falkner--Skan flows, Gaussian bumps, and separation bubbles. To
balance contributions across cases, sample weights are assigned based
on inverse data frequency, and a weighted relative Huber loss is used
during training to mitigate the influence of outliers. The final model
is trained using a four-layer ANN architecture with strong
regularization and early stopping criteria to ensure robust
generalization.

We evaluated the model across a comprehensive suite of validation
cases, distinguishing between internal errors---those intrinsic to the
wall model formulation---and external errors---those arising from the
surrounding LES environment, especially SGS model inaccuracies. This
distinction enables a more precise understanding of the performance of
the model and its limitations in practical WMLES settings. Internal
errors reflect limitations in the physical coverage of the training
dataset or modeling assumptions, while external errors include
imperfections in boundary conditions and incorrect SGS modeling. This
decomposition of errors is central to our evaluation methodology, as
it clarifies whether observed discrepancies stem from the wall model
or from deficiencies elsewhere in the simulation pipeline. We report
results across a rich collection of 140 high-fidelity testing cases (67 training cases included),
including experimental and high-fidelity numerical data. The cases
range from canonical internal flows to highly three-dimensional,
curved, separated, and transitional flows in realistic geometries such
as airfoils and swept wings.

In the \textit{a priori} testing, which isolate wall-model performance
using direct input–output data from high-fidelity sources, the
BFM-WM-v2 consistently outperformed the EQWM across nearly all
cases. These tests covered flows with varying Reynolds numbers,
pressure gradients, wall curvature, and laminar configurations. Even
for cases well outside the training distribution---such as Gaussian
bumps at moderate Reynolds numbers with relaminarization, or
three-dimensional spinning-cylinder boundary layers---the model
maintained resonable predictive accuracy. The overall relative error
across the test sets was reduced substantially compared to EQWM, often
by more than 50\%. Particularly notable was the accuracy of BFM-WM-v2
in the vicinity of incipient separation, where EQWM typically fails
due to its strong equilibrium assumptions. In contrast, BFM-WM-v2
leveraged its non-equilibrium training data to maintain accurate
predictions down to low wall-shear stress magnitudes. The \textit{a
  priori} analysis also revealed that BFM-WM-v2 is robust to
variations in the matching location, capturing the correct trend even
in buffer-layer and logarithmic-layer inputs.

The \textit{a posteriori} evaluation assessed the model performance
within a full WMLES environment, thus accounting for the compounded
effects of SGS modeling, grid resolution, and numerical
discretization. We conducted both \textit{nudged} and \textit{true a
  posteriori} simulations. In nudged simulations, the outer flow is
driven toward a target mean profile to minimize SGS errors. In these
tests, the wall model maintained high fidelity across a wide range of
flow types. In \textit{true a posteriori} simulations, BFM-WM-v2 still
matched or outperformed EQWM, although some degradation was observed
due to large external errors from the SGS model. These results
emphasize that an accurate wall model is a necessary but not
sufficient condition for accurate WMLES
predictions~\cite{lozano-duranMachineLearningBuildingblockflow2023}. Addressing
external errors requires developing SGS models that are numerically
consistent with the wall model---a topic we explore further in
companion work and ongoing development of BFM-SGS-v2.

While the results demonstrate strong performance and generalization,
several limitations remain. First, the model assumes that the
wall-shear stress vector is aligned with the velocity vector at the
first off-wall grid point. Although this assumption is reasonable in
most high–Reynolds-number LES settings, it may introduce bias in flows
with strong velocity skewing. Second, the current training dataset
does not yet cover compressible flows, turbulence with pronounced
mean-flow three-dimensionality, or laminar-to-turbulent transition
scenarios, which may limit accuracy in some applications. The
representation of wall-curvature effects in the training data is also
still very limited.  Third, external errors originating from the SGS
model can degrade overall performance despite the intrinsic accuracy
of the wall model, highlighting the need for co-trained or co-designed
SGS and wall models.

These limitations motivate several directions for future
research. Expanding the building-block training set to include richer
transitional regimes and separated flows with strong
three-dimensionality will improve robustness. Incorporating more
nuanced representations of flow history or path-dependence may also
aid in cases with strong memory effects. Most critically, there is a
need for integrated modeling strategies in which the SGS model and
wall model are developed jointly, ensuring numerical and physical
consistency across the LES solver.

This study represents Part~I of a two-part contribution. In Part~II,
we introduce three additional components that further enhance the
performance and interpretability of the wall model: (i) a flow-regime
classifier that identifies the dominant physics underlying each
prediction and the specific training samples that support it; (ii) a
predictive uncertainty model that decomposes the total error into
epistemic and aleatoric components; and (iii) a confidence-scoring mechanism  using a spectral-normalized
Gaussian process that flags
unreliable predictions and, when desired, corrects them through an
additive error model. These additions render the model more
interpretable and self-aware, enabling practical WMLES applications
with integrated mechanisms for reliability, adaptation, and correction.

\begin{acknowledgments}

This work was supported by the National Science Foundation (NSF) under
grant number \#2317254 and NSF CAREER \#2140775.  The project is also
supported by an Early Career Faculty grant from NASA’s Space
Technology Research Grants Program (grant \#80NSSC23K1498).  The
authors acknowledge the MIT Office of Research Computing and Data for
providing high performance computing resources that have contributed
to the research results reported within this paper. This work also
used the DeltaAI and Delta system at the National Center for
Supercomputing Applications and Bridges-2 at Pittsburgh Supercomputing
Center through allocation PHY250022 from the Advanced
Cyberinfrastructure Coordination Ecosystem: Services and Support
(ACCESS) program, which is supported by National Science Foundation
grants \#2138259, \#2138286, \#2138307, \#2137603, and \#2138296.
  
\end{acknowledgments}

\appendix

\section{DNS of APG/FPG turbulent boundary layers}
\label{sec:app:tbl}

\begin{figure}
    \centering
    \begin{subfigure}[b]{0.8\textwidth}
    \includegraphics[width=\linewidth]{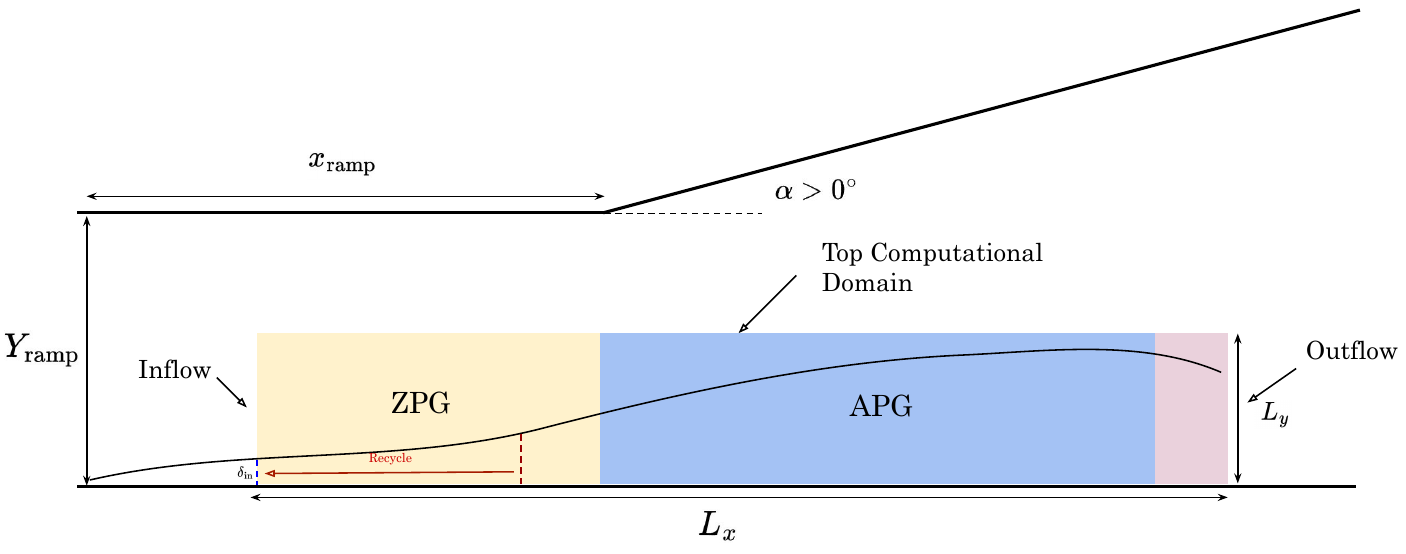}
    \caption{APG TBL.}
    \label{subfig:APG}
    \end{subfigure}
    \begin{subfigure}[b]{0.8\textwidth}
    \includegraphics[width=\linewidth]{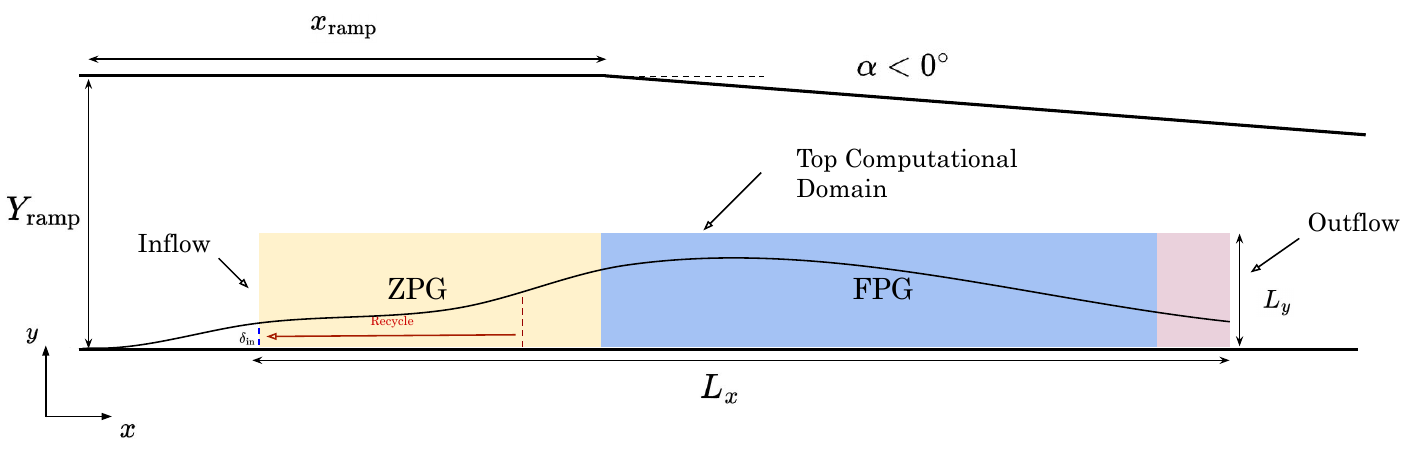}
    \caption{FPG TBL.}
    \label{subfig:FPG}
    \end{subfigure}
    \caption{Schematic of the new DNS of APG/FPG TBLs from
      \citet{Arranz_2025}.}
    \label{fig:TBL_FIG}
\end{figure}

A new set of DNS of APG and FPG turbulent boundary layers was
conducted to fill gaps in the existing literature. The simulation
setup is illustrated in Figure~\ref{fig:TBL_FIG}. We conceptualize a
virtual straight ramp that is deflected either upward or downward to
impose a pressure gradient. An upward deflection ($\alpha >
0^{\circ}$) generates an APG, whereas a downward deflection ($\alpha <
0^{\circ}$) produces an FPG. This two-dimensional ramp geometry is
extruded in the spanwise direction to form a three-dimensional
domain. We investigate a range of deflection angles, $\alpha \in
[-4^{\circ}, -3^{\circ}, -2^{\circ}, -1^{\circ}, 5^{\circ},
  10^{\circ}, 15^{\circ}, 20^{\circ}]$, at two inflow friction
Reynolds numbers, $Re_{\tau, in} = 300$ and $670$, for a total of 16
cases. A detailed analysis of the statistical properties of these TBLs
is provided in \citet{Arranz_2025}; here, we summarize the main
features of the cases.

The domain size is $L_x = 1200\theta_{in}$, $L_y = 200\,\theta_{in}$,
and $L_z = 200\theta_{in}$ in the streamwise, wall-normal, and
spanwise directions, respectively, for the APG cases, and $L_x = 1000
\theta_{in}$, $L_y = 40 \theta_{in}$, and $L_z = 200 \theta_{in}$ for
the FPG cases, where $\theta_{in}$ is the TBL momentum thickness at
the inlet.  The inflow boundary condition is obtained by imposing the
mean velocity profile from DNS of a ZPG TBL at the corresponding
Reynolds number~\cite{sillero2013one}, with superimposed
quasi-periodic velocity fluctuations extracted from a downstream
location at $x_{\text{recycle}} = 100 \theta_{in}$. A convective
outlet boundary condition, $\partial \boldsymbol{u}/\partial t +
U_{\infty} \partial \boldsymbol{u}/\partial x = 0$, is applied at the
domain exit~\citep{Pauley1990}, and small corrections are introduced
to enforce global mass conservation~\citep{Simens2009}.  The spanwise
direction is treated as periodic.  At the top boundary, we apply an
inviscid potential-flow solution obtained by assuming a virtual wall
located further above the domain.  This is achieved by prescribing a
pure potential source for the APG cases or a potential sink for the
FPG cases. The FPG configurations considered here are similar to the
sink-flow simulations of \citet{dixit2008pressure} and
\citet{spalart1986numerical}. The wall-normal velocity is calculated
by
\begin{equation}
    v\big|_{L_y} = \frac{u_{\infty} L_y \left[ (x_{\text{ramp}} -
      x_{\text{source}})^{2} + L_y^{2} \right]}{\left(x_{\text{ramp}}
      - x_{\text{source}}\right) \left[ (x - x_{\text{source}})^{2} +
      L_y^{2} \right]},
\end{equation}
where $x_{\text{source}}$ is determined from the requirement that, at
$x_{\text{ramp}}$, the potential flow must be parallel to the linear
ramp. The source location corresponds to the intersection of the ramp
line with the horizontal bottom wall. It is computed as
\begin{equation}
    x_{\text{source}} = x_{\text{ramp}} - Y_{\text{ramp}}\tan(\alpha).
\end{equation}
The streamwise velocity boundary condition at the top boundary is then
imposed by enforcing
\begin{equation}
    \omega_z \big|_{L_y}
    = \frac{\partial v\big|_{L_y}}{\partial x}
      - \frac{\partial u\big|_{L_y}}{\partial y}
    = 0,
\end{equation}
ensuring that the top boundary remains irrotational. The spanwise
velocity shear, $\partial w/\partial y$, is set to zero at the top of
the computational domain.
\begin{figure}
    \centering
    \begin{subfigure}[b]{0.46\textwidth}
    \includegraphics[width=\linewidth]{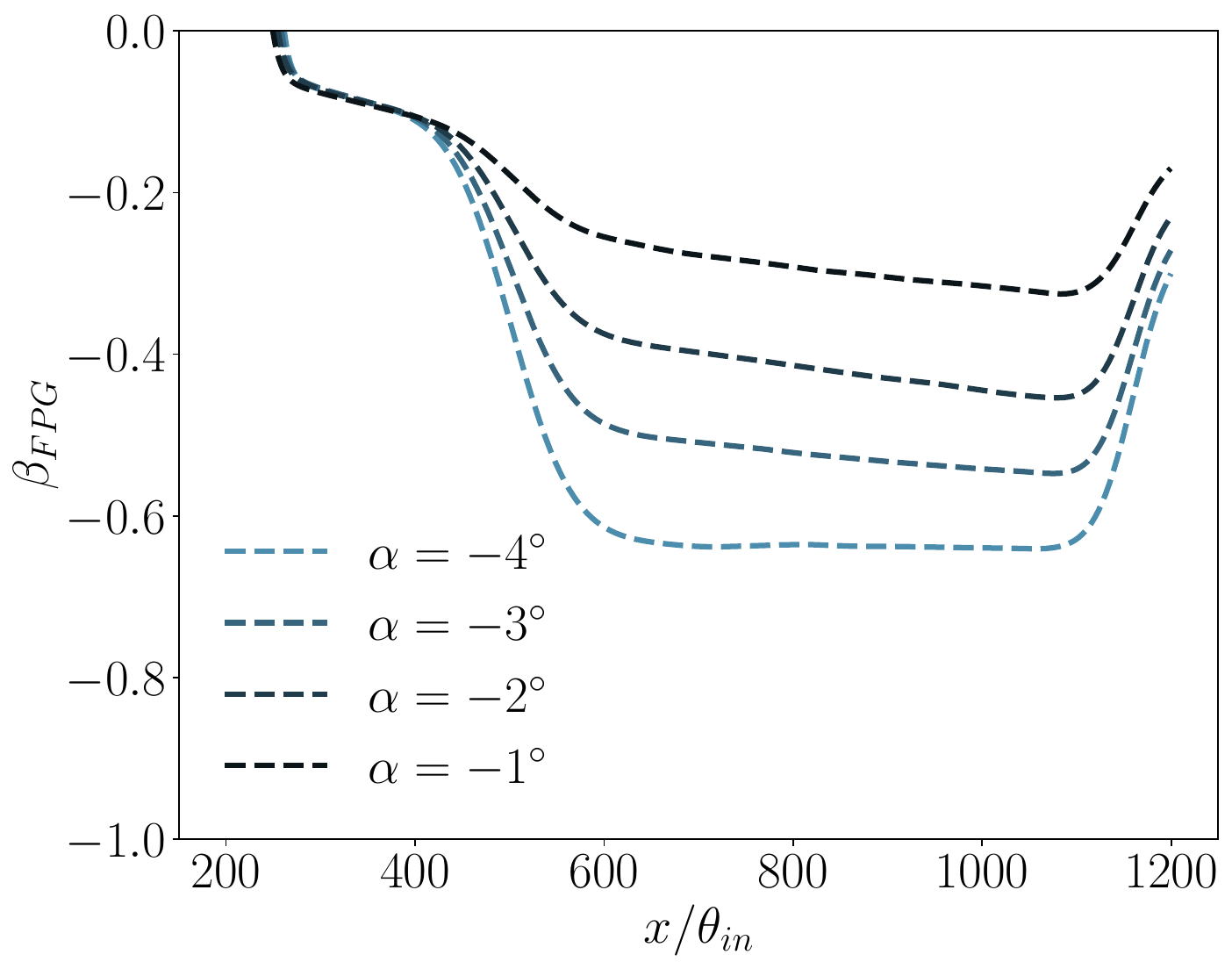}
    \caption{}
    \label{subfig:FPG_beta}
    \end{subfigure}
    \hfill
    \begin{subfigure}[b]{0.455\textwidth}
    \includegraphics[width=\linewidth]{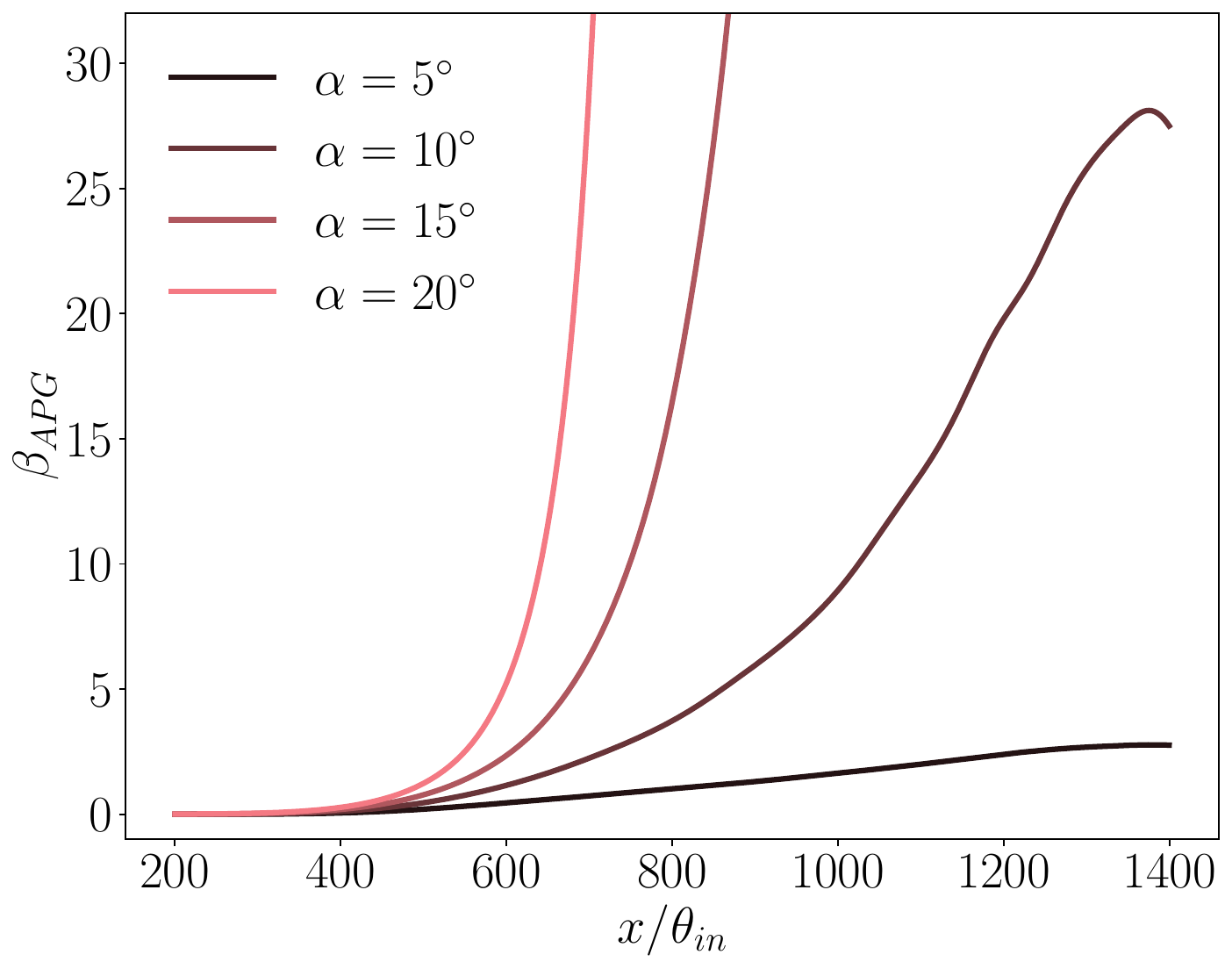}
    \caption{}
    \label{subfig:APG_beta}
    \end{subfigure}
    
    \begin{subfigure}[b]{0.48\textwidth}
    \includegraphics[width=\linewidth]{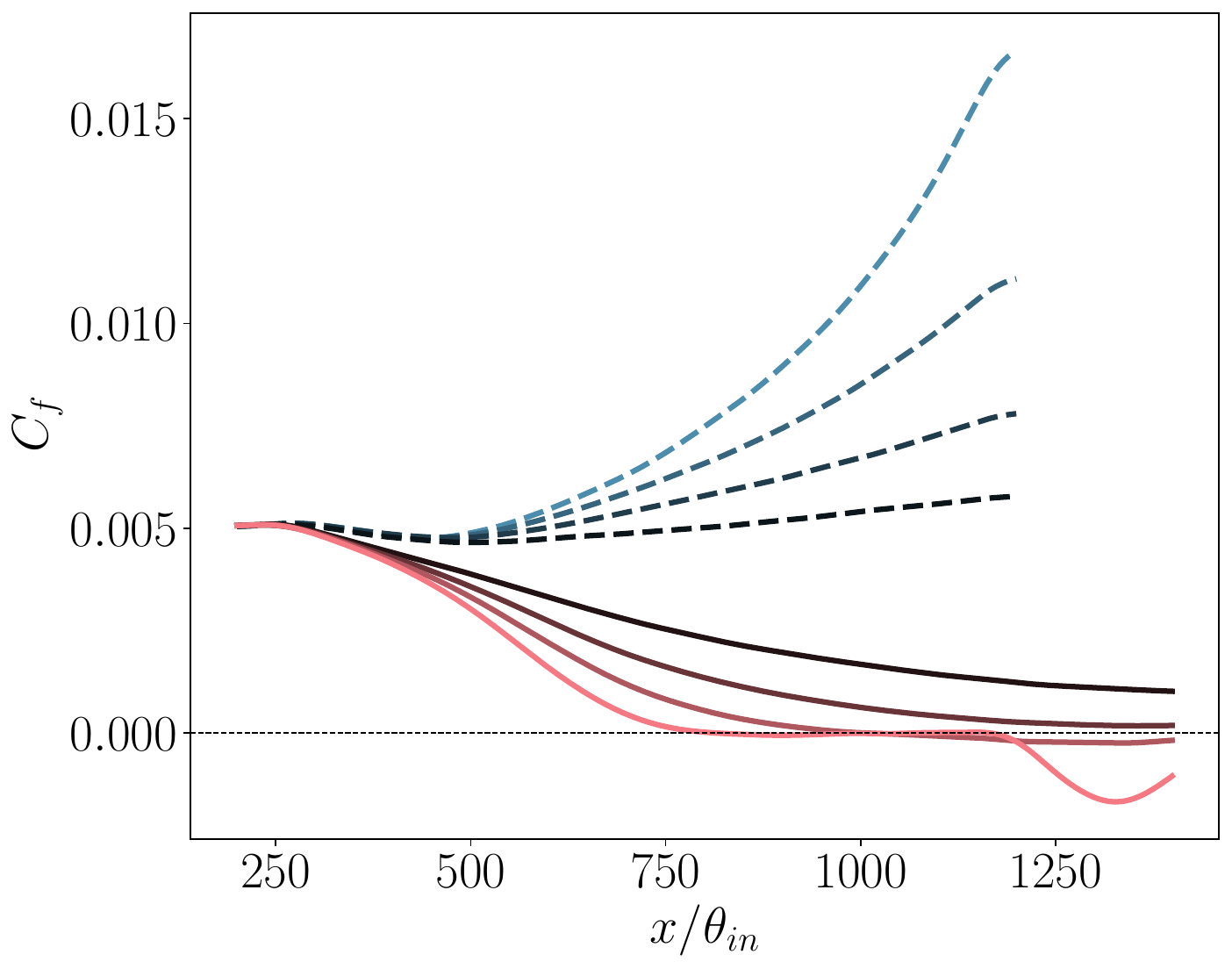}
    \caption{}
    \label{subfig:FPG_retheta}
    \end{subfigure}
    \hfill
    \begin{subfigure}[b]{0.48\textwidth}
    \includegraphics[width=\linewidth]{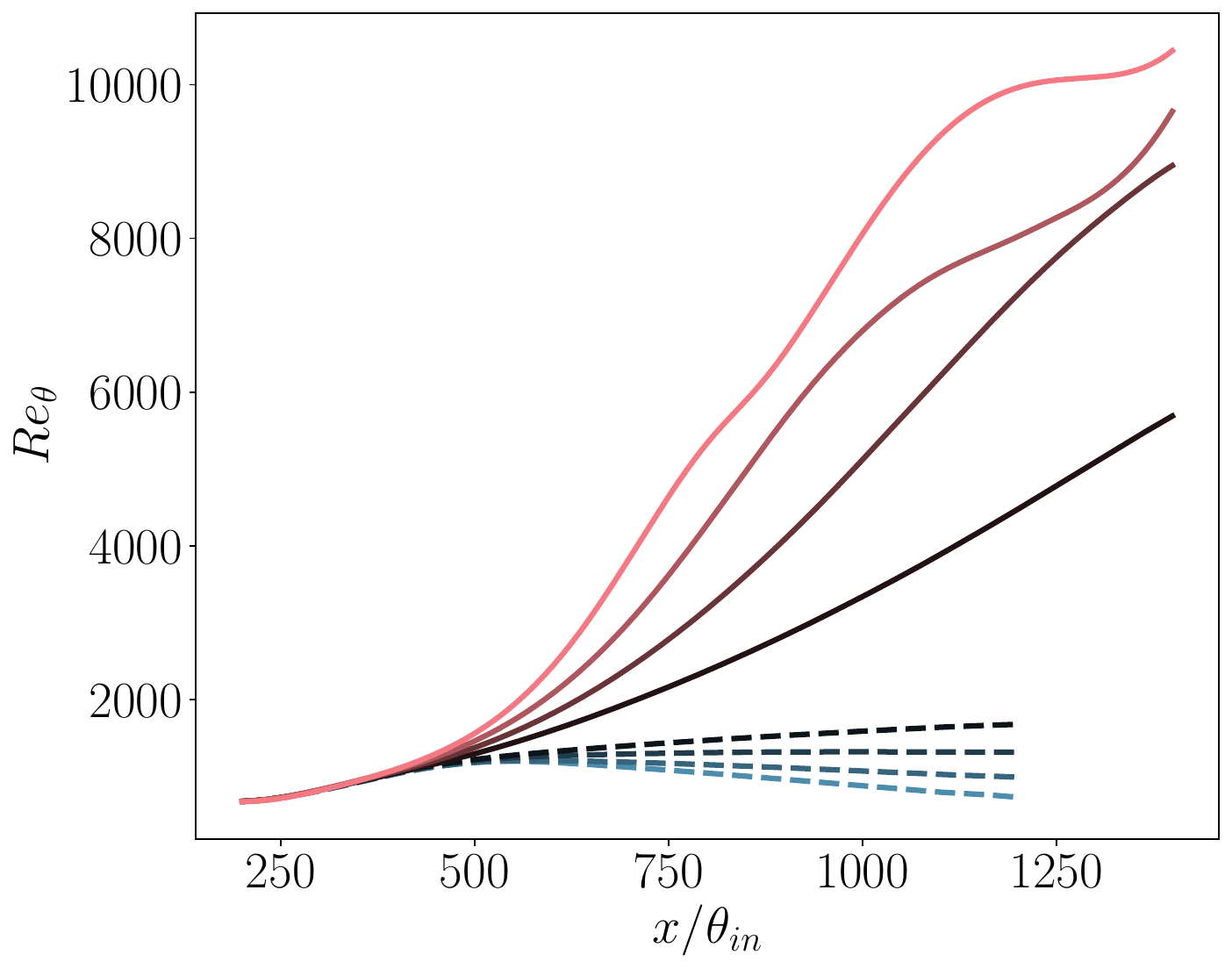}
    \caption{}
    \label{subfig:APG_retheta}
    \end{subfigure}
    \caption{Clauser pressure-gradient parameter $\beta$ for (a) FPG
      and (b) APG cases. (c) Friction coeffcient $C_f$ and (d)
      Momentum-thickness-based Reynolds numbers $Re_{\theta}$ for all
      cases.}
    \label{fig:TBL_stats}
\end{figure}

The solutions are computed by DNS of the incompressible Navier-Stokes
equations. The spatial discretization is a staggered second-order
central finite difference scheme \cite{orlandi2000fluid}. Time
advancement is achieved by a third-order Runge-Kutta scheme
\cite{wray1990minimal} combined with the fractional-step method
\cite{kim1985application}. The code has been validated in previous
studies in turbulent channel flows \cite{bae2018turbulence,
  lozano2020non}, zero-pressure-gradient turbulent boundary layer
\cite{towne2023database} and transitional boundary layers
\cite{Lozano2018_PSE}.  

Some statistical properties of the simulations are shown in
Figure~\ref{fig:TBL_stats}, including the Clauser parameter $\beta$,
the friction coefficient $C_f$, and the momentum-thickness Reynolds
number $Re_\theta$. The momentum thickness is computed following the
method of \citet{spalart1993experimental}. Table~\ref{tab:db_comp}
summarizes the ranges of $\beta$ and $Re_\theta$ in the present TBL
dataset and compares them with those from previous databases in the
literature. The results show that our dataset spans a substantially
wider Reynolds-number range and encompasses pressure-gradient
conditions from strong FPG ($\beta < 0$) to near-incipient separation
($\beta \rightarrow \infty$).
\begin{table}
    \centering
    \caption{Comparison of current and previous APG/FPG TBL databases
      with current cases. $Re_{\theta}=U_{\infty}\theta/\nu$ is the
      momentum-thickness-based Reynolds number and $\beta$ is the
      Clauser pressure-gradient parameter.}
\begin{tabular}{p{2.5cm}p{3.3cm}llll}
\toprule
Case & Source & Type & Reynolds number range & $\beta$ & Case count\\ 
\midrule
APG & \citet{bobkeHistoryEffectsEquilibrium2017} & WRLES & $910<R e_\theta<4320$ & [0.86, 4.07] & 5 \\
\midrule
Wing & \citet{hosseiniDirectNumericalSimulation2016,tanarroEffectAdversePressure2020} & WRLES/DNS & $790<\operatorname{Re}_\theta<3000$ & [0.61, 85] & 5\\
\midrule
ZPG & \citet{schlatter2009turbulent,schlatter2010assessment} & DNS & $500<R e_\theta<4300$ & $\simeq 0$ & 2 \\
\midrule
ZPG & \citet{eitel2014simulation} & WRLES & $600<R e_\theta<8300$ & $\simeq 0$ & 1 \\
\midrule
ZPG & \citet{sillero2013one} & DNS & $2780<R e_\theta<6600$ & $\simeq 0$ & 1 \\
\midrule
APG followed \newline by FPG & \citet{gungor2022energy} & DNS & $1238<R e_\theta<12970$ & [$\sim 0$, $\infty$] & 3 \\
\midrule
FPG and APG & Current & DNS & $300<R e_\theta<10000$ & [-0.6, $\infty$] & 16 \\
\bottomrule
\end{tabular}\label{tab:db_comp}
\end{table}

\section{Effect of using instantaneous inputs instead of averaged inputs}
\label{Sec:App:Instant}

We examine the impact of using instantaneous input values rather than
averaged inputs in BFM. Most wall models, such as EQWM, are calibrated
or trained on mean quantities—i.e., given the mean velocity properties
at the matching location, they return the mean wall-shear stress.
Consider an exact-for-the-mean wall model of the form
\begin{equation}\label{eq:app:true_wm}
    \tau_{w,\text{true}} = f(\overline{\mathbf{q}};\theta,h_{\text{wm}}),
\end{equation}
where $f$ denotes the functional form of the wall model, $\mathbf{q}$
is the vector of inputs at the matching location, $\overline{(\cdot)}$
denotes an ensemble average, $\theta$ represents the model parameters,
and $h_{\text{wm}}$ is the matching height. In \textit{a posteriori}
testing, the model takes as inputs instantaneous values and the
predicted mean wall-shear stress is given by
\begin{equation}\label{eq:app:pred_wm}
    \tau_{w,\text{pred}}
    = \overline{f(\mathbf{q};\theta,h_{\text{wm}})},
\end{equation}
which does not need to coincide with $\tau_{w,\text{true}}$ as the
inputs $\mathbf{q}$ are now instantaneous. The relative error is
defined as~\cite{LarssonBlog}
\begin{equation}
    \varepsilon = \frac{\tau_{w,\text{pred}} -
      \tau_{w,\text{true}}}{\tau_{w,\text{true}}} =
    \frac{\overline{f(\mathbf{q}; \theta, h_{\text{wm}})}}
         {f(\overline{\mathbf{q}}; \theta, h_{\text{wm}})} - 1,
\end{equation}
assuming $\tau_{w,\text{true}} \neq 0$. As shown by
\citet{LarssonBlog}, using instantaneous rather than averaged inputs in
EQWM introduces only very small relative errors (below 2\%). This
result—derived using Reichardt’s law and DNS Reynolds-stress profiles
via a Taylor-expansion analysis—provides indirect justification for the
use of instantaneous flow information in wall models within an
\textit{a posteriori} context.

We estimate the modeling error in BFM under the simplifying assumption
that $\mathbf{q}$ follows a Gaussian distribution. Although
$\mathbf{q}$ is generally not
Gaussian~\cite{dinavahi1995universality}, this exercise is intended to
illustrate why EQWM is comparatively less sensitive to errors induced
by instantaneous inputs. For clarity, we restrict the analysis to a
case with friction Reynolds number $Re_{\tau}=4,200$.

\begin{figure}
    \centering
    \begin{subfigure}[b]{0.48\textwidth}
        \includegraphics[width=\linewidth]{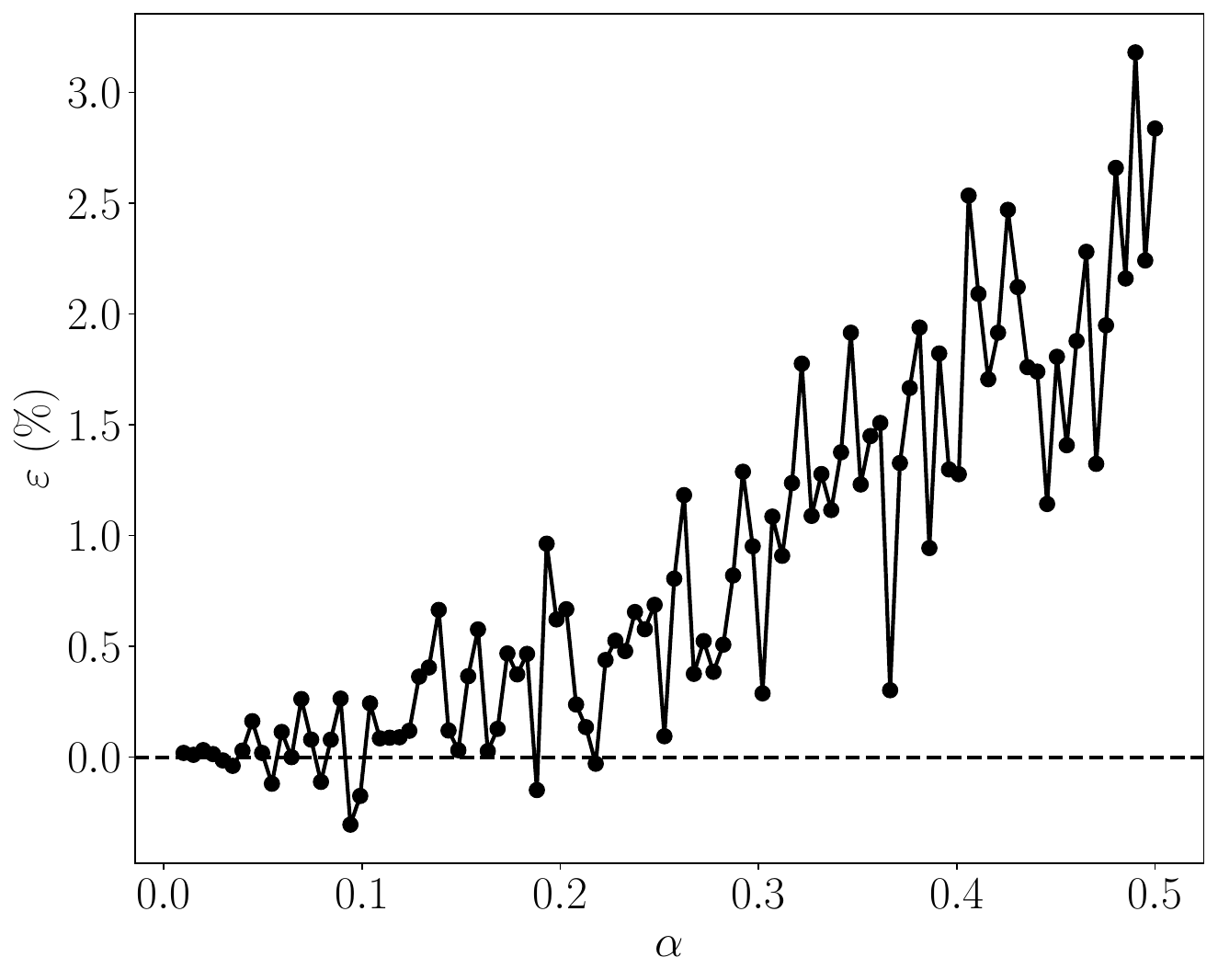}
        \caption{EQWM relative error}
        \label{fig:sub:ins_eqwm}
    \end{subfigure}
    \hfill
    \begin{subfigure}[b]{0.48\textwidth}
        \includegraphics[width=\linewidth]{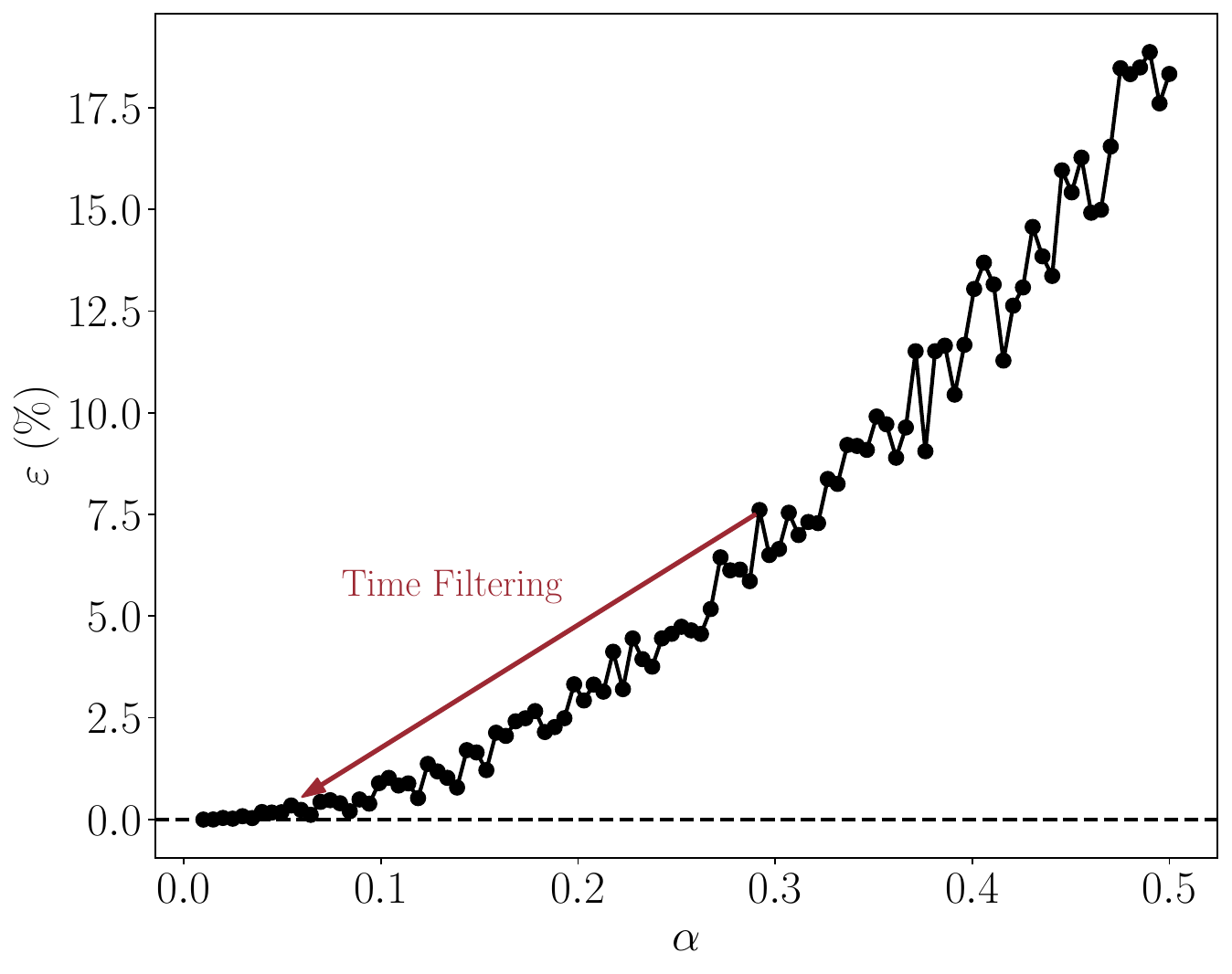}
        \caption{BFM relative error}
        \label{fig:sub:ins_bfm}
    \end{subfigure}
    \caption{Relative error induced by using instantaneous inputs for
      (a) EQWM and (b) BFM. The time filtering is used to reduce the
      scattering of the instantaneous inputs shown by the red arrow.}
    \label{fig:bfm_eqwm_error}
\end{figure}

\begin{figure}
    \centering
    \includegraphics[width=0.5\linewidth]{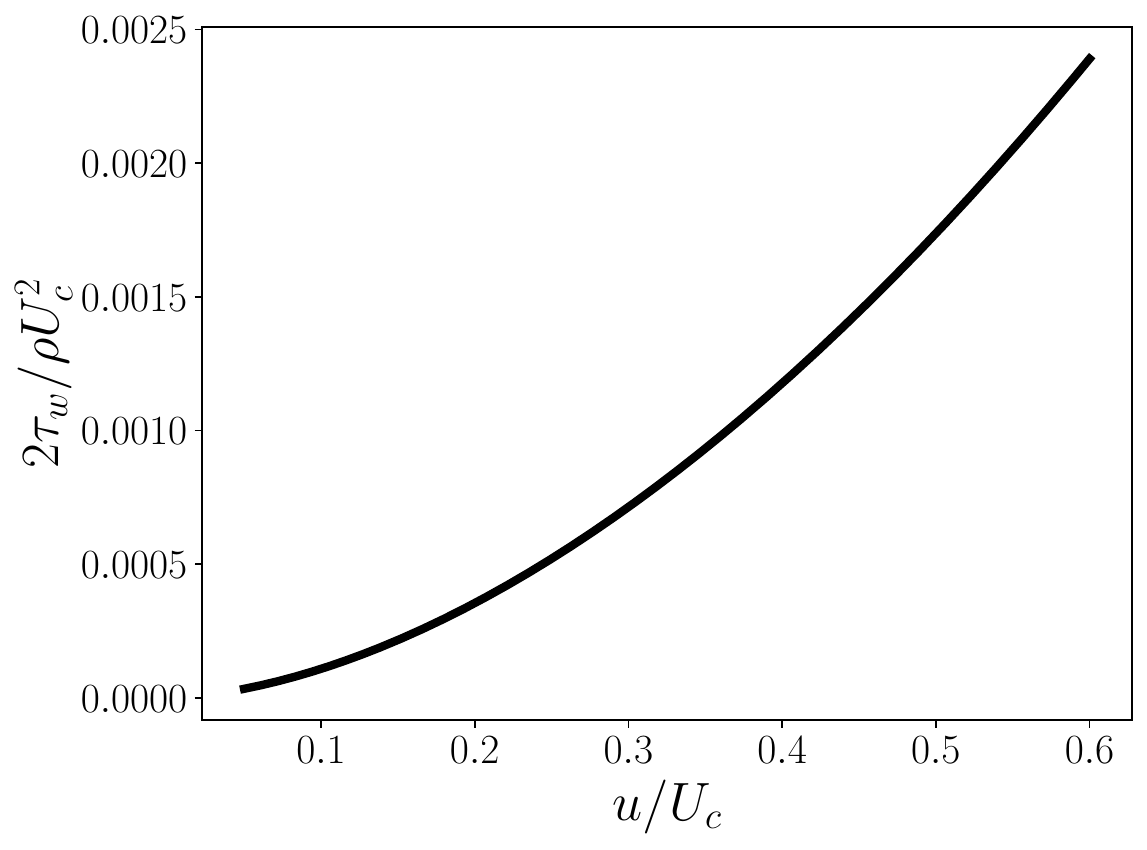}
    \caption{Wall-shear stress predicted by the EQWM as a function of
      input velocity. $U_c$ is the centerline velocity for the
      turbulent channel flow.}
    \label{fig:app:eqwm}
\end{figure}

For EQWM, the input $\mathbf{q}$ reduces to a single scalar $u$, the
wall-parallel velocity magnitude at the matching location. We assume
$u \sim \mathcal{N}(\overline{u}, (\alpha_u \overline{u})^2)$, where
$\alpha_u$ denotes the ratio of the standard deviation to the mean
velocity. For each $\alpha_u \in [0.01, 0.5]$ (incremented by 0.005),
we generated 100,000 samples and evaluated the mean prediction using
Eq.~\ref{eq:eqwm}. As shown in Figure~\ref{fig:sub:ins_eqwm}, the
error introduced by instantaneous inputs remains below 3\%, even when
$\alpha_u$ reaches 0.5. This finding is consistent with prior work
demonstrating that EQWM is relatively insensitive to such
errors~\cite{LarssonBlog}. The primary reason for this robustness is
that although EQWM (Eq.~\ref{eq:eqwm}) is nonlinear, it is monotonic
and contains only mild departures from linearity, as illustrated in
Figure~\ref{fig:app:eqwm}.

We conducted a similar analysis for BFM. For simplicity, we set the
pressure-gradient input to zero and examined how the joint variability
of $u_1$ and $u_2$ affects the predicted output. We assume they follow
the joint distribution
\begin{equation}
\binom{u_1}{u_2} \sim \mathcal{N}\!\left(
\binom{\overline{u_1}}{\overline{u_2}},
\begin{pmatrix}
(\alpha_u\overline{u_1})^2 & \alpha_u^2\overline{u_1}\overline{u_2} \\
\alpha_u^2\overline{u_1}\overline{u_2} & (\alpha_u\overline{u_2})^2
\end{pmatrix}
\right).
\end{equation}
We then evaluate how the error $\varepsilon$ varies with $\alpha_u$.
The results, shown in Figure~\ref{fig:sub:ins_bfm}, reveal that an
$\alpha_u = 0.5$ leads to substantially larger errors—up to 20\%---in
contrast to the much smaller errors observed for EQWM. This indicates
that directly using instantaneous inputs may results in inaccurate
predictions in BFM. As discussed in the main text, a time filter is
applied to reduce $\alpha_u$ and improve the relative error. For
models involving multiple inputs, such filtering is generally
necessary to maintain robust performance. As highlighted by the red
arrow in Figure~\ref{fig:sub:ins_bfm}, the primary effect of the time
filter is to reduce the standard-deviation coefficient $\alpha_u$,
mitigating the induced modeling error.

\section{Effect of timescale used for time filtering}
\label{Sec:App:TimeFilter}

\begin{figure}
    \centering
    \begin{subfigure}[b]{0.48\textwidth}
    \centering
    \includegraphics[width=\textwidth]{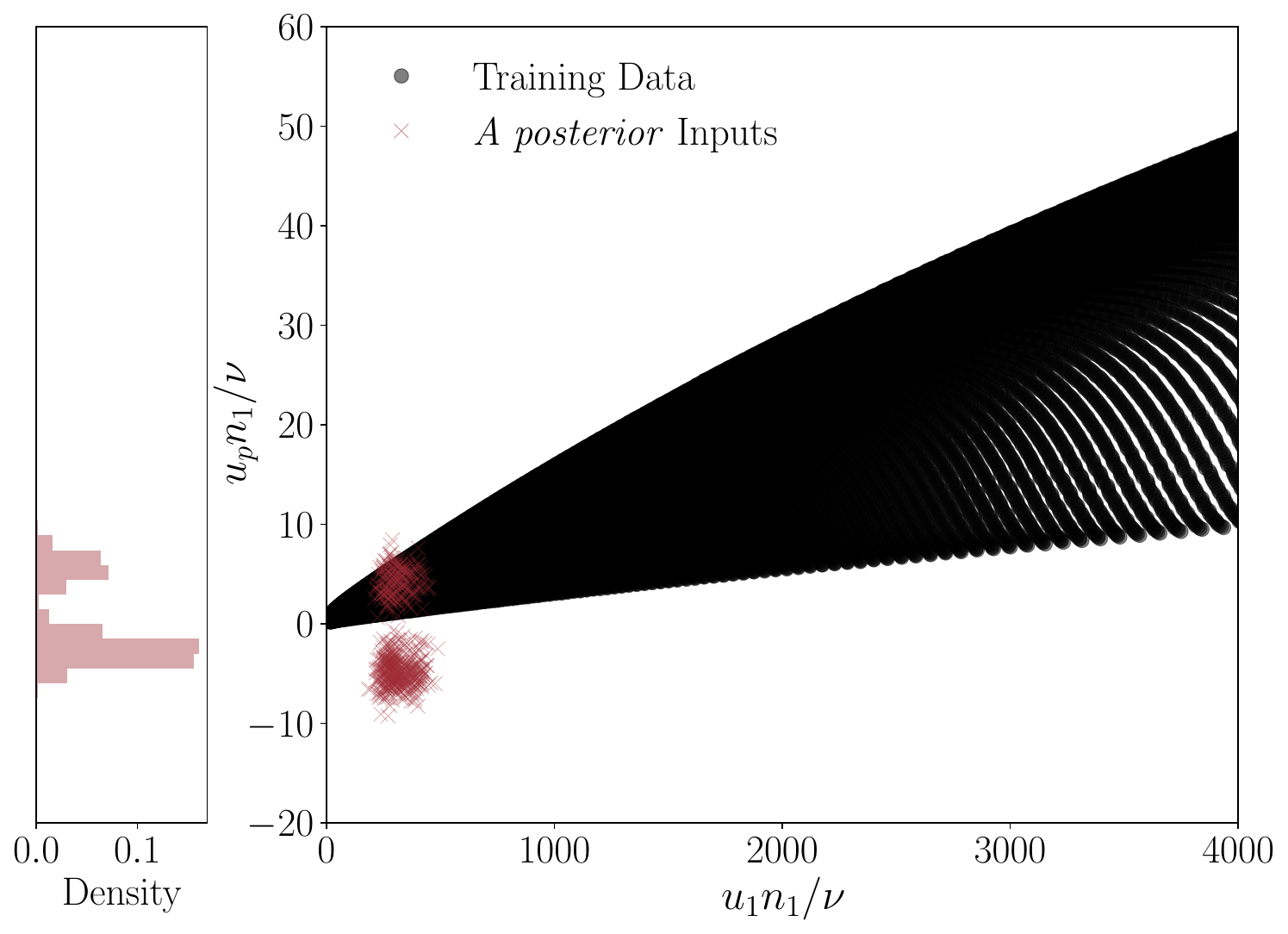}
    \caption{$T_f=0.1/\|\mathbf{S}_1\|$}
    \label{fig:sub:TBL5_small_Tf}
    \end{subfigure}
    \begin{subfigure}[b]{0.48\textwidth}
    \centering
    \includegraphics[width=\textwidth]{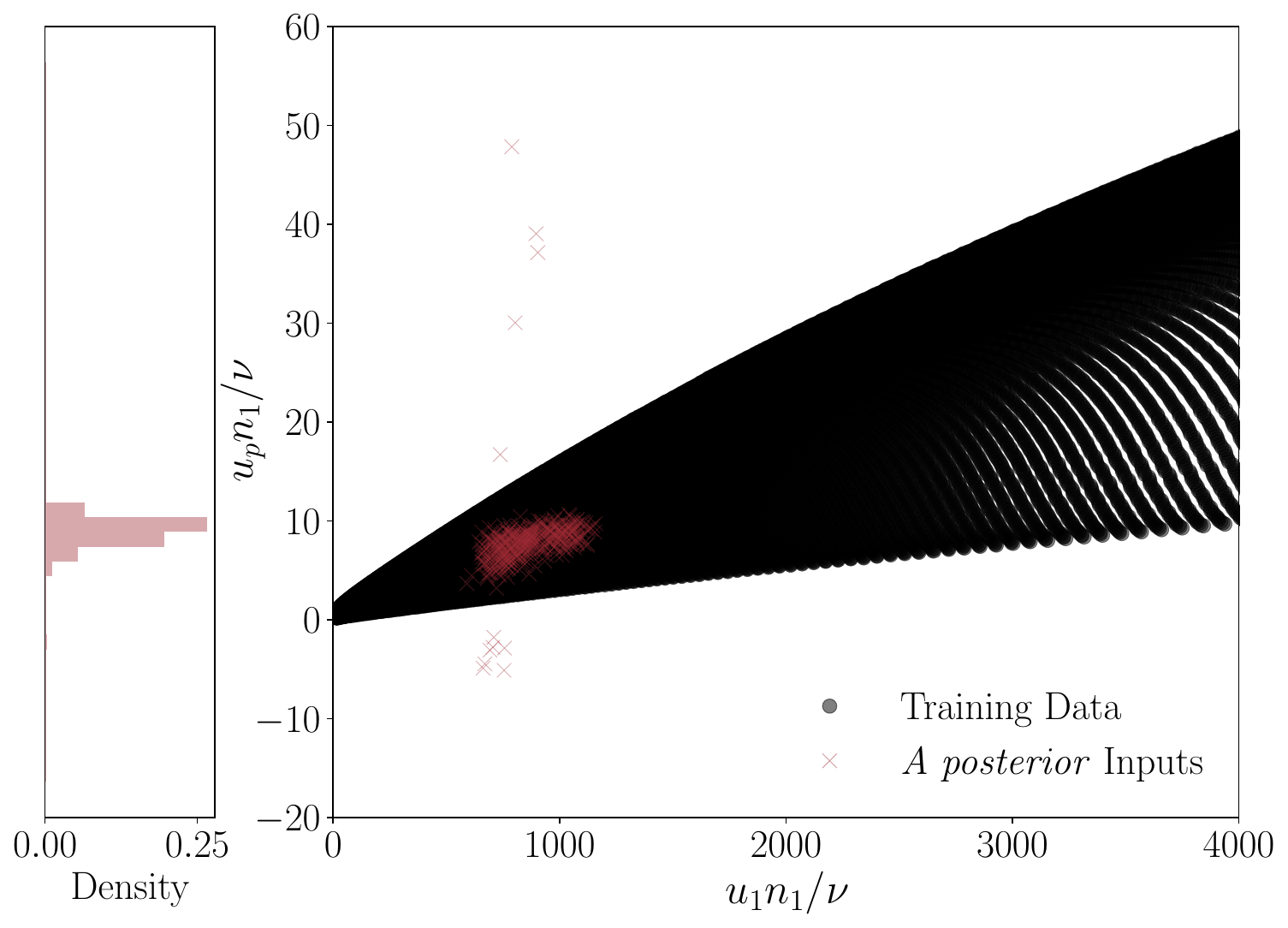}
    \caption{$T_f=100/\|\mathbf{S}_1\|$}
    \label{fig:sub:TBL5_large_Tf}
    \end{subfigure}
    \caption{Comparison between model inputs fed into the model in
      \textit{a posteriori} testing and training data for a turbulent
      channel flow with constant (a) $C=0.1$ and (b) $C=100$.}
    \label{fig:timefilter}
\end{figure}

A time filter is applied to the velocity $u$ and pressure $p$ to
obtain the BFM inputs. The filter uses a timescale $T_f$ defined as
$T_f = C / \|\mathbf{S}_1\|$, where $\|\mathbf{S}_1\|$ is the
magnitude of the rate-of-strain tensor at the first matching
location. The choice of the coefficient $C$ is critical. If $C$ is too
small, the resulting timescale $T_f$ becomes too short, leading to
large oscillations in the pressure gradient and, consequently,
incorrect model inputs. As shown in
Figure~\ref{fig:sub:TBL5_small_Tf}, when $C = 0.1$ the
pressure-gradient input oscillates between positive and negative
values, with more than half of the samples spuriously negative. When
the timescale is increased by a factor of 1000, as shown in
Figure~\ref{fig:sub:TBL5_large_Tf}, nearly all points fall within the
correct training range.

\begin{figure}
    \centering
    \includegraphics[width=0.5\linewidth]{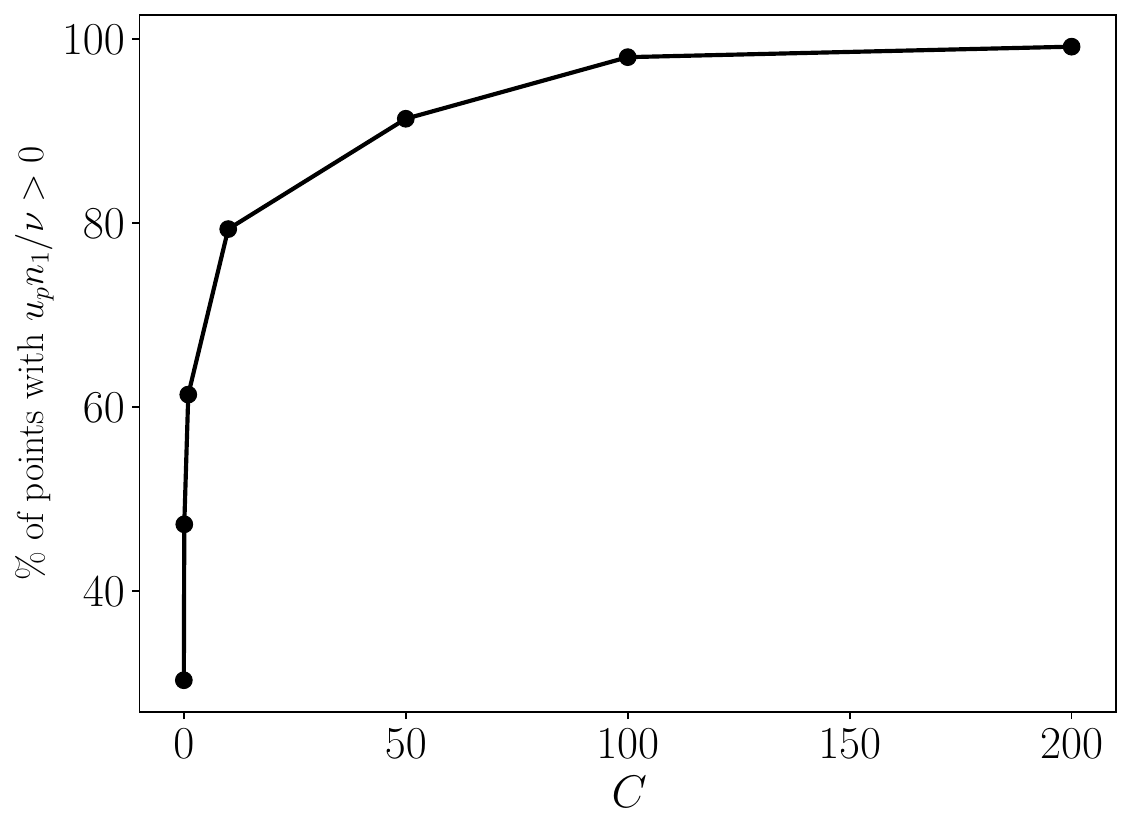}
    \caption{Percentage of points falling into the region $u_p
      n_1/\nu>0$ as a function of $C$.}
    \label{fig:C_vs_ratio}
\end{figure}

There is, however, a trade-off: choosing an excessively large $T_f$
slows the convergence of the simulation. An optimal value of $C$ must
therefore balance input smoothness with computational cost. To
identify this optimum, we varied $C$ and recorded the percentage of
inputs in the \textit{a posteriori} testing that remained within the
region $u_p n_1/\nu>0$, which is physically consistent with the exact
DNS input. As shown in Figure~\ref{fig:C_vs_ratio}, once $C$ reaches
100, more than 98\% of the points lie close to the DNS mean value, and
further increases yield diminishing returns. Accordingly, we set $C =
100$ for this study.

After setting the filtering timescale, a one-sided exponential filter
is applied to the primal variables $u$ and $p$. The filtered variable
$\Phi$ is updated at each time step using the formula:
\begin{equation}
    \Phi(t) =  \epsilon_f \phi(t) + (1-\epsilon_f)\Phi(t-\Delta t),
\end{equation}
where $\phi$ is the instantaneous variable, $\Phi(t-\Delta t)$ is the
filtered variable from the previous time step, $\Delta t$ is the
simulation time step size, and the filtering coefficient $\epsilon_f$
is defined as $\epsilon_f = \Delta t/T_f$.  Finally, these
time-filtered dimensional variables are processed to generate the
model inputs.

\section{Effect of wall-curvature training data}
 \label{Sec:App:Curve}

\begin{figure}
    \centering
    \includegraphics[width=0.7\linewidth]{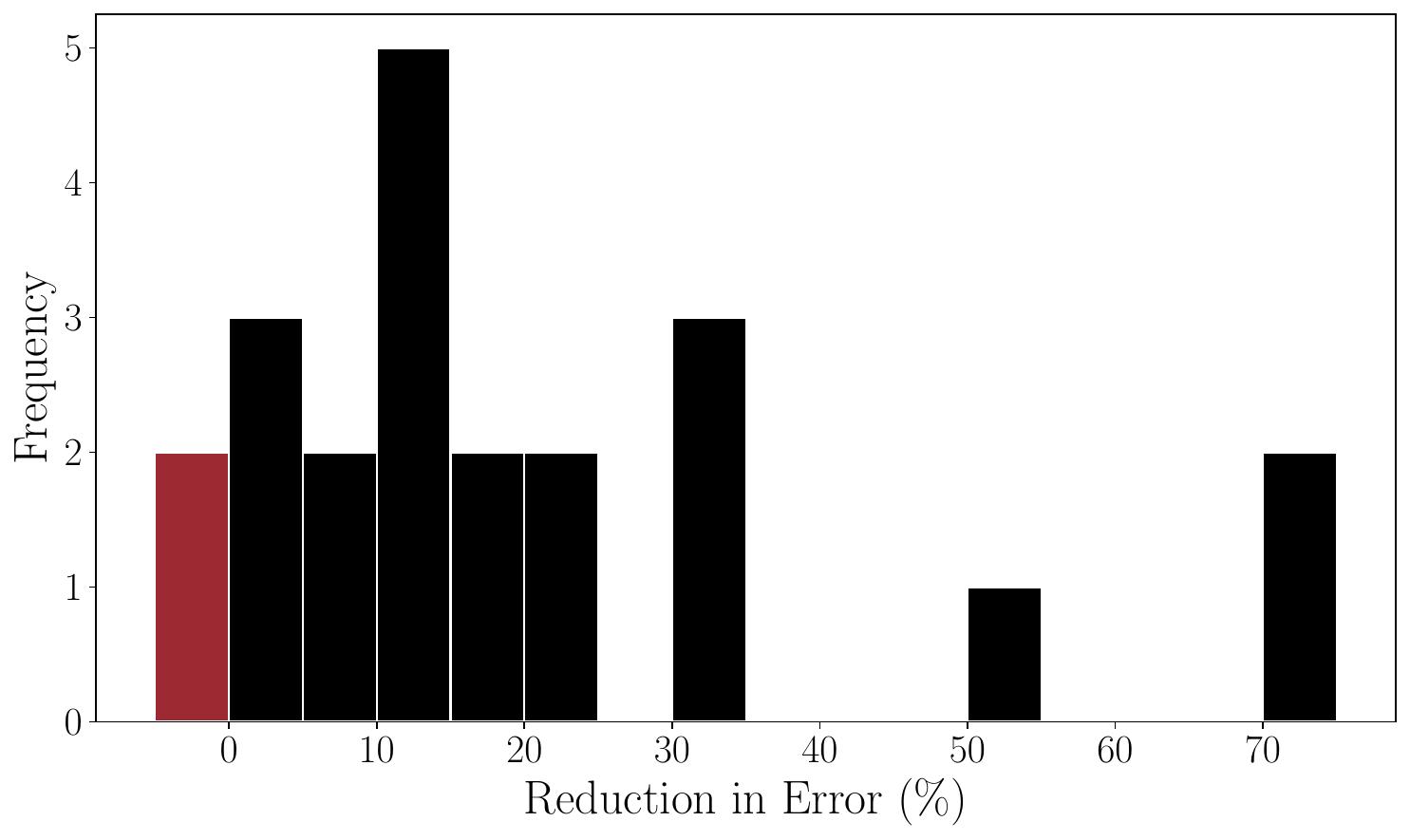}
    \caption{Reduction in relative error for cases with wall curvature
      when curvature-containing training data (i.e., the Gaussian
      bump) is included in the training set. The vertical axis
      indicates the number of cases.  The red bar highlights the two
      cases for which incorporating curved-wall training data resulted
      in reduced prediction accuracy.}
    \label{fig:GP_excl}
\end{figure}

We examine how the model performance is affected when the Gaussian
bump case, which is the only training case that includes
wall-curvature effects, is removed from the training dataset. This
analysis assesses the importance of incorporating turbulence over
curved walls into the training set. A new model was trained using the
same architecture and hyperparameters described in the main text, but
excluding the Gaussian bump data. Its accuracy was then compared with
that of the full model for flows involving wall curvature. As shown in
Figure~\ref{fig:GP_excl}, when the curvature-containing training data
are included, all but two curvature cases exhibit reduced prediction
error. This demonstrates that including curvature effects in the
training dataset is essential for achieving accurate performance
across flow regimes involving wall curvature.

\section{Effect of SGS models on Gaussian bump results}
\label{subsec:SGS}

\begin{figure}
    \centering
    \includegraphics[width=\linewidth]{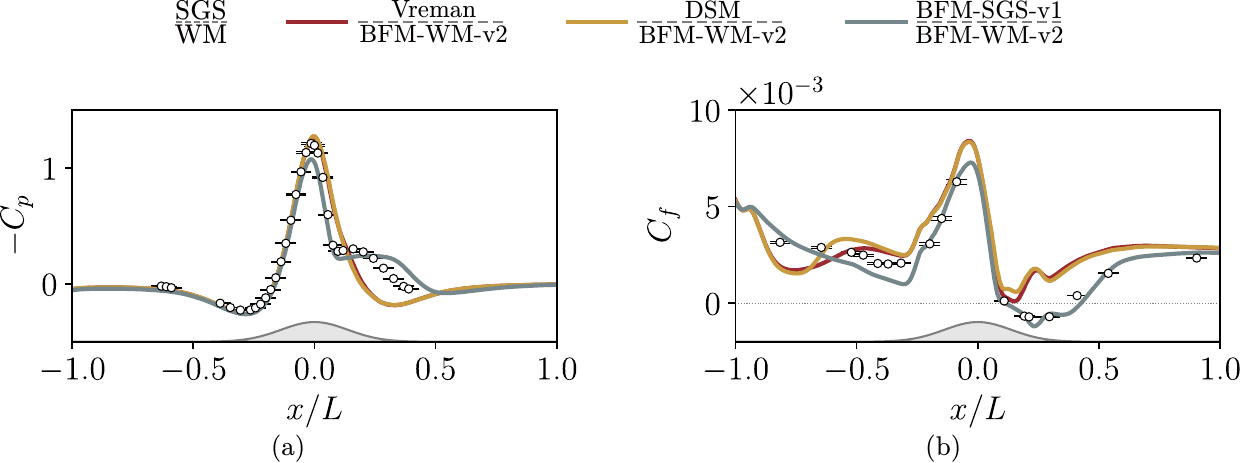}
    \caption{\textit{A posteriori} testing for the Gaussian bump. (a)
      Pressure coefficient and (b) friction coefficient in the plane
      $z/L = 0$. All three curves use BFM-WM-v2 as the wall model. The
      red line corresponds to the Vreman SGS model, the yellow line to
      DSM, and the green line to the BFM-SGS-v1 model from
      \citet{arranzBuildingblockflowComputationalModel2024}.  }
    \label{fig:GP_Results_SameBFWM}
\end{figure}

\begin{figure}
    \centering
    \includegraphics[width=\linewidth]{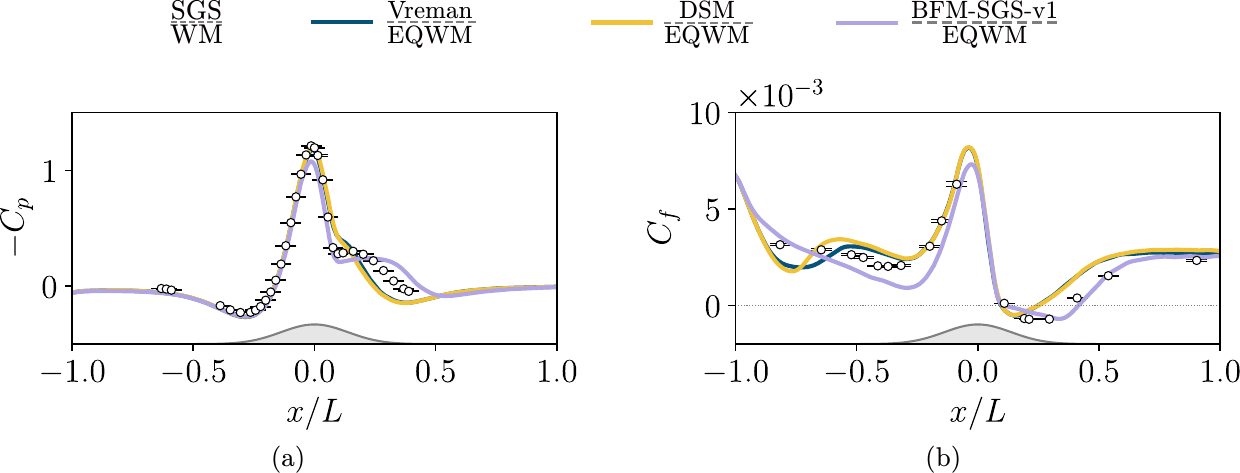}
    \caption{\textit{A posteriori} testing for the Gaussian bump. (a)
      Pressure coefficient and (b) friction coefficient in the plane
      $z/L = 0$. All three curves use EQWM as the wall model. The blue
      line corresponds to the Vreman SGS model, the yellow line to
      DSM, and the purple line to the BFM-SGS-v1 model from
      \citet{arranzBuildingblockflowComputationalModel2024}.}
    \label{fig:GP_Results_SameEQWM}
\end{figure}

The influence of the SGS model on the \textit{a posteriori} performance
for the Gaussian bump case is shown in
Figure~\ref{fig:GP_Results_SameBFWM} and
Figure~\ref{fig:GP_Results_SameEQWM}. The results correspond to
simulations using the same grid described in
Section~\ref{Sec:Apost}. Consistent with the observations of
\citet{whitmore2021large} and \citet{agrawal2022non}, the choice of SGS
model has a significant impact on the overall WMLES performance for this
configuration. For example, at the grid resolution examined here, the
Vreman and DSM models yield comparable accuracy when coupled with
BFM-WM-v2, whereas the BFM-SGS-v1 model provides substantially improved
predictions. A more systematic investigation—such as a full grid
convergence study—is required to assess the convergence properties of
BFM-WM-v2 once the forthcoming BFM-SGS-v2 model, designed to be
numerically consistent with the current BFM wall model, becomes
available.

\section{Investigation of the BFM output landscape}

\begin{figure}
    \centering
    \begin{subfigure}[b]{0.48\textwidth}
        \includegraphics[width=\linewidth]{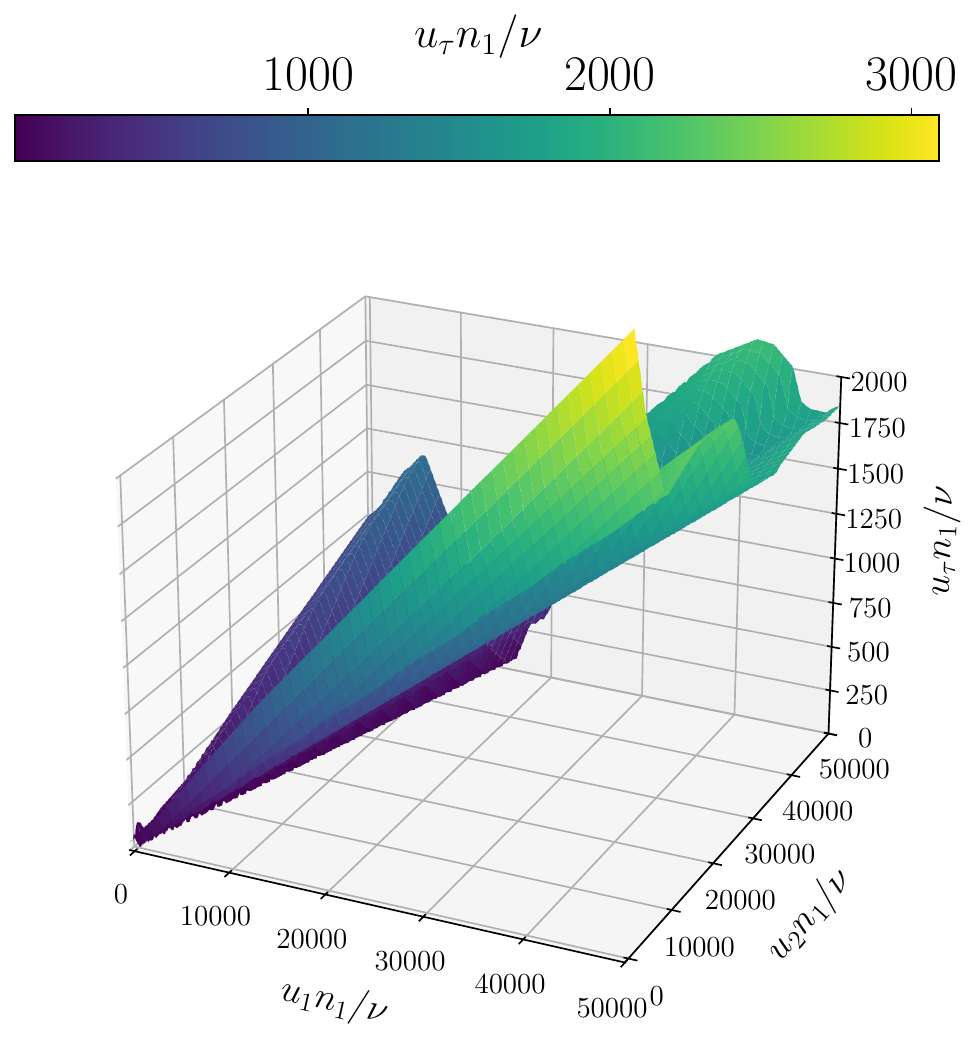}
        \caption{}
        \label{fig:sub:output_p0_3d}
    \end{subfigure}
    \hfill
    \begin{subfigure}[b]{0.48\textwidth}
        \includegraphics[width=\linewidth]{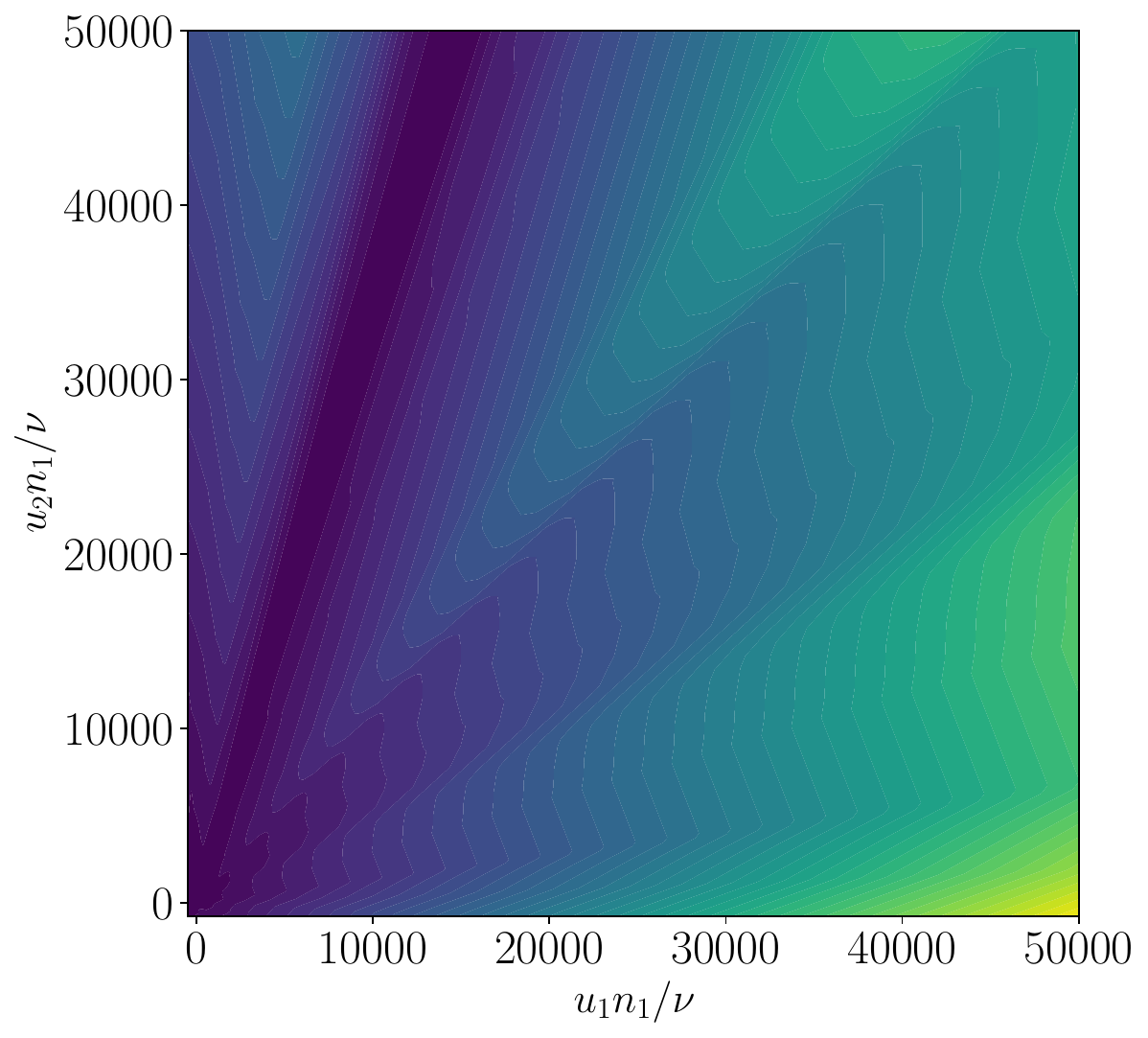}
        \caption{}
        \label{fig:sub:output_p0_2d}
    \end{subfigure}
    \caption{(a) Surface representation of the BFM-WM-v2 output and
      (b) its corresponding 2-D projection, showing the dependence of
      $u_{\tau} n_1 / \nu$ on the inputs $u_1 n_1 / \nu$ and $u_2 n_1
      / \nu$. The third input, $u_p n_1 / \nu$, is held fixed at
      0. The color scale in both panels indicates the output value.}
    \label{fig:output_p0}
\end{figure}

\begin{figure}
    \centering
    \begin{subfigure}[b]{0.48\textwidth}
        \includegraphics[width=\linewidth]{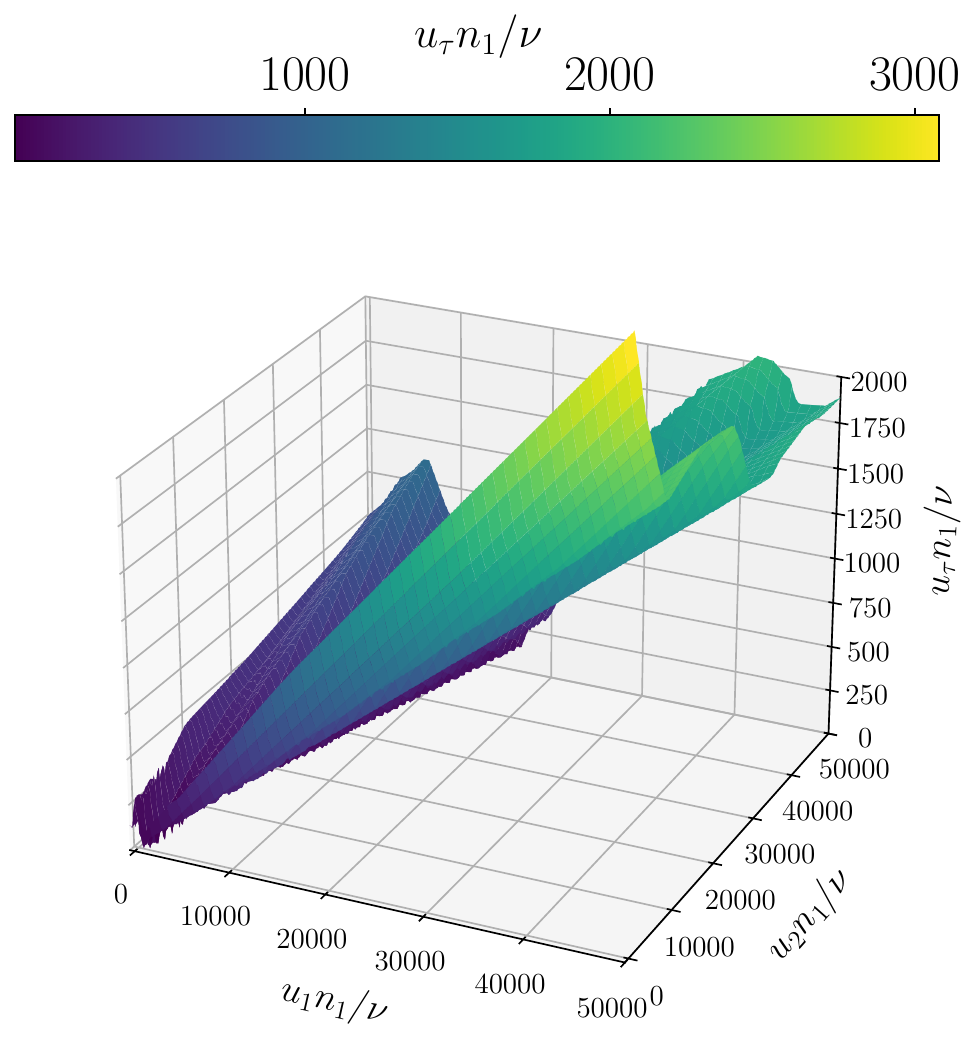}
        \caption{}
        \label{fig:sub:output_p100_3d}
    \end{subfigure}
    \hfill
    \begin{subfigure}[b]{0.48\textwidth}
        \includegraphics[width=\linewidth]{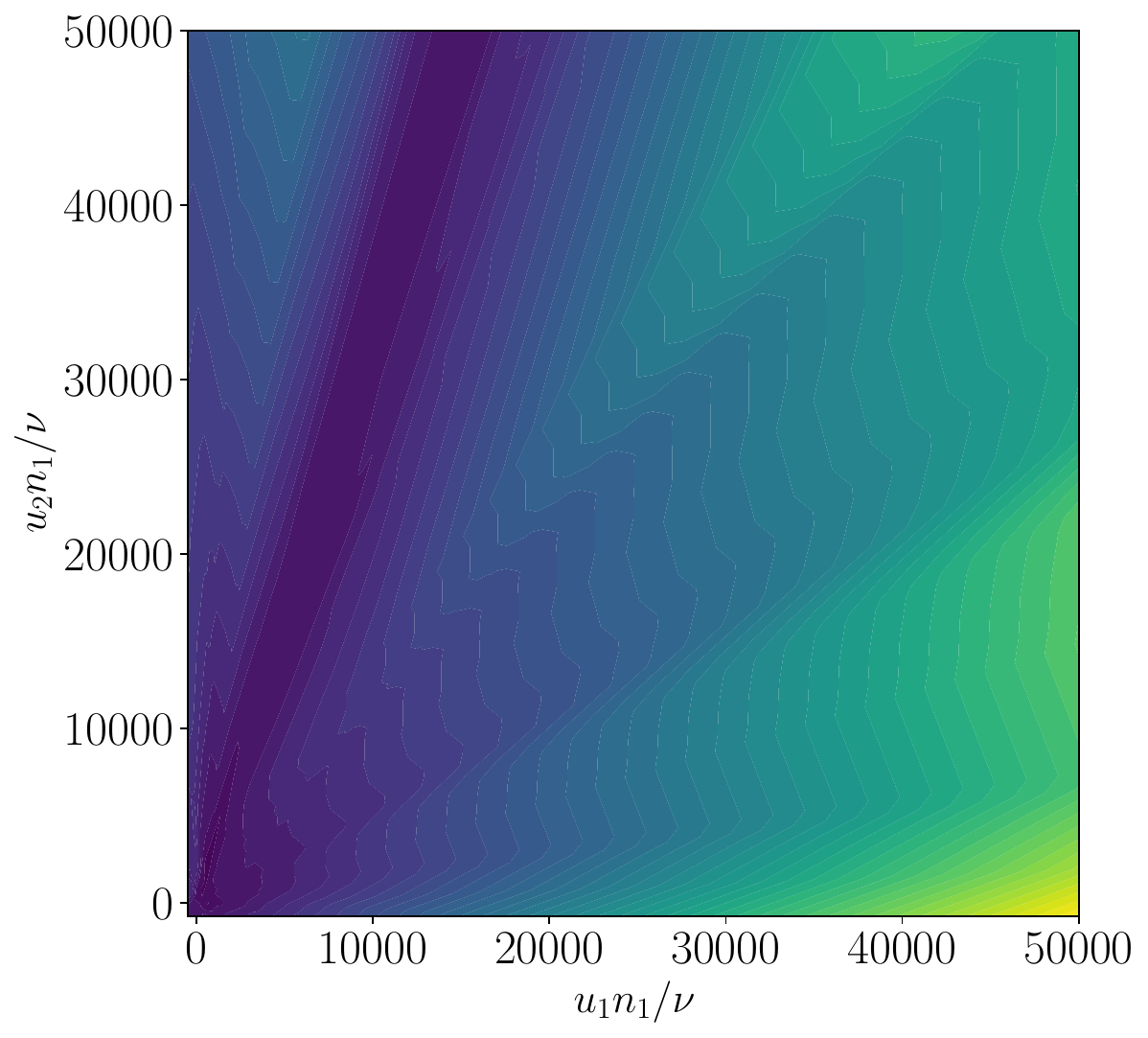}
        \caption{}
        \label{fig:sub:output_p100_2d}
    \end{subfigure}
    \caption{(a) Surface representation of the BFM-WM-v2 output and
      (b) its corresponding 2-D projection, showing the dependence of
      $u_{\tau} n_1 / \nu$ on the inputs $u_1 n_1 / \nu$ and $u_2 n_1
      / \nu$. The third input, $u_p n_1 / \nu$, is held fixed at
      100. The color scale in both panels indicates the output value.}
    \label{fig:output_p100}
\end{figure}

We examine the model behavior in regions not covered by the training
data. Although overfitting was mitigated during training through $L_2$
regularization and early stopping, it is still necessary to assess the
model’s behavior explicitly. Figure~\ref{fig:output_p0} shows the
model output, $u_{\tau} n_1 / \nu$, plotted as a function of $u_1 n_1
/ \nu$ and $u_2 n_1 / \nu$, with the third input $u_p n_1 / \nu$ fixed
at zero, corresponding to the ZPG case. The output surface is smooth
and free of the abrupt peaks or valleys typically associated with
overfitting. A similar plot with $u_p n_1 / \nu = 100$
(Figure~\ref{fig:output_p100}) leads to the same conclusion.

\bibliography{WallModel}

\end{document}

%% file: Figures/Symbols/Right.tex
\tikzset{every picture/.style={line width=0.75pt}} 

\begin{tikzpicture}[yscale=0.5,xscale=0.4]


\draw    (0,0) -- (0.5,0) ;
\draw    (0,0) -- (-0.3,0.4) ;
\draw    (0,0) -- (-0.3,-0.4) ;

\end{tikzpicture}

%% file: Figures/Symbols/ThinDiamond.tex
\tikzset{every picture/.style={line width=0.75pt}} 

\begin{tikzpicture}
  \node[diamond,
    draw = black,
    minimum width = 0.1cm,
    minimum height = 0.6cm] (d) at (0,0) {};
 
\end{tikzpicture}

%% file: Figures/Symbols/Left.tex
\tikzset{every picture/.style={line width=0.75pt}} 

\begin{tikzpicture}[yscale=0.5,xscale=0.5]

\draw    (0,0) -- (-0.5,0) ;
\draw    (0,0) -- (0.3,0.4) ;
\draw    (0,0) -- (0.3,-0.4) ;
\end{tikzpicture}

%% file: Figures/Symbols/Up.tex
\tikzset{every picture/.style={line width=0.75pt}} 

\begin{tikzpicture}[yscale=0.5,xscale=0.5]

\draw    (0,0) -- (0, 0.4) ;
\draw    (0,0) -- (0.3,-0.4) ;
\draw    (0,0) -- (-0.3,-0.4) ;
\end{tikzpicture}

%% file: Figures/Symbols/Down.tex
\tikzset{every picture/.style={line width=0.75pt}} 

\begin{tikzpicture}[yscale=0.5,xscale=0.5]

\draw    (0,0) -- (0, -0.4) ;
\draw    (0,0) -- (0.3,0.4) ;
\draw    (0,0) -- (-0.3,0.4) ;
\end{tikzpicture}

%% file: Figures/Symbols/Cross_Filled.tex
\tikzset{every picture/.style={line width=2pt}}

\begin{tikzpicture}[yscale=0.5,xscale=0.5]

\draw    (-0.25,0.25) -- (0.25, -0.25) ;
\draw    (-0.25,-0.25) -- (0.25, 0.25) ;

\end{tikzpicture}

%% file: Figures/Symbols/Plus_Filled.tex
\tikzset{every picture/.style={line width=2pt}}

\begin{tikzpicture}[yscale=0.5,xscale=0.5]

\draw    (-0.3,0) -- (0.3, 0) ;
\draw    (0,-0.3) -- (0, 0.3) ;

\end{tikzpicture}